\documentclass[aps,prl,reprint,superscriptaddress]{revtex4-2}
\usepackage{graphicx}
\usepackage{dcolumn}
\usepackage{bm}

\begin{document}


  \title{Study of beta spectrum shapes relevant to the prediction of reactor antineutrino spectra}




  \author{G. A. Alcal\'a}
  \email[]{galcala@ific.uv.es}
  \affiliation{Instituto de F\'isica Corpuscular, CSIC-Universitat de Val\`encia, E-46071 Val\`encia, Spain}

  \author{A. Algora}
  \email[]{algora@ific.uv.es}
  \affiliation{Instituto de F\'isica Corpuscular, CSIC-Universitat de Val\`encia, E-46071 Val\`encia, Spain}
  \affiliation{HUN-REN Institute for Nuclear Research (ATOMKI), H-4001 Debrecen, Hungary}

  \author{M. Estienne}
  \affiliation{SUBATECH, CNRS/IN2P3, IMT Atlantique, Nantes Université, F-44307 Nantes, France}

  \author{M. Fallot}
  \affiliation{SUBATECH, CNRS/IN2P3, IMT Atlantique, Nantes Université, F-44307 Nantes, France}

  \author{V. Guadilla}
  \affiliation{SUBATECH, CNRS/IN2P3, IMT Atlantique, Nantes Université, F-44307 Nantes, France}
  \affiliation{Faculty of Physics, University of Warsaw, 02-093 Warsaw, Poland.}
  
  \author{A. Beloeuvre}
  \affiliation{SUBATECH, CNRS/IN2P3, IMT Atlantique, Nantes Université, F-44307 Nantes, France}

  \author{W. Gelletly}
  \affiliation{Department of Physics, University of Surrey, GU27XH Guildford, United Kingdom}
 
  \author{R. Kean}
  \affiliation{SUBATECH, CNRS/IN2P3, IMT Atlantique, Nantes Université, F-44307 Nantes, France}

  \author{A. Porta}
  \affiliation{SUBATECH, CNRS/IN2P3, IMT Atlantique, Nantes Université, F-44307 Nantes, France}

  \author{S. Bouvier}
  \affiliation{SUBATECH, CNRS/IN2P3, IMT Atlantique, Nantes Université, F-44307 Nantes, France}

  \author{J.-S. Stutzmann}
  \affiliation{SUBATECH, CNRS/IN2P3, IMT Atlantique, Nantes Université, F-44307 Nantes, France}

  \author{E. Bonnet}
  \affiliation{SUBATECH, CNRS/IN2P3, IMT Atlantique, Nantes Université, F-44307 Nantes, France}

  \author{T. Eronen}
  \affiliation{Department of Physics, University of Jyväskylä, P.O. Box 35, FI-40014 Jyväskylä, Finland}

 \author{D. Etasse}
  \affiliation{LPC Caen, ENSICAEN, Université de Caen, CNRS/IN2P3, F-Caen, France}

  \author{J. Agramunt}
  \affiliation{Instituto de F\'isica Corpuscular, CSIC-Universitat de Val\`encia, E-46071 Val\`encia, Spain}

   \author{J. L. Tain}
  \affiliation{Instituto de F\'isica Corpuscular, CSIC-Universitat de Val\`encia, E-46071 Val\`encia, Spain}

  \author{H. Garcia Cabrera}
  \affiliation{Instituto de F\'isica Corpuscular, CSIC-Universitat de Val\`encia, E-46071 Val\`encia, Spain}
  \affiliation{Universidad Europea de Madrid, Department of Biosciences, Faculty of Biomedical and Health Sciences, E-28670 Madrid, Spain}

  \author{L. Giot}
  \affiliation{SUBATECH, CNRS/IN2P3, IMT Atlantique, Nantes Université, F-44307 Nantes, France}

  \author{A. Laureau}
  \affiliation{SUBATECH, CNRS/IN2P3, IMT Atlantique, Nantes Université, F-44307 Nantes, France}

  \author{J. A. Victoria}
  \affiliation{Instituto de F\'isica Corpuscular, CSIC-Universitat de Val\`encia, E-46071 Val\`encia, Spain}

  \author{Y. Molla}
  \affiliation{SUBATECH, CNRS/IN2P3, IMT Atlantique, Nantes Université, F-44307 Nantes, France}

  \author{A. Jaries}
  \affiliation{Department of Physics, University of Jyväskylä, P.O. Box 35, FI-40014 Jyväskylä, Finland}

  \author{L. Al Ayoubi}
  \affiliation{Department of Physics, University of Jyväskylä, P.O. Box 35, FI-40014 Jyväskylä, Finland}

  \author{O. Beliuskina}
  \affiliation{Department of Physics, University of Jyväskylä, P.O. Box 35, FI-40014 Jyväskylä, Finland}

  \author{W. Gins}
  \affiliation{Department of Physics, University of Jyväskylä, P.O. Box 35, FI-40014 Jyväskylä, Finland}

  \author{M. Hukkanen}
  \affiliation{Department of Physics, University of Jyväskylä, P.O. Box 35, FI-40014 Jyväskylä, Finland}

  \author{A. Illana}
  \affiliation{Department of Physics, University of Jyväskylä, P.O. Box 35, FI-40014 Jyväskylä, Finland}
  \affiliation{Grupo de Física Nuclear, EMFTEL, and IPARCOS, Universidad Complutense de Madrid, CEI Moncloa, E-28040 Madrid, Spain.}

  \author{A. Kankainen}
  \affiliation{Department of Physics, University of Jyväskylä, P.O. Box 35, FI-40014 Jyväskylä, Finland}

  \author{S. Kujanpää}
  \affiliation{Department of Physics, University of Jyväskylä, P.O. Box 35, FI-40014 Jyväskylä, Finland}

  \author{I. Moore}
  \affiliation{Department of Physics, University of Jyväskylä, P.O. Box 35, FI-40014 Jyväskylä, Finland}

  \author{I. Pohjalainen}
  \affiliation{Department of Physics, University of Jyväskylä, P.O. Box 35, FI-40014 Jyväskylä, Finland}

  \author{D. Pitman}
  \affiliation{Department of Physics, University of Jyväskylä, P.O. Box 35, FI-40014 Jyväskylä, Finland}
  \affiliation{Department of Physics and Astronomy, University of Manchester, Manchester M13 9PL, United Kingdom}

  \author{A. Raggio}
  \affiliation{Department of Physics, University of Jyväskylä, P.O. Box 35, FI-40014 Jyväskylä, Finland}

  \author{M. Reponen}
  \affiliation{Department of Physics, University of Jyväskylä, P.O. Box 35, FI-40014 Jyväskylä, Finland}

  \author{J. Romero}
  \affiliation{Department of Physics, University of Jyväskylä, P.O. Box 35, FI-40014 Jyväskylä, Finland}
  \affiliation{Oliver Lodge Laboratory, University of Liverpool, Liverpool, L69 7ZE, United Kingdom}

  \author{J. Ruotsalainen}
  \affiliation{Department of Physics, University of Jyväskylä, P.O. Box 35, FI-40014 Jyväskylä, Finland}

  \author{M. Stryjczyk}
  \affiliation{Department of Physics, University of Jyväskylä, P.O. Box 35, FI-40014 Jyväskylä, Finland}
  \affiliation{Institut Laue-Langevin, 71 Avenue des Martyrs, F-38042 Grenoble, France}

  \author{V. Virtanen}
  \affiliation{Department of Physics, University of Jyväskylä, P.O. Box 35, FI-40014 Jyväskylä, Finland}


  \date{\today}

  \begin{abstract}
The shapes of the beta spectra of $^{92}$Rb and $^{142}$Cs, two of the beta decays most relevant for the prediction of the antineutrino spectrum in reactors, have been measured. A new setup composed of two $\Delta$E-E telescopes has been used. High purity radioactive beams of the isotopes of interest were provided by the IGISOL facility using the JYFLTRAP double Penning trap. The resulting beta spectra have been compared with model predictions using beta decay feedings from total absorption gamma spectroscopy measurements and shape corrections employed in the calculation of the antineutrino spectrum, validating both further. The procedure can be extended to other relevant nuclei in the future, providing solid ground for the prediction of the antineutrino spectrum in reactors.   
  
  \end{abstract}


  \maketitle

Measurements of the shape of the emitted spectrum in beta decay have always played a key role in weak interaction research. In the early days of beta decay theory, it was already clear that a measurement of the shape of the beta spectrum could be used to distinguish between the possible theories available (see the debate between the Fermi and Konopinski-Uhlenbeck descriptions of the shape of the spectrum \cite{EVANS,Konopinski1,Konopinski2}). Beta spectrum shape measurements are relevant for a variety of practical and fundamental applications including metrology \cite{Mougeot}, biologically targeted radiotherapy \cite{Boswell}, the measurement of the mean beta energy released in a beta decays \cite{TENGBLAD} of relevance for reactor decay heat,  and more recently in the search for complementary constrains on physics beyond the electro-weak standard model \cite{Cirgiliano,Dekeukeleere,Rozpedzik}. Recent studies have also considered exploiting the sensitivity of the shape of the beta spectra of forbidden decays to the quenching of the $g_A$ coupling constant in the nuclear medium \cite{Kostensalo,Pagnanini}. The shape of the beta spectrum close to the end point of the decay is also a key observable for constraining the neutrino mass \cite{Katrin,Ashtari1,Ashtari2}.
    
There is also a renewed interest in beta spectrum shape studies from the perspective of the description of the reactor antineutrino spectrum. The interest here is manifold. Above all, these studies provide means for validating the different corrections applied to both beta and neutrino spectra of individual decays, improving the precision of the reactor antineutrino spectrum calculations. In this context, it is worth noting that there are limited options to calculate the antineutrino spectrum of reactors. One possibility is based on the conversion into the antineutrino spectrum of the measured compound beta spectrum of all fission products created by the fission of the $^{235,238}$U, $^{239,241}$Pu fissile nuclei. Such compound beta spectrum measurements were performed by Schreckenbach \textit{et al.} at the ILL reactor \cite{SCHRECKENBACH1,SCHRECKENBACH2,SCHRECKENBACH3} and later by Haag \textit{et al.} at the FRM II in Garching \cite{HAAG}.

Based on energy and momentum conservation, the conversion from beta to antineutrino spectrum is a straightforward procedure for individual beta transitions, but it is a challenging calculation for the integral spectrum from a fissile material. Conversion requires the introduction of ``effective (or virtual)'' beta branches to approximate the mixture of beta decays coming from different parent nuclei with similar endpoints and assumptions about their shapes. The conversion procedure implemented by Schreckenbach \textit{et al.} \cite{SCHRECKENBACH1,SCHRECKENBACH2,SCHRECKENBACH3} was revisited by Huber \cite{HUBER}, who added additional corrections and profited from the improvement in nuclear databases since the measurements of Schreckenbach \textit{et al.} \cite{SCHRECKENBACH1,SCHRECKENBACH2,SCHRECKENBACH3}.

An alternative method for calculating the antineutrino spectrum is the summation method (or {\it ab initio}) (see for example \cite{MUELLER,FALLOT,ESTIENNE,SONZOGNI2,PERISSE}). This approach is based on summing the contributions from the beta decays of all the fission products produced in a working reactor, weighted by their cumulative fission yields. In this method, the beta decay data are taken conventionally from nuclear databases. This approach was explored by Mueller {\it et al.} in their work \cite{MUELLER}. Their results showed the limitations of the available nuclear data in 2011. The comparison of the results of their summation method with the Schreckenbach \textit{et al.} \cite{SCHRECKENBACH1,SCHRECKENBACH2,SCHRECKENBACH3} data showed discrepancies of the order of 10\%. To further refine their antineutrino model calculations, Mueller {\it et al.} corrected the remaining differences using an additional conversion procedure with a reduced number of effective beta branches \cite{MUELLER}. The Huber and Mueller \textit{et al.} models provide similar parameterizations of the antineutrino spectrum, since they are ultimately constrained by the Schreckenbach \textit{et al.} \cite{SCHRECKENBACH1,SCHRECKENBACH2,SCHRECKENBACH3} data, and are considered a reference in the field.

An alternative approach is based on a pure summation method, such as that presented by Fallot {\it et al.} \cite{FALLOT} and Estienne {\it et al.} \cite{ESTIENNE}. In these works, considerable effort was made to select data free from the Pandemonium effect on a priority basis, including a systematic study of the beta decays thought to be most relevant \cite{FALLOT}. An independent effort to dissect the antineutrino spectrum and to identify the most relevant decays was published by Sonzogni {\it et al.} \cite{SONZOGNI1}. Please note that in the nuclear physics context, Pandemonium means that the beta decay intensity distribution is affected by a systematic error that occurs in some studies where setups composed of high-purity Ge (HPGe) detectors are employed \cite{HARDY}.

  Shortly after the Huber-Mueller revisions \cite{HUBER,MUELLER}, Mention {\it et al.} \cite{MENTION} showed a systematic discrepancy between the detected and calculated antineutrino flux at short baselines regarding reactor antineutrino oscillation experiments \cite{DCHOOZ1,DCHOOZ2,RENO,DAYABAY}. The discrepancy, referred to as the ``reactor antineutrino anomaly (RAA)'', has attracted considerable  attention. One potential explanation for the missing flux is the possibility of oscillations into sterile antineutrinos \cite{MENTION}. Alternative explanations have also been proposed, including the potential contribution of first-forbidden beta decays and inadequacies in other corrections applied in the conversion method \cite{HAYES2,HAYEN_1F}. Even though there are indications that the anomaly is probably caused by a normalization problem of Schreckenbach \textit{et al.} data for the compound beta spectrum of $^{235}$U \cite{DAYABAY,KOPEIKIN}, the question is not completely settled. In this context, measurements of the shape of first-forbidden decays, and how well they are reproduced by the approaches used by Huber and Mueller {\it et al.} and the different summation models, remain of great interest. It should be noted that the summation model of Estienne {\it et al.} reproduces the measured flux within a 2\% limit \cite{ESTIENNE}, questioning the existence of the RAA and showing the relevance of a proper selection of beta decay data.  

  Comparisons of the measured antineutrino spectrum with Huber-Mueller model predictions have also shown that there is a shape distortion in the energy spectrum \cite{DOUBLECHOOZ_bump,DAYABAY,RENO}. The distortion, located between 5 and 7 MeV of the measured antineutrino spectrum, is not yet understood. As in the case of the RAA, it has been proposed that it can be caused by first-forbidden corrections not taken into account in the antineutrino spectrum calculations \cite{HAYEN_1F}. 

  In both the conversion and the summation method, the starting point is a calculation of the shape of the spectrum, either for an effective or a real beta endpoint. A better understanding of the RAA and the shape distortion calls for direct beta spectrum shape measurements of the most relevant beta decays and a validation of the approaches used for the corrections. The need for these measurements has been highlighted in several publications and recent meetings \cite{KOPEIKIN,FALLOT,MEETING1,MEETING2,MEETING3}. A recent article by the EXO-200 collaboration addressed this problem, studying the beta decay of $^{137}$Xe \cite{EXO200}. In this Letter, we report on the measurement of the beta spectrum shape of $^{92}$Rb and $^{142}$Cs decays, two of the most relevant contributors to the high energy part of the antineutrino spectrum, using for the first time radioactive beams of high purity produced by trap-assisted spectroscopy. 

  Beta spectrum shape measurements are challenging. They require the production of very pure radioactive sources to minimize distortions in the spectrum induced by possible contamination of the beam. They also require very well-characterized setups for the analysis. In these experiments, the shape of the true, original or undistorted spectrum is obtained based on a solution of an inverse problem, similar to the total absorption one \cite{TAIN}, given by 
  \begin{equation}
    \label{eq_Response}
    D_i = \displaystyle\sum_{j=1}^{j_{max}}\mathcal{R}_{ij}O_j  + C_i \;,
  \end{equation}
  {\noindent}where $D_i$ is the content of channel $i$ of the measured beta spectrum, $\mathcal{R}_{ij}$ is the response matrix of the setup to monoenergetic electrons, $O_j$ is the original or true beta spectrum (our goal) and $C_i$ is the contribution of possible spectrum distortions such as pile-up, daughter activity, contamination of the radioactive beam, etc., that have to be identified and accounted for. The $j$ index accounts for the dimension of the response matrix and for the dimension of the original beta spectrum. 

\begin{figure}[htb]
    \resizebox{1.00\columnwidth}{!}{\rotatebox{0}{\includegraphics[clip=]{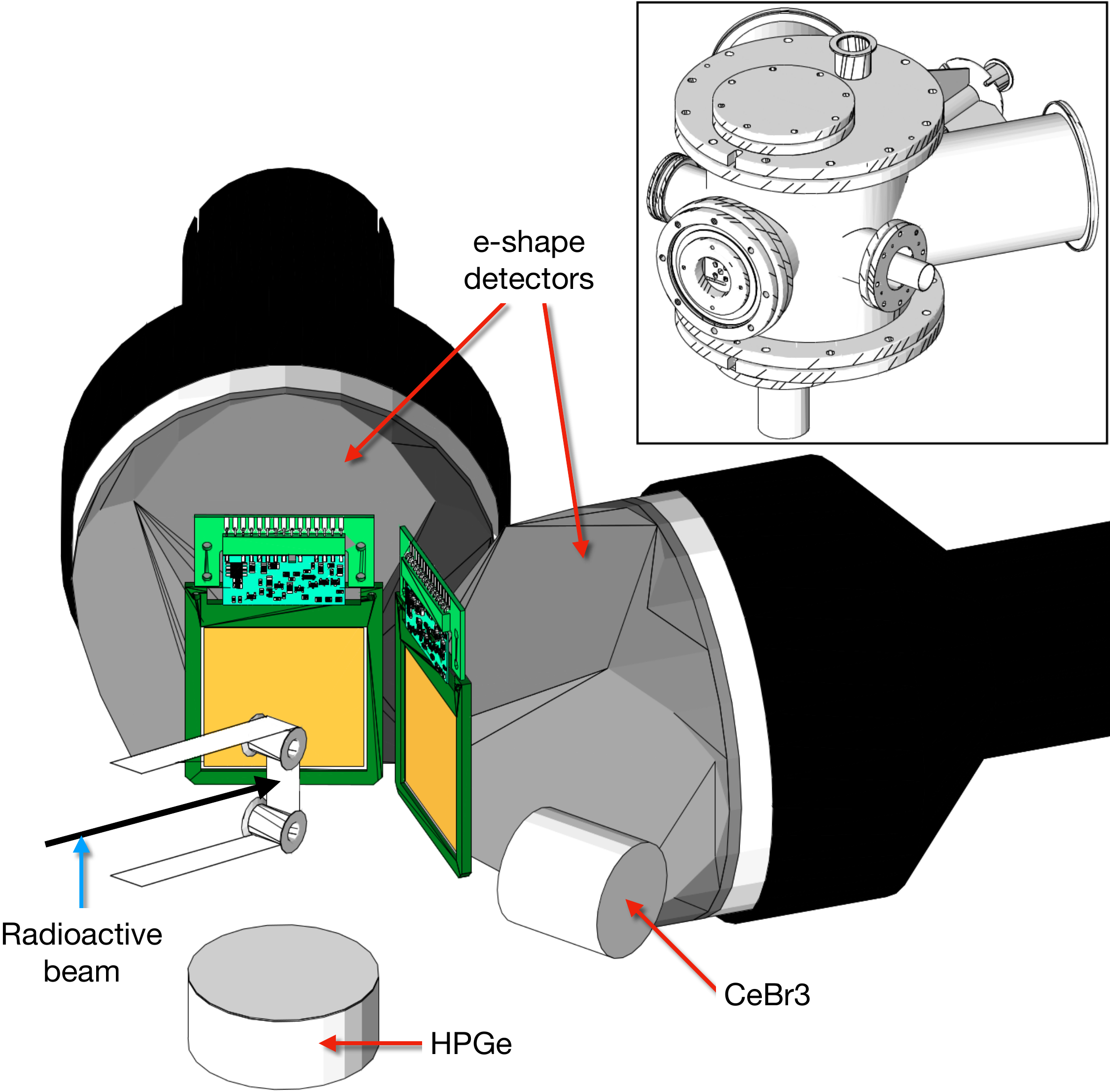}}}
    \caption{Geometry of the {\it e-shape} setup implemented in Geant4 simulations. The upper right inset shows the vacuum chamber of the setup fully implemented in the simulations. The figure depicts the geometry of the relevant detectors: the {\it e-shape} telescopes placed inside the vacuum chamber, an HPGe below the implantation point, and a CeBr3 scintillator placed at the right side of the implantation point. These two detectors were placed outside the vacuum chamber and used for beta delay gamma identification. Please note that for the sake of simplicity, only the active volume of the gamma detectors is shown.}
    \label{fig_Geometry}
  \end{figure}

  Our beta spectra measurements were conducted at the IGISOL facility of the University of Jyv\"askyl\"a, Finland \cite{MOORE}. 
  In this facility, a radioactive beam is produced by protons impinging on a $^{\text{nat}}$U target. The fission-product radioactive beam is then extracted using the ion guide technique \cite{AYSTO}.  
  The beam is first separated in mass ($A$) using a separator magnet of moderate mass resolving power ($m/\Delta m \sim$ 500). Isotopic, or, depending on the case, isomeric separation is obtained via the JYFLTRAP double Penning trap \cite{ERONEN} before the beam reaches our experimental setup. JYFLTRAP has a mass resolving power of the order of 10$^5$ or better. 
  After extraction from the Penning trap, the selected isotope is implanted on a magnetic tape 
  (see Fig. \ref{fig_Geometry}). As is typical in this kind of measurement, the measuring cycles of the tape system are optimized based on the half-life of the isotope of interest and the half-lives of its descendants. 

  For the beta spectrum shape measurements, a new setup called {\it e-shape} was developed (see Fig. \ref{fig_Geometry}). The setup is composed of two $\Delta$E-E telescopes in vacuum. They are positioned at 35$^\circ$ to the incoming beam direction around the implantation position. A detailed description of the {\it e-shape} setup and the results of its first characterization and test can be found in Guadilla {\it et al.} \cite{GUADILLA}. In the measurements presented here, each of the two telescopes is composed of a 527 $\mu$m thick silicon $\Delta$E detector combined with a 7.5 cm thick tapered plastic E detector. The thickness of the plastic was optimized using Monte Carlo (MC) simulations to measure beta particles of energies up to 10 MeV. The setup has an electron detection efficiency of the order of 6\% for one telescope and a sensitivity to gamma rays of less than 0.06\%, both in coincidence mode. Please note that the thickness of the silicon detector limits the lowest energy accessible in coincidence mode.  During our measurements, the resolution was 24 keV (FWHM) at 0.715 MeV for the silicon detectors and 110 keV at 1 MeV for the plastic detectors. 
  Two ancillary gamma detectors, a HPGe and a CeBr3 detector were also included in the setup to continuously monitor the beam purity. 
  In particular, the HPGe detector was used in the analysis to determine the amount of contamination from the daughter activity. In the measurements, a new digital data acquisition system based on FASTER cards \cite{FASTER} was used. 

\begin{figure}[htb]
    \resizebox{1.0\columnwidth}{!}{\rotatebox{0}{\includegraphics[clip=]{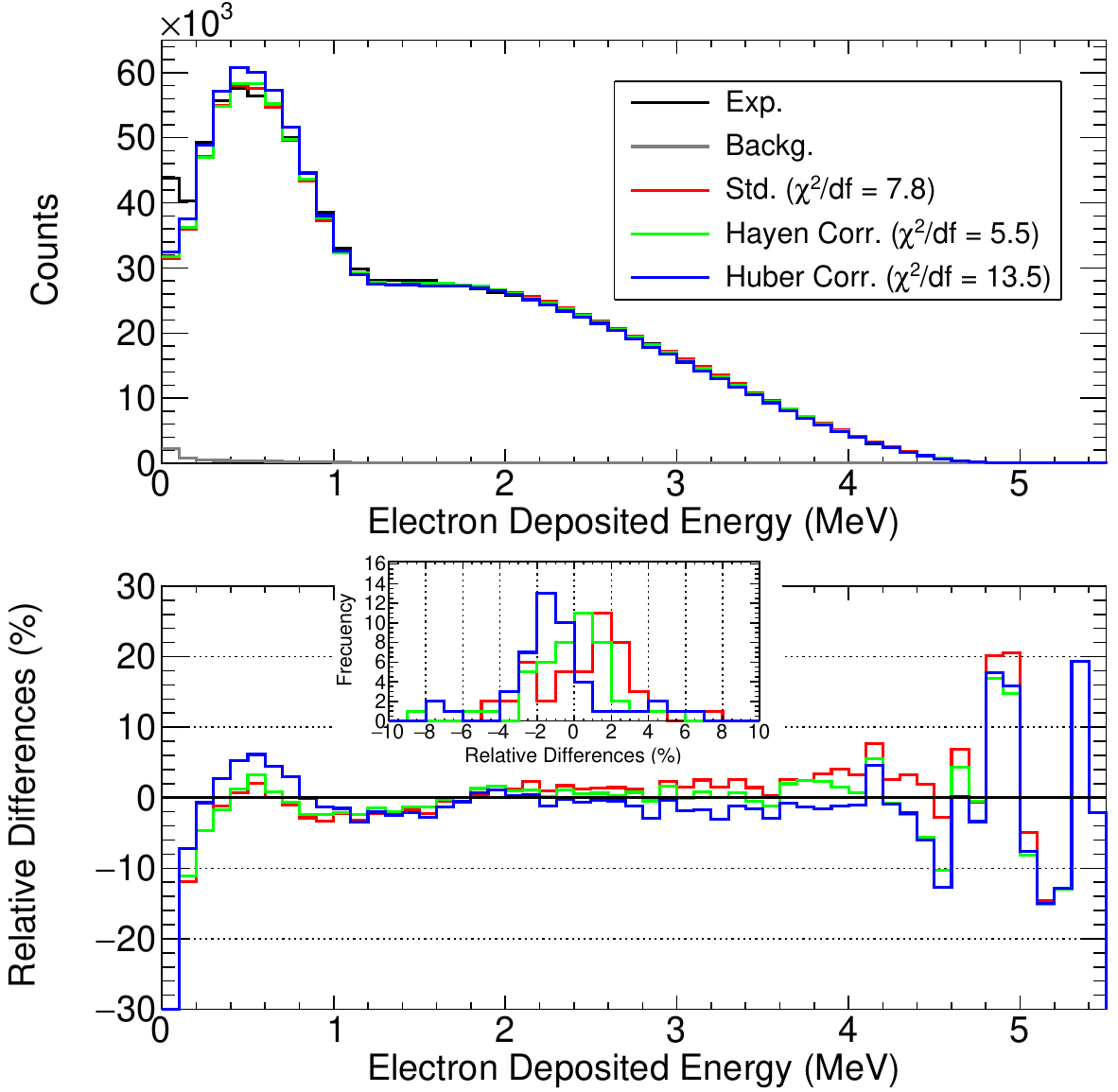}}}
    \caption{Upper panel. Comparison of the measured spectrum of the combined $^{114}$Pd $\rightarrow$ $^{114}$Ag $\rightarrow$ $^{114}$Cd beta decays with Monte Carlo simulations employing different corrections. The background has been added to the simulations for the comparison. Std. means the classical Fermi approximation. Hayen Corr. and Huber Corr. stand for the Fermi approach plus the corrections introduced by Hayen {\it et al.} \cite{HAYEN_Allowed} and Huber \cite{HUBER} (see the End Matter section for more details). The weak magnetism term of Huber \cite{HUBER} was included in the corrections of Hayen {\it et al.} \cite{HAYEN_Allowed}. Lower panel. Relative differences between the experimental and simulated spectra. In the insert, a frequency plot of the differences is presented for the full energy range.}
    \label{fig_validation}
  \end{figure}

  As mentioned earlier, the analysis of this kind of experiment requires the determination of the response matrix of the setup to mono-energetic electrons ($\mathcal{R}_{ij}$ in formula (\ref{eq_Response})). This can only be done with MC  simulations, which should be previously validated. In the validation process, the beta decay of $^{114}$Ag into $^{114}$Cd was used. This decay is expected to be dominated by a strong ground state to ground state transition of allowed character \cite{ENSDF_114Ag}. Unfortunately, at the IGISOL facility the production of $^{114}$Ag is not high (a consequence of the direct fission yield), so $^{114}$Ag was produced through the decay of the higher-yield parent $^{114}$Pd, and the decay chain $^{114}$Pd $\rightarrow$ $^{114}$Ag $\rightarrow$ $^{114}$Cd was measured and characterized as a whole. For the comparison of the measured spectra with MC simulations, proper energy and resolution calibrations are needed. This is discussed in more detail in the next paragraph about the analysis procedure. The comparison of experiment vs MC for the decay chain $^{114}$Pd $\rightarrow$ $^{114}$Ag $\rightarrow$ $^{114}$Cd validation case is shown in Fig. \ref{fig_validation}. The shape of the experimental $\Delta E-E$ coincidence spectrum is well reproduced at the 2$\%$ level in the energy range between 0.4 and 4.5 MeV. In the MC simulation of the decay chain, it is assumed that every $^{114}$Pd decay is followed by a $^{114}$Ag decay. High-resolution data available in the Evaluated Nuclear Structure Data File (ENSDF) for both decays were used for the simulations \cite{ENSDF_114Ag}. In this work, the Geant4 MC code (version 10.5.1) was employed. As part of the validation, different physics interaction models were tested.  Very similar results were obtained with the Livermore and Penelope physics models. In this work, the Livermore physics model was chosen since it was slightly faster compared with the other available options. To implement the geometry of the setup in full detail, the CADMesh library was used \cite{CADMESH1,CADMESH2}. This library allows the importation of computer-aided designs from technical drawings. 

  Both the Monte Carlo responses and the experimental spectra must be calibrated in energy and width (resolution), since their energy and energy resolution should be matched for the analysis procedure. For the experimental calibration, several beta decays in the range from 1.440 to 8.095 MeV and conversion electrons, like the E0 transition in $^{98}$Mo, were used. 
  The beta endpoints were fitted using a function with the standard shape of an allowed beta decay. Conversion electron peaks were fitted using Gaussian functions with low- and high-energy exponential tails. The quality of the width calibration is essential for the correct generation of the mono-energetic responses needed for the response matrix. Further details on the procedure will be provided in an upcoming article \cite{ALCALA2}.

  In this Letter, we present the results from the measurement of the beta decays of $^{92}$Rb and $^{142}$Cs. $^{92}$Rb decay provides the most important contribution to the antineutrino spectrum in Pressurized Water Reactors (PWR) at high energies, amounting to 16\% of the spectrum in the 5–8 MeV range \cite{SONZOGNI2,ZAKARI}. $^{142}$Cs decay is the third most important contributor at the end of a typical fuel cycle. Both decays are characterized by strong ground state to ground state first-forbidden decays of the 0$^-$ to 0$^+$ type \cite{SONZOGNI2,HAYEN_1F}. 

  To show the analysis procedure followed, we discuss in detail the $^{92}$Rb case. This decay was measured with an ion implantation cycle of 22.49 s to minimize the contribution of the $^{92}$Sr daughter decay. Note that the half-life of $^{92}$Rb is 4.48(3)s and the half-life of $^{92}$Sr is 2.611(17)h \cite{ENSDF_92Rb}. Even though the daughter contribution is expected to be small, the contamination of the beta spectrum was determined from the gamma spectrum collected with the HPGe detector. The pile-up distortion of the spectrum was also considered. In the reconstruction of the events for the analysis, the pile-up flag of the digital data acquisition system \cite{FASTER} was employed to reject events affected by this distortion. Typical event rates during our experiment were of the order of 80 counts/s, leading to a reduced impact of pile-up in general. 

\begin{figure}[htb]
    \resizebox{1.0\columnwidth}{!}{\rotatebox{0}{\includegraphics[clip=]{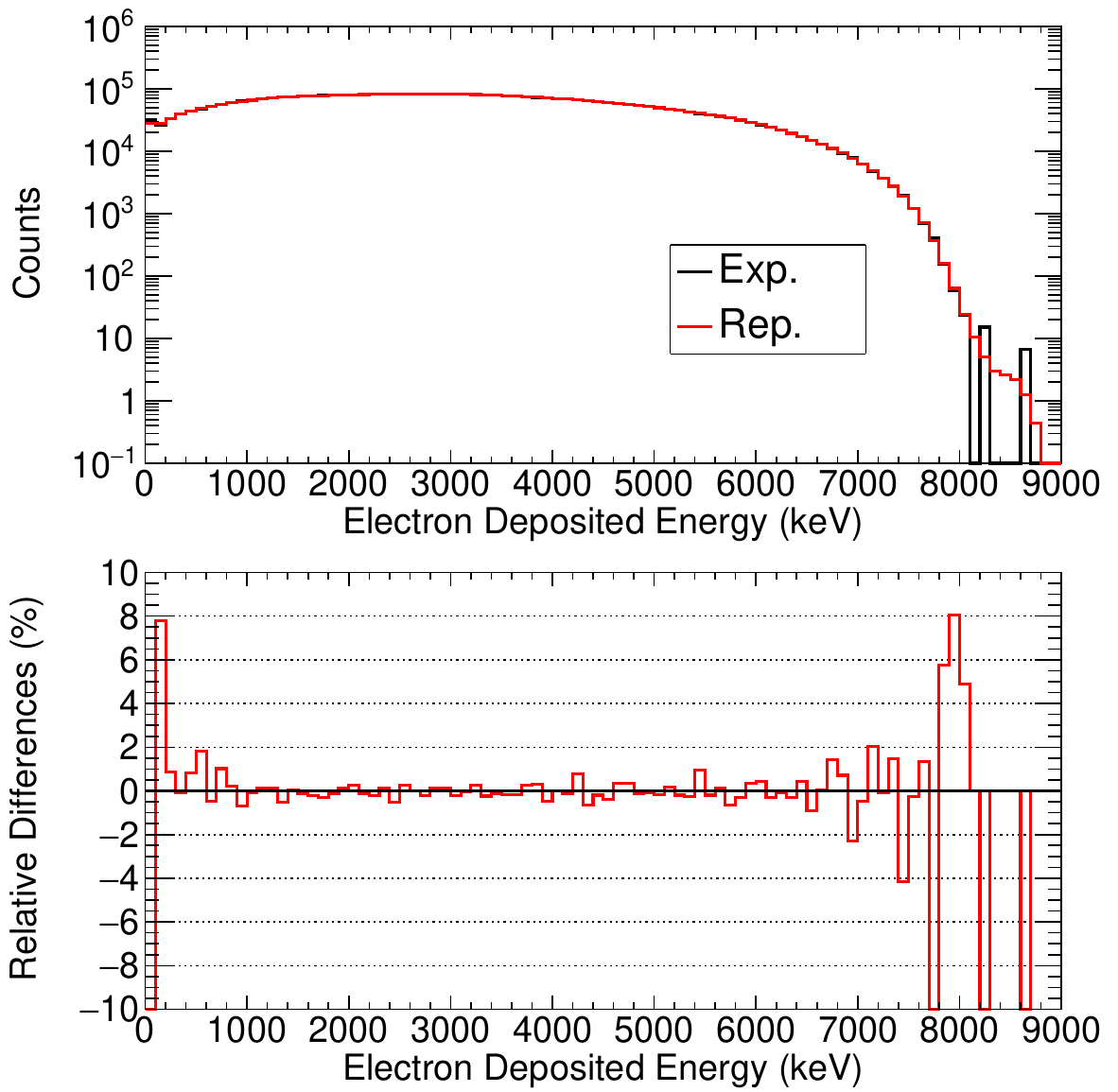}}}
    \caption{Comparison of the measured $^{92}$Rb experimental spectrum (Exp., corresponding to $D_i$ in formula (\ref{eq_Response})) with the reproduction of the experiment obtained from the result of the expectation-maximization deconvolution algorithm (Rep., corresponding to $\sum_j\mathcal{R}_{ij}O_j$, on right side of formula (\ref{eq_Response})), obtained after the analysis. The relative differences are presented in the lower panel.}
    \label{fig_ExpRepro}
  \end{figure}

  For the analysis of the beta spectra, the maximum entropy and expectation-maximization deconvolution algorithms were used \cite{TAIN}. Since both methods provided similar results, the expectation-maximization algorithm was chosen for the final analysis because it showed fewer oscillations. The following uncertainties were considered for the final results: the statistical uncertainty per bin, the corresponding covariance matrix values obtained in the analyses 
  and the impact of the contributions of the possible contaminants to the analyzed spectra varied in the range of the uncertainty of their normalization. Fig. \ref{fig_ExpRepro} shows the experimental $^{92}$Rb spectrum 
  along with the reproduction of the spectrum after the analysis using the expectation-maximization algorithm. 
 This last spectrum is obtained as the product of the response matrix and the deduced original beta spectrum $O_j$ (see the right side of formula (\ref{eq_Response})). The relative differences between the experimental and the reconstructed spectra are shown in the lower panel of Fig. \ref{fig_ExpRepro}.

\begin{figure}[htb]
    \resizebox{1.0\columnwidth}{!}{\rotatebox{0}{\includegraphics[clip=]{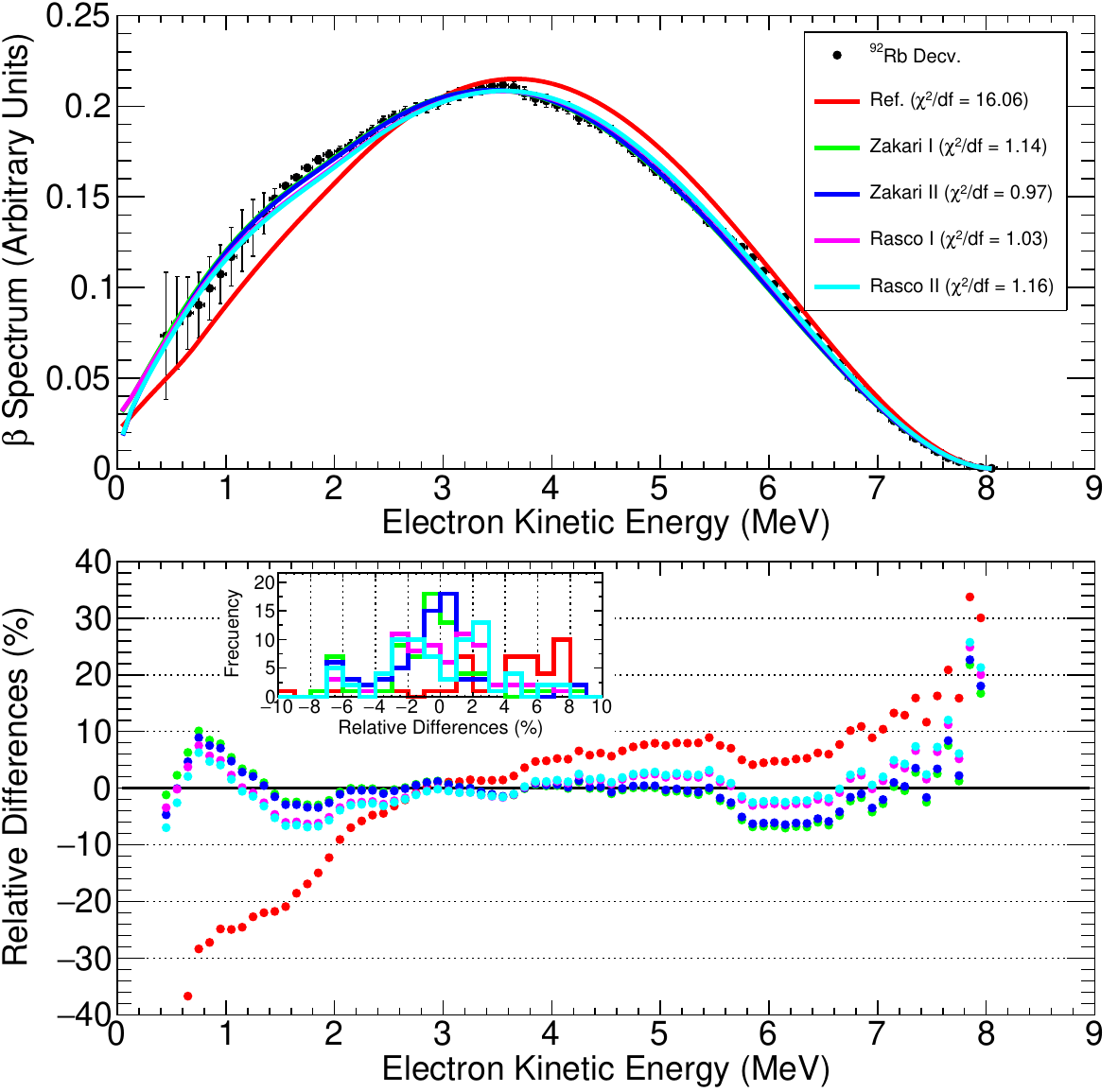}}}
    \caption{Comparison of the deduced (or true) beta spectrum of the ${}^{92}\text{Rb} \rightarrow {}^{92}\text{Sr}$ beta decay ($^{92}$Rb Decay) with various beta decay feedings and beta shape correction models. Ref. corresponds to the prediction obtained using ENSDF high-resolution feedings \cite{ENSDF_92Rb}, employing only the Fermi function. Zakari I and II refer to predictions based on Zakari-Issoufou \textit{et al.} TAGS feedings \cite{ZAKARI}. For Zakari I, the allowed shape corrections from Hayen \textit{et al.} \cite{HAYEN_Allowed}, the weak magnetism term from Huber \cite{HUBER}, and the ground state to ground state first-forbidden shape correction from Hayen \textit{et al.} \cite{HAYEN_1F} were applied. For Zakari II, the allowed shape corrections from Huber \cite{HUBER} were used. Rasco I and II denote predictions based on Rasco \textit{et al.} TAGS feedings \cite{RASCO}, with the same respective shape corrections applied.}
    \label{fig_Decv92Rb}
  \end{figure}

\begin{figure}[htb]
    \resizebox{1.0\columnwidth}{!}{\rotatebox{0}{\includegraphics[clip=]{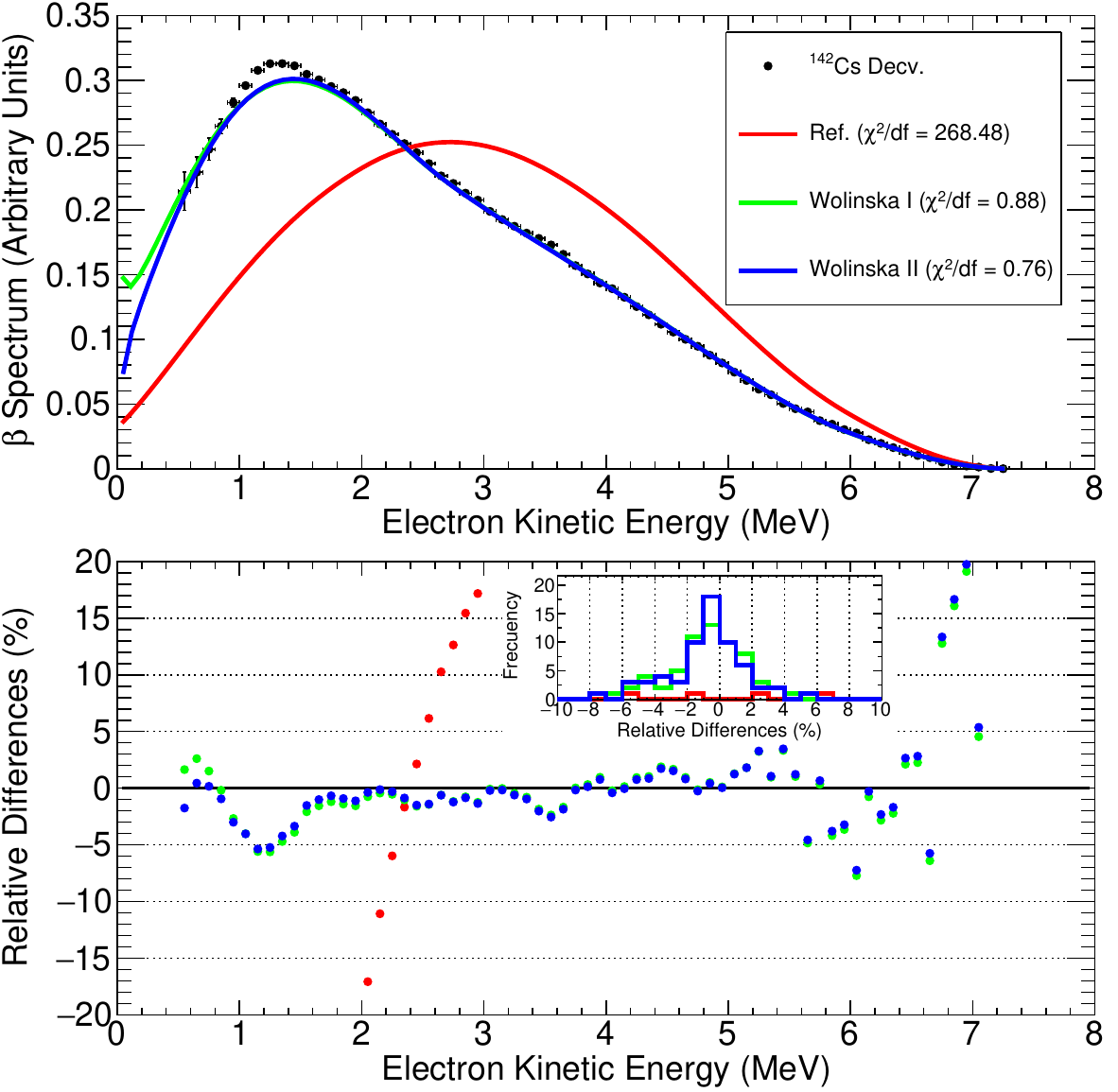}}}
    \caption{Comparison of the deduced (or true) beta spectrum of the $^{142}\text{Cs}\rightarrow{}^{142}\text{Ba}$ beta decay ($^{142}$Cs Decv.) with various beta decay feedings and beta shape correction models. Ref. corresponds to the prediction obtained using ENSDF high-resolution feedings \cite{ENSDF_142Cs}, employing only the Fermi function. Wolinska I and II refer to predictions based on Woli\'nska-Cichocka \textit{et al.} TAGS feedings \cite{WOLINSKA}. For Wolinska I, the allowed shape corrections from Hayen \textit{et al.} \cite{HAYEN_Allowed}, the weak magnetism term from Huber \cite{HUBER}, and the ground state to ground state first-forbidden shape correction from Hayen \textit{et al.} \cite{HAYEN_1F} were applied. For Wolinska II, the allowed shape corrections from Huber \cite{HUBER} were used.}
    \label{fig_Decv142Cs}
  \end{figure}

  The true beta spectrum obtained in the analysis ($O_j$ in formula (\ref{eq_Response})) can then be used to test beta feedings and beta shape correction models employed in antineutrino spectrum calculations. To do so, a beta spectrum generator was developed. It takes as input beta decay feedings from existing measurements and integrates different beta shape model corrections to be tested. $^{92}$Rb has been previously studied using the total absorption gamma spectroscopy (TAGS) technique \cite{ZAKARI,RASCO}. Fig. \ref{fig_Decv92Rb} shows the comparison of the deduced spectrum ($O_j$) with predictions generated with different feeding models \cite{ZAKARI,RASCO}, including the allowed and first-forbidden shape corrections of Hayen {\it et al.} \cite{HAYEN_Allowed,HAYEN_1F} and the shape corrections of Huber \cite{HUBER}. Similarly, Fig. \ref{fig_Decv142Cs} presents the final result of the analysis of $^{142}$Cs decay compared with the respective predictions with different feedings and shape correction models (see additional details in the End Matter section). 

  The results obtained from the $^{92}$Rb and $^{142}$Cs cases show that the shapes of the deduced spectra are nicely reproduced by the feeding models obtained from TAGS measurements including the shape corrections employed by Huber \cite{HUBER} and Hayen \textit{et al.} \cite{HAYEN_Allowed, HAYEN_1F}. The reproductions of the deduced beta spectra are similar in quality in both models. It must be highlighted that the inclusion of the tested shape correction factors for first-forbidden transitions \cite{HAYEN_1F} had a negligible effect on the shape of the spectra, confirming that they are of limited interest for $0^-$ to $0^+$ transitions \cite{HAYES1} (see End Matter section). Compared with an older direct measurement of the beta spectrum shape \cite{TENGBLAD}, our data do not show irregularities at around 1 MeV, and show a smooth behaviour. 
  The new true beta spectra can provide data for an independent way to calculate the antineutrino spectrum for these important beta decays. 
  
  The deduced beta spectra can also be used to constrain the ground state to ground state feedings of these relevant decays based on a comparison with MC simulations, providing an additional method to that presented in \cite{GUADILLA3,GUADILLA_GS}. To achieve this, different beta shape spectra were generated by changing the ground state to ground state transition feedings of the TAGS feedings obtained in \cite{ZAKARI,RASCO} 
  in continuous steps of 0.1$\%$. For a given ground state to ground state value, the remaining feedings were proportionally adjusted to preserve their total normalization to 100$\%$. A standard $\chi^2$ test was employed to determine the optimized ground state feedings. For the $^{92}$Rb case, Zakari-Issoufou \textit{et al.} \cite{ZAKARI} reported a ground state to ground state feeding value of 87.5(25)\%. The optimized values obtained in this work are 89.0(26)\% and 88.5(25)\% using the shape corrections of Hayen \textit{et al.} \cite{HAYEN_Allowed, HAYEN_1F} plus the weak magnetism term of Huber \cite{HUBER} and the shape corrections of Huber \cite{HUBER}, respectively. Rasco \textit{et al.} \cite{RASCO} reported a ground state to ground state feeding value of 91(3)\%, while the optimized values are 91.1(21)\% and 90.7(21)\% using the same shape corrections respectively. 
  All optimized values are within the uncertainties of the corresponding reference feedings from the TAGS measurements.

It is worth mentioning that TAGS and beta spectrum shape measurements constitute the only two existing ways of determining beta decay data free from the Pandemonium effect \cite{HARDY}. The nice agreement obtained in Figs. \ref{fig_Decv92Rb} and \ref{fig_Decv142Cs} show the consistency of the results obtained from both techniques, validating them further. The comparisons also provide a method for testing the overall consistency of beta decay data, when the two types of measurements are available.  
  
In summary, the results obtained represent the first measurements of the shapes of beta spectra of fission products relevant to the prediction of the antineutrino spectrum in reactors employing isotopically pure beams (trap-assisted beta shape spectroscopy). The deduced spectra provide a way to test the shape corrections used to calculate the antineutrino spectrum for reactors. Please note that $\Delta J^\pi=0^-$ beta transitions of the type addressed in this work, represent the majority of the approximately 60$\%$ forbidden decay contributions to the antineutrino spectrum in reactors (see Fig. 3 in \cite{HAYEN_1F} for the $^{235}U$ fuel). The analyses of the measured spectra have been performed using state-of-the-art analysis and Monte Carlo techniques. The comparison of the results obtained with model calculations using TAGS beta decay feedings and the theoretical approaches presented in Huber and Hayen et al. \cite{HUBER,HAYEN_Allowed,HAYEN_1F} validate those approaches for two of the most relevant beta decays for the prediction of the antineutrino spectrum in reactors, and in particular for $0^-$ to $0^+$ transitions.
The performed measurements can also provide additional information to constrain ground state to ground state feedings, or the feedings of dominant beta branches, and provide the needed sensitivity to beta shape factors. It can also provide a method to obtain beta decay data free from the Pandemonium effect that can be used to generate directly antineutrino spectra and mean beta decay energies relevant to reactor applications. 

Future measurements, to test the shape corrections of other type of forbidden decays relevant for the antineutrino spectrum calculations are highly recommended. 
  

This work has been supported by the Spanish Ministerio de Economía y Competitividad Grants No. FPA2017-83946-C2-1-P and No. RTI2018-098868-B-I00, by the Spanish Ministerio de Ciencia e Innovación Grants No. PID2019-104714GB-C21 and PID2022-138297NB-C21 and by the Generalitat Valenciana Prometeo Grant CIPROM/2022/9. This work has also been supported by the CNRS challenge NEEDS and the associated NACRE project, which as co-founded a part of the experimental setup and A. Beloeuvre's PhD grant, the CNRS/IN2P3 e-shape and CNRS/IN2P3 PICS TAGS between Subatech and IFIC, and the CNRS/IN2P3 Master projects Jyväskylä and OPALE. Part of the e-shape detector and R. Kean's postdoc contract were funded by a grant from the Pays de Loire region. Authors also acknowledge the financial support from the Ministerio de Ciencia e Innovación with funding from the European Union NextGenerationEU and Generalitat Valenciana in the call Programa de Planes Complementarios de I+D+i (PRTR 2022) under Project DETCOM, reference ASFAE/2022/027. G. A acknowledges the support of the Santiago Grisolia Program of the Comunitat Valenciana. W.G. acknowledges the support of the U.K. Science and Technology Facilities Council grant ST/P005314. V.G. acknowledges the support of the National Science Center, Poland, under Contract No. 2019/35/D/ST2/02081. The support of the EU Horizon 2020 research and innovation program under Grant No. 771036 (ERC CoG MAIDEN) is also acknowledged. Authors would also like to thank Neil Clark from MICRON, Massimo Volpi from CAEN, and Paul Schotanus from Scionix for fruitful discussions, as well as Leendert Hayen for sharing the beta-shape theoretical corrections with our collaboration. The technical support of the electronic and mechanical services of Subatech, IFIC, and Univ. of Jyv\"askyl\"a are also acknowledged. Enlightening discussions with Alejandro Sonzogni, Jouni Suhonen, and Marlom De Oliveira Ramalho are also acknowledged. We also thank B. C. Rasco for providing us with the published $^{92}$Rb beta decay feeding data for the comparisons. 

  \bibliography{bib_prl}{}

\providecommand{\noopsort}[1]{}\providecommand{\singleletter}[1]{#1}%
\begin{thebibliography}{58}%
\makeatletter
\providecommand \@ifxundefined [1]{%
 \@ifx{#1\undefined}
}%
\providecommand \@ifnum [1]{%
 \ifnum #1\expandafter \@firstoftwo
 \else \expandafter \@secondoftwo
 \fi
}%
\providecommand \@ifx [1]{%
 \ifx #1\expandafter \@firstoftwo
 \else \expandafter \@secondoftwo
 \fi
}%
\providecommand \natexlab [1]{#1}%
\providecommand \enquote  [1]{``#1''}%
\providecommand \bibnamefont  [1]{#1}%
\providecommand \bibfnamefont [1]{#1}%
\providecommand \citenamefont [1]{#1}%
\providecommand \href@noop [0]{\@secondoftwo}%
\providecommand \href [0]{\begingroup \@sanitize@url \@href}%
\providecommand \@href[1]{\@@startlink{#1}\@@href}%
\providecommand \@@href[1]{\endgroup#1\@@endlink}%
\providecommand \@sanitize@url [0]{\catcode `\\12\catcode `\$12\catcode
  `\&12\catcode `\#12\catcode `\^12\catcode `\_12\catcode `\%12\relax}%
\providecommand \@@startlink[1]{}%
\providecommand \@@endlink[0]{}%
\providecommand \url  [0]{\begingroup\@sanitize@url \@url }%
\providecommand \@url [1]{\endgroup\@href {#1}{\urlprefix }}%
\providecommand \urlprefix  [0]{URL }%
\providecommand \Eprint [0]{\href }%
\providecommand \doibase [0]{https://doi.org/}%
\providecommand \selectlanguage [0]{\@gobble}%
\providecommand \bibinfo  [0]{\@secondoftwo}%
\providecommand \bibfield  [0]{\@secondoftwo}%
\providecommand \translation [1]{[#1]}%
\providecommand \BibitemOpen [0]{}%
\providecommand \bibitemStop [0]{}%
\providecommand \bibitemNoStop [0]{.\EOS\space}%
\providecommand \EOS [0]{\spacefactor3000\relax}%
\providecommand \BibitemShut  [1]{\csname bibitem#1\endcsname}%
\let\auto@bib@innerbib\@empty
\bibitem [{\citenamefont {Evans}(1955)}]{EVANS}%
  \BibitemOpen
  \bibfield  {author} {\bibinfo {author} {\bibfnamefont {R.~D.}\ \bibnamefont
  {Evans}},\ }\href@noop {} {\emph {\bibinfo {title} {The Atomic Nucleus}}}\
  (\bibinfo  {publisher} {McGraw-Hill Publishing Company LTD},\ \bibinfo {year}
  {1955})\BibitemShut {NoStop}%
\bibitem [{\citenamefont {Konopinski}\ and\ \citenamefont
  {Uhlenbeck}(1935)}]{Konopinski1}%
  \BibitemOpen
  \bibfield  {author} {\bibinfo {author} {\bibfnamefont {E.~J.}\ \bibnamefont
  {Konopinski}}\ and\ \bibinfo {author} {\bibfnamefont {G.~E.}\ \bibnamefont
  {Uhlenbeck}},\ }\href {https://doi.org/10.1103/PhysRev.48.7} {\bibfield
  {journal} {\bibinfo  {journal} {Phys. Rev.}\ }\textbf {\bibinfo {volume}
  {48}},\ \bibinfo {pages} {7} (\bibinfo {year} {1935})}\BibitemShut {NoStop}%
\bibitem [{\citenamefont {Konopinski}(1943)}]{Konopinski2}%
  \BibitemOpen
  \bibfield  {author} {\bibinfo {author} {\bibfnamefont {E.~J.}\ \bibnamefont
  {Konopinski}},\ }\href {https://doi.org/10.1103/RevModPhys.15.209} {\bibfield
   {journal} {\bibinfo  {journal} {Rev. Mod. Phys.}\ }\textbf {\bibinfo
  {volume} {15}},\ \bibinfo {pages} {209} (\bibinfo {year} {1943})}\BibitemShut
  {NoStop}%
\bibitem [{\citenamefont {{Mougeot, X.}}\ \emph {et~al.}(2014)\citenamefont
  {{Mougeot, X.}}, \citenamefont {{Bé, M.-M.}},\ and\ \citenamefont {{Bisch,
  C.}}}]{Mougeot}%
  \BibitemOpen
  \bibfield  {author} {\bibinfo {author} {\bibnamefont {{Mougeot, X.}}},
  \bibinfo {author} {\bibnamefont {{Bé, M.-M.}}},\ and\ \bibinfo {author}
  {\bibnamefont {{Bisch, C.}}},\ }\href
  {https://doi.org/10.1051/radiopro/2014017} {\bibfield  {journal} {\bibinfo
  {journal} {Radioprotection}\ }\textbf {\bibinfo {volume} {49}},\ \bibinfo
  {pages} {269} (\bibinfo {year} {2014})}\BibitemShut {NoStop}%
\bibitem [{\citenamefont {Boswell}\ and\ \citenamefont
  {Brechbiel}(2007)}]{Boswell}%
  \BibitemOpen
  \bibfield  {author} {\bibinfo {author} {\bibfnamefont {C.~A.}\ \bibnamefont
  {Boswell}}\ and\ \bibinfo {author} {\bibfnamefont {M.~W.}\ \bibnamefont
  {Brechbiel}},\ }\href
  {https://doi.org/https://doi.org/10.1016/j.nucmedbio.2007.04.001} {\bibfield
  {journal} {\bibinfo  {journal} {Nucl. Med. Biol.}\ }\textbf {\bibinfo
  {volume} {34}},\ \bibinfo {pages} {757} (\bibinfo {year} {2007})}\BibitemShut
  {NoStop}%
\bibitem [{\citenamefont {Tengblad}\ \emph {et~al.}(1989)\citenamefont
  {Tengblad}, \citenamefont {Aleklett}, \citenamefont {{Von Dincklage}},
  \citenamefont {Lund}, \citenamefont {Nyman},\ and\ \citenamefont
  {Rudstam}}]{TENGBLAD}%
  \BibitemOpen
  \bibfield  {author} {\bibinfo {author} {\bibfnamefont {O.}~\bibnamefont
  {Tengblad}}, \bibinfo {author} {\bibfnamefont {K.}~\bibnamefont {Aleklett}},
  \bibinfo {author} {\bibfnamefont {R.}~\bibnamefont {{Von Dincklage}}},
  \bibinfo {author} {\bibfnamefont {E.}~\bibnamefont {Lund}}, \bibinfo {author}
  {\bibfnamefont {G.}~\bibnamefont {Nyman}},\ and\ \bibinfo {author}
  {\bibfnamefont {G.}~\bibnamefont {Rudstam}},\ }\href
  {https://doi.org/https://doi.org/10.1016/0375-9474(89)90258-3} {\bibfield
  {journal} {\bibinfo  {journal} {Nucl. Phys. A}\ }\textbf {\bibinfo {volume}
  {503}},\ \bibinfo {pages} {136} (\bibinfo {year} {1989})}\BibitemShut
  {NoStop}%
\bibitem [{\citenamefont {Cirgiliano}\ \emph {et~al.}(2019)\citenamefont
  {Cirgiliano}, \citenamefont {Garcia}, \citenamefont {Gazit}, \citenamefont
  {Naviliat-Cuncic}, \citenamefont {Savard},\ and\ \citenamefont
  {Young}}]{Cirgiliano}%
  \BibitemOpen
  \bibfield  {author} {\bibinfo {author} {\bibfnamefont {V.}~\bibnamefont
  {Cirgiliano}}, \bibinfo {author} {\bibfnamefont {A.}~\bibnamefont {Garcia}},
  \bibinfo {author} {\bibfnamefont {D.}~\bibnamefont {Gazit}}, \bibinfo
  {author} {\bibfnamefont {O.}~\bibnamefont {Naviliat-Cuncic}}, \bibinfo
  {author} {\bibfnamefont {G.}~\bibnamefont {Savard}},\ and\ \bibinfo {author}
  {\bibfnamefont {A.}~\bibnamefont {Young}},\ }\href
  {https://arxiv.org/abs/1907.02164} {\bibinfo {title} {Precision beta decay as
  a probe of new physics}} (\bibinfo {year} {2019}),\ \Eprint
  {https://arxiv.org/abs/1907.02164} {arXiv:1907.02164 [nucl-ex]} \BibitemShut
  {NoStop}%
\bibitem [{\citenamefont {Keukeleere}\ \emph {et~al.}(2024)\citenamefont
  {Keukeleere}, \citenamefont {Rozpedzik}, \citenamefont {Severijns},
  \citenamefont {Bodek}, \citenamefont {Hayen}, \citenamefont {Lojek},
  \citenamefont {Perkowski},\ and\ \citenamefont
  {Vanlangendonck}}]{Dekeukeleere}%
  \BibitemOpen
  \bibfield  {author} {\bibinfo {author} {\bibfnamefont {L.~D.}\ \bibnamefont
  {Keukeleere}}, \bibinfo {author} {\bibfnamefont {D.}~\bibnamefont
  {Rozpedzik}}, \bibinfo {author} {\bibfnamefont {N.}~\bibnamefont
  {Severijns}}, \bibinfo {author} {\bibfnamefont {K.}~\bibnamefont {Bodek}},
  \bibinfo {author} {\bibfnamefont {L.}~\bibnamefont {Hayen}}, \bibinfo
  {author} {\bibfnamefont {K.}~\bibnamefont {Lojek}}, \bibinfo {author}
  {\bibfnamefont {M.}~\bibnamefont {Perkowski}},\ and\ \bibinfo {author}
  {\bibfnamefont {S.}~\bibnamefont {Vanlangendonck}},\ }\href
  {https://arxiv.org/abs/2404.03140} {\bibinfo {title} {A first extraction of
  the weak magnetism form factor and fierz interference term from the
  $^{114}$in $\rightarrow$ $^{114}$sn gamow-teller transition}} (\bibinfo
  {year} {2024}),\ \Eprint {https://arxiv.org/abs/2404.03140} {arXiv:2404.03140
  [nucl-ex]} \BibitemShut {NoStop}%
\bibitem [{\citenamefont {Rozpedzik}\ \emph {et~al.}(2023)\citenamefont
  {Rozpedzik}, \citenamefont {De~Keukeleere}, \citenamefont {Bodek},
  \citenamefont {Hayen}, \citenamefont {Lojek}, \citenamefont {Perkowski},\
  and\ \citenamefont {Severijns}}]{Rozpedzik}%
  \BibitemOpen
  \bibfield  {author} {\bibinfo {author} {\bibfnamefont {D.}~\bibnamefont
  {Rozpedzik}}, \bibinfo {author} {\bibfnamefont {L.}~\bibnamefont
  {De~Keukeleere}}, \bibinfo {author} {\bibfnamefont {K.}~\bibnamefont
  {Bodek}}, \bibinfo {author} {\bibfnamefont {L.}~\bibnamefont {Hayen}},
  \bibinfo {author} {\bibfnamefont {K.}~\bibnamefont {Lojek}}, \bibinfo
  {author} {\bibfnamefont {M.}~\bibnamefont {Perkowski}},\ and\ \bibinfo
  {author} {\bibfnamefont {N.}~\bibnamefont {Severijns}},\ }\href
  {https://doi.org/10.1088/1742-6596/2586/1/012141} {\bibfield  {journal}
  {\bibinfo  {journal} {J. Phys. Conf. Ser.}\ }\textbf {\bibinfo {volume}
  {2586}},\ \bibinfo {pages} {012141} (\bibinfo {year} {2023})}\BibitemShut
  {NoStop}%
\bibitem [{\citenamefont {Kostensalo}\ \emph {et~al.}(2023)\citenamefont
  {Kostensalo}, \citenamefont {Lisi}, \citenamefont {Marrone},\ and\
  \citenamefont {Suhonen}}]{Kostensalo}%
  \BibitemOpen
  \bibfield  {author} {\bibinfo {author} {\bibfnamefont {J.}~\bibnamefont
  {Kostensalo}}, \bibinfo {author} {\bibfnamefont {E.}~\bibnamefont {Lisi}},
  \bibinfo {author} {\bibfnamefont {A.}~\bibnamefont {Marrone}},\ and\ \bibinfo
  {author} {\bibfnamefont {J.}~\bibnamefont {Suhonen}},\ }\href
  {https://doi.org/10.1103/PhysRevC.107.055502} {\bibfield  {journal} {\bibinfo
   {journal} {Phys. Rev. C}\ }\textbf {\bibinfo {volume} {107}},\ \bibinfo
  {pages} {055502} (\bibinfo {year} {2023})}\BibitemShut {NoStop}%
\bibitem [{\citenamefont {Pagnanini}\ \emph {et~al.}(2024)\citenamefont
  {Pagnanini}, \citenamefont {Benato}, \citenamefont {Carniti}, \citenamefont
  {Celi}, \citenamefont {Chiesa}, \citenamefont {Corbett}, \citenamefont
  {Dafinei}, \citenamefont {Di~Domizio}, \citenamefont {Di~Stefano},
  \citenamefont {Ghislandi}, \citenamefont {Gotti}, \citenamefont {Helis},
  \citenamefont {Knobel}, \citenamefont {Kostensalo}, \citenamefont {Kotila},
  \citenamefont {Nagorny}, \citenamefont {Pessina}, \citenamefont {Pirro},
  \citenamefont {Pozzi}, \citenamefont {Puiu}, \citenamefont {Quitadamo},
  \citenamefont {Sisti}, \citenamefont {Suhonen},\ and\ \citenamefont
  {Kuznetsov}}]{Pagnanini}%
  \BibitemOpen
  \bibfield  {author} {\bibinfo {author} {\bibfnamefont {L.}~\bibnamefont
  {Pagnanini}}, \bibinfo {author} {\bibfnamefont {G.}~\bibnamefont {Benato}},
  \bibinfo {author} {\bibfnamefont {P.}~\bibnamefont {Carniti}}, \bibinfo
  {author} {\bibfnamefont {E.}~\bibnamefont {Celi}}, \bibinfo {author}
  {\bibfnamefont {D.}~\bibnamefont {Chiesa}}, \bibinfo {author} {\bibfnamefont
  {J.}~\bibnamefont {Corbett}}, \bibinfo {author} {\bibfnamefont
  {I.}~\bibnamefont {Dafinei}}, \bibinfo {author} {\bibfnamefont
  {S.}~\bibnamefont {Di~Domizio}}, \bibinfo {author} {\bibfnamefont
  {P.}~\bibnamefont {Di~Stefano}}, \bibinfo {author} {\bibfnamefont
  {S.}~\bibnamefont {Ghislandi}}, \bibinfo {author} {\bibfnamefont
  {C.}~\bibnamefont {Gotti}}, \bibinfo {author} {\bibfnamefont {D.~L.}\
  \bibnamefont {Helis}}, \bibinfo {author} {\bibfnamefont {R.}~\bibnamefont
  {Knobel}}, \bibinfo {author} {\bibfnamefont {J.}~\bibnamefont {Kostensalo}},
  \bibinfo {author} {\bibfnamefont {J.}~\bibnamefont {Kotila}}, \bibinfo
  {author} {\bibfnamefont {S.}~\bibnamefont {Nagorny}}, \bibinfo {author}
  {\bibfnamefont {G.}~\bibnamefont {Pessina}}, \bibinfo {author} {\bibfnamefont
  {S.}~\bibnamefont {Pirro}}, \bibinfo {author} {\bibfnamefont
  {S.}~\bibnamefont {Pozzi}}, \bibinfo {author} {\bibfnamefont
  {A.}~\bibnamefont {Puiu}}, \bibinfo {author} {\bibfnamefont {S.}~\bibnamefont
  {Quitadamo}}, \bibinfo {author} {\bibfnamefont {M.}~\bibnamefont {Sisti}},
  \bibinfo {author} {\bibfnamefont {J.}~\bibnamefont {Suhonen}},\ and\ \bibinfo
  {author} {\bibfnamefont {S.}~\bibnamefont {Kuznetsov}},\ }\href
  {https://doi.org/10.1103/PhysRevLett.133.122501} {\bibfield  {journal}
  {\bibinfo  {journal} {Phys. Rev. Lett.}\ }\textbf {\bibinfo {volume} {133}},\
  \bibinfo {pages} {122501} (\bibinfo {year} {2024})}\BibitemShut {NoStop}%
\bibitem [{\citenamefont {Aker}\ \emph {et~al.}(2019)\citenamefont {Aker},
  \citenamefont {Altenm\"uller}, \citenamefont {Arenz}, \citenamefont
  {Babutzka}, \citenamefont {Barrett}, \citenamefont {Bauer}, \citenamefont
  {Beck}, \citenamefont {Beglarian}, \citenamefont {Behrens}, \citenamefont
  {Bergmann}, \citenamefont {Besserer}, \citenamefont {Blaum}, \citenamefont
  {Block}, \citenamefont {Bobien}, \citenamefont {Bokeloh}, \citenamefont
  {Bonn}, \citenamefont {Bornschein}, \citenamefont {Bornschein}, \citenamefont
  {Bouquet}, \citenamefont {Brunst}, \citenamefont {Caldwell}, \citenamefont
  {La~Cascio}, \citenamefont {Chilingaryan}, \citenamefont {Choi},
  \citenamefont {Corona}, \citenamefont {Debowski}, \citenamefont {Deffert},
  \citenamefont {Descher}, \citenamefont {Doe}, \citenamefont {Dragoun},
  \citenamefont {Drexlin}, \citenamefont {Dunmore}, \citenamefont {Dyba},
  \citenamefont {Edzards}, \citenamefont {Eisenbl\"atter}, \citenamefont
  {Eitel}, \citenamefont {Ellinger}, \citenamefont {Engel}, \citenamefont
  {Enomoto}, \citenamefont {Erhard}, \citenamefont {Eversheim}, \citenamefont
  {Fedkevych}, \citenamefont {Felden}, \citenamefont {Fischer}, \citenamefont
  {Flatt}, \citenamefont {Formaggio}, \citenamefont {Fr\"ankle}, \citenamefont
  {Franklin}, \citenamefont {Frankrone}, \citenamefont {Friedel}, \citenamefont
  {Fuchs}, \citenamefont {Fulst}, \citenamefont {Furse}, \citenamefont {Gauda},
  \citenamefont {Gemmeke}, \citenamefont {Gil}, \citenamefont {Gl\"uck},
  \citenamefont {G\"orhardt}, \citenamefont {Groh}, \citenamefont {Grohmann},
  \citenamefont {Gr\"ossle}, \citenamefont {Gumbsheimer}, \citenamefont
  {Ha~Minh}, \citenamefont {Hackenjos}, \citenamefont {Hannen}, \citenamefont
  {Harms}, \citenamefont {Hartmann}, \citenamefont {Hau\ss{}mann},
  \citenamefont {Heizmann}, \citenamefont {Helbing}, \citenamefont {Hickford},
  \citenamefont {Hilk}, \citenamefont {Hillen}, \citenamefont {Hillesheimer},
  \citenamefont {Hinz}, \citenamefont {H\"ohn}, \citenamefont {Holzapfel},
  \citenamefont {Holzmann}, \citenamefont {Houdy}, \citenamefont {Howe},
  \citenamefont {Huber}, \citenamefont {James}, \citenamefont {Jansen},
  \citenamefont {Kaboth}, \citenamefont {Karl}, \citenamefont {Kazachenko},
  \citenamefont {Kellerer}, \citenamefont {Kernert}, \citenamefont
  {Kippenbrock}, \citenamefont {Kleesiek}, \citenamefont {Klein}, \citenamefont
  {K\"ohler}, \citenamefont {K\"ollenberger}, \citenamefont {Kopmann},
  \citenamefont {Korzeczek}, \citenamefont {Kosmider}, \citenamefont
  {Koval\'{\i}k}, \citenamefont {Krasch}, \citenamefont {Kraus}, \citenamefont
  {Krause}, \citenamefont {Kuckert}, \citenamefont {Kuffner}, \citenamefont
  {Kunka}, \citenamefont {Lasserre}, \citenamefont {Le}, \citenamefont
  {Lebeda}, \citenamefont {Leber}, \citenamefont {Lehnert}, \citenamefont
  {Letnev}, \citenamefont {Leven}, \citenamefont {Lichter}, \citenamefont
  {Lobashev}, \citenamefont {Lokhov}, \citenamefont {Machatschek},
  \citenamefont {Malcherek}, \citenamefont {M\"uller}, \citenamefont {Mark},
  \citenamefont {Marsteller}, \citenamefont {Martin}, \citenamefont {Melzer},
  \citenamefont {Menshikov}, \citenamefont {Mertens}, \citenamefont {Minter},
  \citenamefont {Mirz}, \citenamefont {Monreal}, \citenamefont
  {Morales~Guzm\'an}, \citenamefont {M\"uller}, \citenamefont {Naumann},
  \citenamefont {Ndeke}, \citenamefont {Neumann}, \citenamefont {Niemes},
  \citenamefont {Noe}, \citenamefont {Oblath}, \citenamefont {Ortjohann},
  \citenamefont {Osipowicz}, \citenamefont {Ostrick}, \citenamefont {Otten},
  \citenamefont {Parno}, \citenamefont {Phillips}, \citenamefont {Plischke},
  \citenamefont {Pollithy}, \citenamefont {Poon}, \citenamefont {Pouryamout},
  \citenamefont {Prall}, \citenamefont {Priester}, \citenamefont {R\"ollig},
  \citenamefont {R\"ottele}, \citenamefont {Ranitzsch}, \citenamefont {Rest},
  \citenamefont {Rinderspacher}, \citenamefont {Robertson}, \citenamefont
  {Rodenbeck}, \citenamefont {Rohr}, \citenamefont {Roll}, \citenamefont
  {Rupp}, \citenamefont {Ry\ifmmode~\check{s}\else \v{s}\fi{}av\'y},
  \citenamefont {Sack}, \citenamefont {Saenz}, \citenamefont {Sch\"afer},
  \citenamefont {Schimpf}, \citenamefont {Schl\"osser}, \citenamefont
  {Schl\"osser}, \citenamefont {Schl\"uter}, \citenamefont {Sch\"on},
  \citenamefont {Sch\"onung}, \citenamefont {Schrank}, \citenamefont {Schulz},
  \citenamefont {Schwarz}, \citenamefont {Seitz-Moskaliuk}, \citenamefont
  {Seller}, \citenamefont {Sibille}, \citenamefont {Siegmann}, \citenamefont
  {Skasyrskaya}, \citenamefont {Slez\'ak}, \citenamefont
  {\ifmmode~\check{S}\else \v{S}\fi{}palek}, \citenamefont {Spanier},
  \citenamefont {Steidl}, \citenamefont {Steinbrink}, \citenamefont {Sturm},
  \citenamefont {Suesser}, \citenamefont {Sun}, \citenamefont
  {Tcherniakhovski}, \citenamefont {Telle}, \citenamefont {Th\"ummler},
  \citenamefont {Thorne}, \citenamefont {Titov}, \citenamefont {Tkachev},
  \citenamefont {Trost}, \citenamefont {Urban}, \citenamefont {V\'enos},
  \citenamefont {Valerius}, \citenamefont {VanDevender}, \citenamefont
  {Vianden}, \citenamefont {Vizcaya~Hern\'andez}, \citenamefont {Wall},
  \citenamefont {W\"ustling}, \citenamefont {Weber}, \citenamefont
  {Weinheimer}, \citenamefont {Weiss}, \citenamefont {Welte}, \citenamefont
  {Wendel}, \citenamefont {Wierman}, \citenamefont {Wilkerson}, \citenamefont
  {Wolf}, \citenamefont {Xu}, \citenamefont {Yen}, \citenamefont {Zacher},
  \citenamefont {Zadorozhny}, \citenamefont {Zbo\ifmmode~\check{r}\else
  \v{r}\fi{}il},\ and\ \citenamefont {Zeller}}]{Katrin}%
  \BibitemOpen
  \bibfield  {author} {\bibinfo {author} {\bibfnamefont {M.}~\bibnamefont
  {Aker}}, \bibinfo {author} {\bibfnamefont {K.}~\bibnamefont {Altenm\"uller}},
  \bibinfo {author} {\bibfnamefont {M.}~\bibnamefont {Arenz}}, \bibinfo
  {author} {\bibfnamefont {M.}~\bibnamefont {Babutzka}}, \bibinfo {author}
  {\bibfnamefont {J.}~\bibnamefont {Barrett}}, \bibinfo {author} {\bibfnamefont
  {S.}~\bibnamefont {Bauer}}, \bibinfo {author} {\bibfnamefont
  {M.}~\bibnamefont {Beck}}, \bibinfo {author} {\bibfnamefont {A.}~\bibnamefont
  {Beglarian}}, \bibinfo {author} {\bibfnamefont {J.}~\bibnamefont {Behrens}},
  \bibinfo {author} {\bibfnamefont {T.}~\bibnamefont {Bergmann}}, \bibinfo
  {author} {\bibfnamefont {U.}~\bibnamefont {Besserer}}, \bibinfo {author}
  {\bibfnamefont {K.}~\bibnamefont {Blaum}}, \bibinfo {author} {\bibfnamefont
  {F.}~\bibnamefont {Block}}, \bibinfo {author} {\bibfnamefont
  {S.}~\bibnamefont {Bobien}}, \bibinfo {author} {\bibfnamefont
  {K.}~\bibnamefont {Bokeloh}}, \bibinfo {author} {\bibfnamefont
  {J.}~\bibnamefont {Bonn}}, \bibinfo {author} {\bibfnamefont {B.}~\bibnamefont
  {Bornschein}}, \bibinfo {author} {\bibfnamefont {L.}~\bibnamefont
  {Bornschein}}, \bibinfo {author} {\bibfnamefont {H.}~\bibnamefont {Bouquet}},
  \bibinfo {author} {\bibfnamefont {T.}~\bibnamefont {Brunst}}, \bibinfo
  {author} {\bibfnamefont {T.~S.}\ \bibnamefont {Caldwell}}, \bibinfo {author}
  {\bibfnamefont {L.}~\bibnamefont {La~Cascio}}, \bibinfo {author}
  {\bibfnamefont {S.}~\bibnamefont {Chilingaryan}}, \bibinfo {author}
  {\bibfnamefont {W.}~\bibnamefont {Choi}}, \bibinfo {author} {\bibfnamefont
  {T.~J.}\ \bibnamefont {Corona}}, \bibinfo {author} {\bibfnamefont
  {K.}~\bibnamefont {Debowski}}, \bibinfo {author} {\bibfnamefont
  {M.}~\bibnamefont {Deffert}}, \bibinfo {author} {\bibfnamefont
  {M.}~\bibnamefont {Descher}}, \bibinfo {author} {\bibfnamefont {P.~J.}\
  \bibnamefont {Doe}}, \bibinfo {author} {\bibfnamefont {O.}~\bibnamefont
  {Dragoun}}, \bibinfo {author} {\bibfnamefont {G.}~\bibnamefont {Drexlin}},
  \bibinfo {author} {\bibfnamefont {J.~A.}\ \bibnamefont {Dunmore}}, \bibinfo
  {author} {\bibfnamefont {S.}~\bibnamefont {Dyba}}, \bibinfo {author}
  {\bibfnamefont {F.}~\bibnamefont {Edzards}}, \bibinfo {author} {\bibfnamefont
  {L.}~\bibnamefont {Eisenbl\"atter}}, \bibinfo {author} {\bibfnamefont
  {K.}~\bibnamefont {Eitel}}, \bibinfo {author} {\bibfnamefont
  {E.}~\bibnamefont {Ellinger}}, \bibinfo {author} {\bibfnamefont
  {R.}~\bibnamefont {Engel}}, \bibinfo {author} {\bibfnamefont
  {S.}~\bibnamefont {Enomoto}}, \bibinfo {author} {\bibfnamefont
  {M.}~\bibnamefont {Erhard}}, \bibinfo {author} {\bibfnamefont
  {D.}~\bibnamefont {Eversheim}}, \bibinfo {author} {\bibfnamefont
  {M.}~\bibnamefont {Fedkevych}}, \bibinfo {author} {\bibfnamefont
  {A.}~\bibnamefont {Felden}}, \bibinfo {author} {\bibfnamefont
  {S.}~\bibnamefont {Fischer}}, \bibinfo {author} {\bibfnamefont
  {B.}~\bibnamefont {Flatt}}, \bibinfo {author} {\bibfnamefont {J.~A.}\
  \bibnamefont {Formaggio}}, \bibinfo {author} {\bibfnamefont {F.~M.}\
  \bibnamefont {Fr\"ankle}}, \bibinfo {author} {\bibfnamefont {G.~B.}\
  \bibnamefont {Franklin}}, \bibinfo {author} {\bibfnamefont {H.}~\bibnamefont
  {Frankrone}}, \bibinfo {author} {\bibfnamefont {F.}~\bibnamefont {Friedel}},
  \bibinfo {author} {\bibfnamefont {D.}~\bibnamefont {Fuchs}}, \bibinfo
  {author} {\bibfnamefont {A.}~\bibnamefont {Fulst}}, \bibinfo {author}
  {\bibfnamefont {D.}~\bibnamefont {Furse}}, \bibinfo {author} {\bibfnamefont
  {K.}~\bibnamefont {Gauda}}, \bibinfo {author} {\bibfnamefont
  {H.}~\bibnamefont {Gemmeke}}, \bibinfo {author} {\bibfnamefont
  {W.}~\bibnamefont {Gil}}, \bibinfo {author} {\bibfnamefont {F.}~\bibnamefont
  {Gl\"uck}}, \bibinfo {author} {\bibfnamefont {S.}~\bibnamefont {G\"orhardt}},
  \bibinfo {author} {\bibfnamefont {S.}~\bibnamefont {Groh}}, \bibinfo {author}
  {\bibfnamefont {S.}~\bibnamefont {Grohmann}}, \bibinfo {author}
  {\bibfnamefont {R.}~\bibnamefont {Gr\"ossle}}, \bibinfo {author}
  {\bibfnamefont {R.}~\bibnamefont {Gumbsheimer}}, \bibinfo {author}
  {\bibfnamefont {M.}~\bibnamefont {Ha~Minh}}, \bibinfo {author} {\bibfnamefont
  {M.}~\bibnamefont {Hackenjos}}, \bibinfo {author} {\bibfnamefont
  {V.}~\bibnamefont {Hannen}}, \bibinfo {author} {\bibfnamefont
  {F.}~\bibnamefont {Harms}}, \bibinfo {author} {\bibfnamefont
  {J.}~\bibnamefont {Hartmann}}, \bibinfo {author} {\bibfnamefont
  {N.}~\bibnamefont {Hau\ss{}mann}}, \bibinfo {author} {\bibfnamefont
  {F.}~\bibnamefont {Heizmann}}, \bibinfo {author} {\bibfnamefont
  {K.}~\bibnamefont {Helbing}}, \bibinfo {author} {\bibfnamefont
  {S.}~\bibnamefont {Hickford}}, \bibinfo {author} {\bibfnamefont
  {D.}~\bibnamefont {Hilk}}, \bibinfo {author} {\bibfnamefont {B.}~\bibnamefont
  {Hillen}}, \bibinfo {author} {\bibfnamefont {D.}~\bibnamefont
  {Hillesheimer}}, \bibinfo {author} {\bibfnamefont {D.}~\bibnamefont {Hinz}},
  \bibinfo {author} {\bibfnamefont {T.}~\bibnamefont {H\"ohn}}, \bibinfo
  {author} {\bibfnamefont {B.}~\bibnamefont {Holzapfel}}, \bibinfo {author}
  {\bibfnamefont {S.}~\bibnamefont {Holzmann}}, \bibinfo {author}
  {\bibfnamefont {T.}~\bibnamefont {Houdy}}, \bibinfo {author} {\bibfnamefont
  {M.~A.}\ \bibnamefont {Howe}}, \bibinfo {author} {\bibfnamefont
  {A.}~\bibnamefont {Huber}}, \bibinfo {author} {\bibfnamefont {T.~M.}\
  \bibnamefont {James}}, \bibinfo {author} {\bibfnamefont {A.}~\bibnamefont
  {Jansen}}, \bibinfo {author} {\bibfnamefont {A.}~\bibnamefont {Kaboth}},
  \bibinfo {author} {\bibfnamefont {C.}~\bibnamefont {Karl}}, \bibinfo {author}
  {\bibfnamefont {O.}~\bibnamefont {Kazachenko}}, \bibinfo {author}
  {\bibfnamefont {J.}~\bibnamefont {Kellerer}}, \bibinfo {author}
  {\bibfnamefont {N.}~\bibnamefont {Kernert}}, \bibinfo {author} {\bibfnamefont
  {L.}~\bibnamefont {Kippenbrock}}, \bibinfo {author} {\bibfnamefont
  {M.}~\bibnamefont {Kleesiek}}, \bibinfo {author} {\bibfnamefont
  {M.}~\bibnamefont {Klein}}, \bibinfo {author} {\bibfnamefont
  {C.}~\bibnamefont {K\"ohler}}, \bibinfo {author} {\bibfnamefont
  {L.}~\bibnamefont {K\"ollenberger}}, \bibinfo {author} {\bibfnamefont
  {A.}~\bibnamefont {Kopmann}}, \bibinfo {author} {\bibfnamefont
  {M.}~\bibnamefont {Korzeczek}}, \bibinfo {author} {\bibfnamefont
  {A.}~\bibnamefont {Kosmider}}, \bibinfo {author} {\bibfnamefont
  {A.}~\bibnamefont {Koval\'{\i}k}}, \bibinfo {author} {\bibfnamefont
  {B.}~\bibnamefont {Krasch}}, \bibinfo {author} {\bibfnamefont
  {M.}~\bibnamefont {Kraus}}, \bibinfo {author} {\bibfnamefont
  {H.}~\bibnamefont {Krause}}, \bibinfo {author} {\bibfnamefont
  {L.}~\bibnamefont {Kuckert}}, \bibinfo {author} {\bibfnamefont
  {B.}~\bibnamefont {Kuffner}}, \bibinfo {author} {\bibfnamefont
  {N.}~\bibnamefont {Kunka}}, \bibinfo {author} {\bibfnamefont
  {T.}~\bibnamefont {Lasserre}}, \bibinfo {author} {\bibfnamefont {T.~L.}\
  \bibnamefont {Le}}, \bibinfo {author} {\bibfnamefont {O.}~\bibnamefont
  {Lebeda}}, \bibinfo {author} {\bibfnamefont {M.}~\bibnamefont {Leber}},
  \bibinfo {author} {\bibfnamefont {B.}~\bibnamefont {Lehnert}}, \bibinfo
  {author} {\bibfnamefont {J.}~\bibnamefont {Letnev}}, \bibinfo {author}
  {\bibfnamefont {F.}~\bibnamefont {Leven}}, \bibinfo {author} {\bibfnamefont
  {S.}~\bibnamefont {Lichter}}, \bibinfo {author} {\bibfnamefont {V.~M.}\
  \bibnamefont {Lobashev}}, \bibinfo {author} {\bibfnamefont {A.}~\bibnamefont
  {Lokhov}}, \bibinfo {author} {\bibfnamefont {M.}~\bibnamefont {Machatschek}},
  \bibinfo {author} {\bibfnamefont {E.}~\bibnamefont {Malcherek}}, \bibinfo
  {author} {\bibfnamefont {K.}~\bibnamefont {M\"uller}}, \bibinfo {author}
  {\bibfnamefont {M.}~\bibnamefont {Mark}}, \bibinfo {author} {\bibfnamefont
  {A.}~\bibnamefont {Marsteller}}, \bibinfo {author} {\bibfnamefont {E.~L.}\
  \bibnamefont {Martin}}, \bibinfo {author} {\bibfnamefont {C.}~\bibnamefont
  {Melzer}}, \bibinfo {author} {\bibfnamefont {A.}~\bibnamefont {Menshikov}},
  \bibinfo {author} {\bibfnamefont {S.}~\bibnamefont {Mertens}}, \bibinfo
  {author} {\bibfnamefont {L.~I.}\ \bibnamefont {Minter}}, \bibinfo {author}
  {\bibfnamefont {S.}~\bibnamefont {Mirz}}, \bibinfo {author} {\bibfnamefont
  {B.}~\bibnamefont {Monreal}}, \bibinfo {author} {\bibfnamefont {P.~I.}\
  \bibnamefont {Morales~Guzm\'an}}, \bibinfo {author} {\bibfnamefont
  {K.}~\bibnamefont {M\"uller}}, \bibinfo {author} {\bibfnamefont
  {U.}~\bibnamefont {Naumann}}, \bibinfo {author} {\bibfnamefont
  {W.}~\bibnamefont {Ndeke}}, \bibinfo {author} {\bibfnamefont
  {H.}~\bibnamefont {Neumann}}, \bibinfo {author} {\bibfnamefont
  {S.}~\bibnamefont {Niemes}}, \bibinfo {author} {\bibfnamefont
  {M.}~\bibnamefont {Noe}}, \bibinfo {author} {\bibfnamefont {N.~S.}\
  \bibnamefont {Oblath}}, \bibinfo {author} {\bibfnamefont {H.-W.}\
  \bibnamefont {Ortjohann}}, \bibinfo {author} {\bibfnamefont {A.}~\bibnamefont
  {Osipowicz}}, \bibinfo {author} {\bibfnamefont {B.}~\bibnamefont {Ostrick}},
  \bibinfo {author} {\bibfnamefont {E.}~\bibnamefont {Otten}}, \bibinfo
  {author} {\bibfnamefont {D.~S.}\ \bibnamefont {Parno}}, \bibinfo {author}
  {\bibfnamefont {D.~G.}\ \bibnamefont {Phillips}}, \bibinfo {author}
  {\bibfnamefont {P.}~\bibnamefont {Plischke}}, \bibinfo {author}
  {\bibfnamefont {A.}~\bibnamefont {Pollithy}}, \bibinfo {author}
  {\bibfnamefont {A.~W.~P.}\ \bibnamefont {Poon}}, \bibinfo {author}
  {\bibfnamefont {J.}~\bibnamefont {Pouryamout}}, \bibinfo {author}
  {\bibfnamefont {M.}~\bibnamefont {Prall}}, \bibinfo {author} {\bibfnamefont
  {F.}~\bibnamefont {Priester}}, \bibinfo {author} {\bibfnamefont
  {M.}~\bibnamefont {R\"ollig}}, \bibinfo {author} {\bibfnamefont
  {C.}~\bibnamefont {R\"ottele}}, \bibinfo {author} {\bibfnamefont {P.~C.-O.}\
  \bibnamefont {Ranitzsch}}, \bibinfo {author} {\bibfnamefont {O.}~\bibnamefont
  {Rest}}, \bibinfo {author} {\bibfnamefont {R.}~\bibnamefont {Rinderspacher}},
  \bibinfo {author} {\bibfnamefont {R.~G.~H.}\ \bibnamefont {Robertson}},
  \bibinfo {author} {\bibfnamefont {C.}~\bibnamefont {Rodenbeck}}, \bibinfo
  {author} {\bibfnamefont {P.}~\bibnamefont {Rohr}}, \bibinfo {author}
  {\bibfnamefont {C.}~\bibnamefont {Roll}}, \bibinfo {author} {\bibfnamefont
  {S.}~\bibnamefont {Rupp}}, \bibinfo {author} {\bibfnamefont {M.}~\bibnamefont
  {Ry\ifmmode~\check{s}\else \v{s}\fi{}av\'y}}, \bibinfo {author}
  {\bibfnamefont {R.}~\bibnamefont {Sack}}, \bibinfo {author} {\bibfnamefont
  {A.}~\bibnamefont {Saenz}}, \bibinfo {author} {\bibfnamefont
  {P.}~\bibnamefont {Sch\"afer}}, \bibinfo {author} {\bibfnamefont
  {L.}~\bibnamefont {Schimpf}}, \bibinfo {author} {\bibfnamefont
  {K.}~\bibnamefont {Schl\"osser}}, \bibinfo {author} {\bibfnamefont
  {M.}~\bibnamefont {Schl\"osser}}, \bibinfo {author} {\bibfnamefont
  {L.}~\bibnamefont {Schl\"uter}}, \bibinfo {author} {\bibfnamefont
  {H.}~\bibnamefont {Sch\"on}}, \bibinfo {author} {\bibfnamefont
  {K.}~\bibnamefont {Sch\"onung}}, \bibinfo {author} {\bibfnamefont
  {M.}~\bibnamefont {Schrank}}, \bibinfo {author} {\bibfnamefont
  {B.}~\bibnamefont {Schulz}}, \bibinfo {author} {\bibfnamefont
  {J.}~\bibnamefont {Schwarz}}, \bibinfo {author} {\bibfnamefont
  {H.}~\bibnamefont {Seitz-Moskaliuk}}, \bibinfo {author} {\bibfnamefont
  {W.}~\bibnamefont {Seller}}, \bibinfo {author} {\bibfnamefont
  {V.}~\bibnamefont {Sibille}}, \bibinfo {author} {\bibfnamefont
  {D.}~\bibnamefont {Siegmann}}, \bibinfo {author} {\bibfnamefont
  {A.}~\bibnamefont {Skasyrskaya}}, \bibinfo {author} {\bibfnamefont
  {M.}~\bibnamefont {Slez\'ak}}, \bibinfo {author} {\bibfnamefont
  {A.}~\bibnamefont {\ifmmode~\check{S}\else \v{S}\fi{}palek}}, \bibinfo
  {author} {\bibfnamefont {F.}~\bibnamefont {Spanier}}, \bibinfo {author}
  {\bibfnamefont {M.}~\bibnamefont {Steidl}}, \bibinfo {author} {\bibfnamefont
  {N.}~\bibnamefont {Steinbrink}}, \bibinfo {author} {\bibfnamefont
  {M.}~\bibnamefont {Sturm}}, \bibinfo {author} {\bibfnamefont
  {M.}~\bibnamefont {Suesser}}, \bibinfo {author} {\bibfnamefont
  {M.}~\bibnamefont {Sun}}, \bibinfo {author} {\bibfnamefont {D.}~\bibnamefont
  {Tcherniakhovski}}, \bibinfo {author} {\bibfnamefont {H.~H.}\ \bibnamefont
  {Telle}}, \bibinfo {author} {\bibfnamefont {T.}~\bibnamefont {Th\"ummler}},
  \bibinfo {author} {\bibfnamefont {L.~A.}\ \bibnamefont {Thorne}}, \bibinfo
  {author} {\bibfnamefont {N.}~\bibnamefont {Titov}}, \bibinfo {author}
  {\bibfnamefont {I.}~\bibnamefont {Tkachev}}, \bibinfo {author} {\bibfnamefont
  {N.}~\bibnamefont {Trost}}, \bibinfo {author} {\bibfnamefont
  {K.}~\bibnamefont {Urban}}, \bibinfo {author} {\bibfnamefont
  {D.}~\bibnamefont {V\'enos}}, \bibinfo {author} {\bibfnamefont
  {K.}~\bibnamefont {Valerius}}, \bibinfo {author} {\bibfnamefont {B.~A.}\
  \bibnamefont {VanDevender}}, \bibinfo {author} {\bibfnamefont
  {R.}~\bibnamefont {Vianden}}, \bibinfo {author} {\bibfnamefont {A.~P.}\
  \bibnamefont {Vizcaya~Hern\'andez}}, \bibinfo {author} {\bibfnamefont
  {B.~L.}\ \bibnamefont {Wall}}, \bibinfo {author} {\bibfnamefont
  {S.}~\bibnamefont {W\"ustling}}, \bibinfo {author} {\bibfnamefont
  {M.}~\bibnamefont {Weber}}, \bibinfo {author} {\bibfnamefont
  {C.}~\bibnamefont {Weinheimer}}, \bibinfo {author} {\bibfnamefont
  {C.}~\bibnamefont {Weiss}}, \bibinfo {author} {\bibfnamefont
  {S.}~\bibnamefont {Welte}}, \bibinfo {author} {\bibfnamefont
  {J.}~\bibnamefont {Wendel}}, \bibinfo {author} {\bibfnamefont {K.~J.}\
  \bibnamefont {Wierman}}, \bibinfo {author} {\bibfnamefont {J.~F.}\
  \bibnamefont {Wilkerson}}, \bibinfo {author} {\bibfnamefont {J.}~\bibnamefont
  {Wolf}}, \bibinfo {author} {\bibfnamefont {W.}~\bibnamefont {Xu}}, \bibinfo
  {author} {\bibfnamefont {Y.-R.}\ \bibnamefont {Yen}}, \bibinfo {author}
  {\bibfnamefont {M.}~\bibnamefont {Zacher}}, \bibinfo {author} {\bibfnamefont
  {S.}~\bibnamefont {Zadorozhny}}, \bibinfo {author} {\bibfnamefont
  {M.}~\bibnamefont {Zbo\ifmmode~\check{r}\else \v{r}\fi{}il}},\ and\ \bibinfo
  {author} {\bibfnamefont {G.}~\bibnamefont {Zeller}} (\bibinfo {collaboration}
  {KATRIN Collaboration}),\ }\href
  {https://doi.org/10.1103/PhysRevLett.123.221802} {\bibfield  {journal}
  {\bibinfo  {journal} {Phys. Rev. Lett.}\ }\textbf {\bibinfo {volume} {123}},\
  \bibinfo {pages} {221802} (\bibinfo {year} {2019})}\BibitemShut {NoStop}%
\bibitem [{\citenamefont {Ashtari~Esfahani}\ \emph {et~al.}(2023)\citenamefont
  {Ashtari~Esfahani}, \citenamefont {B\"oser}, \citenamefont {Buzinsky},
  \citenamefont {Carmona-Benitez}, \citenamefont {Claessens}, \citenamefont
  {de~Viveiros}, \citenamefont {Doe}, \citenamefont {Fertl}, \citenamefont
  {Formaggio}, \citenamefont {Gaison}, \citenamefont {Gladstone}, \citenamefont
  {Grando}, \citenamefont {Guigue}, \citenamefont {Hartse}, \citenamefont
  {Heeger}, \citenamefont {Huyan}, \citenamefont {Johnston}, \citenamefont
  {Jones}, \citenamefont {Kazkaz}, \citenamefont {LaRoque}, \citenamefont {Li},
  \citenamefont {Lindman}, \citenamefont {Machado}, \citenamefont {Marsteller},
  \citenamefont {Matth\'e}, \citenamefont {Mohiuddin}, \citenamefont {Monreal},
  \citenamefont {Mueller}, \citenamefont {Nikkel}, \citenamefont {Novitski},
  \citenamefont {Oblath}, \citenamefont {Pe\~na}, \citenamefont {Pettus},
  \citenamefont {Reimann}, \citenamefont {Robertson}, \citenamefont {Rosa
  De~Jes\'us}, \citenamefont {Rybka}, \citenamefont {Salda\~na}, \citenamefont
  {Schram}, \citenamefont {Slocum}, \citenamefont {Stachurska}, \citenamefont
  {Sun}, \citenamefont {Surukuchi}, \citenamefont {Tedeschi}, \citenamefont
  {Telles}, \citenamefont {Thomas}, \citenamefont {Thomas}, \citenamefont
  {Thorne}, \citenamefont {Th\"ummler}, \citenamefont {Tvrznikova},
  \citenamefont {Van De~Pontseele}, \citenamefont {VanDevender}, \citenamefont
  {Weintroub}, \citenamefont {Weiss}, \citenamefont {Wendler}, \citenamefont
  {Young}, \citenamefont {Zayas},\ and\ \citenamefont {Ziegler}}]{Ashtari1}%
  \BibitemOpen
  \bibfield  {author} {\bibinfo {author} {\bibfnamefont {A.}~\bibnamefont
  {Ashtari~Esfahani}}, \bibinfo {author} {\bibfnamefont {S.}~\bibnamefont
  {B\"oser}}, \bibinfo {author} {\bibfnamefont {N.}~\bibnamefont {Buzinsky}},
  \bibinfo {author} {\bibfnamefont {M.~C.}\ \bibnamefont {Carmona-Benitez}},
  \bibinfo {author} {\bibfnamefont {C.}~\bibnamefont {Claessens}}, \bibinfo
  {author} {\bibfnamefont {L.}~\bibnamefont {de~Viveiros}}, \bibinfo {author}
  {\bibfnamefont {P.~J.}\ \bibnamefont {Doe}}, \bibinfo {author} {\bibfnamefont
  {M.}~\bibnamefont {Fertl}}, \bibinfo {author} {\bibfnamefont {J.~A.}\
  \bibnamefont {Formaggio}}, \bibinfo {author} {\bibfnamefont {J.~K.}\
  \bibnamefont {Gaison}}, \bibinfo {author} {\bibfnamefont {L.}~\bibnamefont
  {Gladstone}}, \bibinfo {author} {\bibfnamefont {M.}~\bibnamefont {Grando}},
  \bibinfo {author} {\bibfnamefont {M.}~\bibnamefont {Guigue}}, \bibinfo
  {author} {\bibfnamefont {J.}~\bibnamefont {Hartse}}, \bibinfo {author}
  {\bibfnamefont {K.~M.}\ \bibnamefont {Heeger}}, \bibinfo {author}
  {\bibfnamefont {X.}~\bibnamefont {Huyan}}, \bibinfo {author} {\bibfnamefont
  {J.}~\bibnamefont {Johnston}}, \bibinfo {author} {\bibfnamefont {A.~M.}\
  \bibnamefont {Jones}}, \bibinfo {author} {\bibfnamefont {K.}~\bibnamefont
  {Kazkaz}}, \bibinfo {author} {\bibfnamefont {B.~H.}\ \bibnamefont {LaRoque}},
  \bibinfo {author} {\bibfnamefont {M.}~\bibnamefont {Li}}, \bibinfo {author}
  {\bibfnamefont {A.}~\bibnamefont {Lindman}}, \bibinfo {author} {\bibfnamefont
  {E.}~\bibnamefont {Machado}}, \bibinfo {author} {\bibfnamefont
  {A.}~\bibnamefont {Marsteller}}, \bibinfo {author} {\bibfnamefont
  {C.}~\bibnamefont {Matth\'e}}, \bibinfo {author} {\bibfnamefont
  {R.}~\bibnamefont {Mohiuddin}}, \bibinfo {author} {\bibfnamefont
  {B.}~\bibnamefont {Monreal}}, \bibinfo {author} {\bibfnamefont
  {R.}~\bibnamefont {Mueller}}, \bibinfo {author} {\bibfnamefont {J.~A.}\
  \bibnamefont {Nikkel}}, \bibinfo {author} {\bibfnamefont {E.}~\bibnamefont
  {Novitski}}, \bibinfo {author} {\bibfnamefont {N.~S.}\ \bibnamefont
  {Oblath}}, \bibinfo {author} {\bibfnamefont {J.~I.}\ \bibnamefont {Pe\~na}},
  \bibinfo {author} {\bibfnamefont {W.}~\bibnamefont {Pettus}}, \bibinfo
  {author} {\bibfnamefont {R.}~\bibnamefont {Reimann}}, \bibinfo {author}
  {\bibfnamefont {R.~G.~H.}\ \bibnamefont {Robertson}}, \bibinfo {author}
  {\bibfnamefont {D.}~\bibnamefont {Rosa De~Jes\'us}}, \bibinfo {author}
  {\bibfnamefont {G.}~\bibnamefont {Rybka}}, \bibinfo {author} {\bibfnamefont
  {L.}~\bibnamefont {Salda\~na}}, \bibinfo {author} {\bibfnamefont
  {M.}~\bibnamefont {Schram}}, \bibinfo {author} {\bibfnamefont {P.~L.}\
  \bibnamefont {Slocum}}, \bibinfo {author} {\bibfnamefont {J.}~\bibnamefont
  {Stachurska}}, \bibinfo {author} {\bibfnamefont {Y.-H.}\ \bibnamefont {Sun}},
  \bibinfo {author} {\bibfnamefont {P.~T.}\ \bibnamefont {Surukuchi}}, \bibinfo
  {author} {\bibfnamefont {J.~R.}\ \bibnamefont {Tedeschi}}, \bibinfo {author}
  {\bibfnamefont {A.~B.}\ \bibnamefont {Telles}}, \bibinfo {author}
  {\bibfnamefont {F.}~\bibnamefont {Thomas}}, \bibinfo {author} {\bibfnamefont
  {M.}~\bibnamefont {Thomas}}, \bibinfo {author} {\bibfnamefont {L.~A.}\
  \bibnamefont {Thorne}}, \bibinfo {author} {\bibfnamefont {T.}~\bibnamefont
  {Th\"ummler}}, \bibinfo {author} {\bibfnamefont {L.}~\bibnamefont
  {Tvrznikova}}, \bibinfo {author} {\bibfnamefont {W.}~\bibnamefont {Van
  De~Pontseele}}, \bibinfo {author} {\bibfnamefont {B.~A.}\ \bibnamefont
  {VanDevender}}, \bibinfo {author} {\bibfnamefont {J.}~\bibnamefont
  {Weintroub}}, \bibinfo {author} {\bibfnamefont {T.~E.}\ \bibnamefont
  {Weiss}}, \bibinfo {author} {\bibfnamefont {T.}~\bibnamefont {Wendler}},
  \bibinfo {author} {\bibfnamefont {A.}~\bibnamefont {Young}}, \bibinfo
  {author} {\bibfnamefont {E.}~\bibnamefont {Zayas}},\ and\ \bibinfo {author}
  {\bibfnamefont {A.}~\bibnamefont {Ziegler}} (\bibinfo {collaboration}
  {Project 8 Collaboration}),\ }\href
  {https://doi.org/10.1103/PhysRevLett.131.102502} {\bibfield  {journal}
  {\bibinfo  {journal} {Phys. Rev. Lett.}\ }\textbf {\bibinfo {volume} {131}},\
  \bibinfo {pages} {102502} (\bibinfo {year} {2023})}\BibitemShut {NoStop}%
\bibitem [{\citenamefont {Ashtari~Esfahani}\ \emph {et~al.}(2024)\citenamefont
  {Ashtari~Esfahani}, \citenamefont {B\"oser}, \citenamefont {Buzinsky},
  \citenamefont {Carmona-Benitez}, \citenamefont {Claessens}, \citenamefont
  {de~Viveiros}, \citenamefont {Doe}, \citenamefont {Fertl}, \citenamefont
  {Formaggio}, \citenamefont {Gaison}, \citenamefont {Gladstone}, \citenamefont
  {Guigue}, \citenamefont {Hartse}, \citenamefont {Heeger}, \citenamefont
  {Huyan}, \citenamefont {Jones}, \citenamefont {Kazkaz}, \citenamefont
  {LaRoque}, \citenamefont {Li}, \citenamefont {Lindman}, \citenamefont
  {Machado}, \citenamefont {Marsteller}, \citenamefont {Matth\'e},
  \citenamefont {Mohiuddin}, \citenamefont {Monreal}, \citenamefont {Mueller},
  \citenamefont {Nikkel}, \citenamefont {Novitski}, \citenamefont {Oblath},
  \citenamefont {Pe\~na}, \citenamefont {Pettus}, \citenamefont {Reimann},
  \citenamefont {Robertson}, \citenamefont {Rosa De~Jes\'us}, \citenamefont
  {Rybka}, \citenamefont {Salda\~na}, \citenamefont {Schram}, \citenamefont
  {Slocum}, \citenamefont {Stachurska}, \citenamefont {Sun}, \citenamefont
  {Surukuchi}, \citenamefont {Tedeschi}, \citenamefont {Telles}, \citenamefont
  {Thomas}, \citenamefont {Thomas}, \citenamefont {Thorne}, \citenamefont
  {Th\"ummler}, \citenamefont {Tvrznikova}, \citenamefont {Van De~Pontseele},
  \citenamefont {VanDevender}, \citenamefont {Weintroub}, \citenamefont
  {Weiss}, \citenamefont {Wendler}, \citenamefont {Young}, \citenamefont
  {Zayas},\ and\ \citenamefont {Ziegler}}]{Ashtari2}%
  \BibitemOpen
  \bibfield  {author} {\bibinfo {author} {\bibfnamefont {A.}~\bibnamefont
  {Ashtari~Esfahani}}, \bibinfo {author} {\bibfnamefont {S.}~\bibnamefont
  {B\"oser}}, \bibinfo {author} {\bibfnamefont {N.}~\bibnamefont {Buzinsky}},
  \bibinfo {author} {\bibfnamefont {M.~C.}\ \bibnamefont {Carmona-Benitez}},
  \bibinfo {author} {\bibfnamefont {C.}~\bibnamefont {Claessens}}, \bibinfo
  {author} {\bibfnamefont {L.}~\bibnamefont {de~Viveiros}}, \bibinfo {author}
  {\bibfnamefont {P.~J.}\ \bibnamefont {Doe}}, \bibinfo {author} {\bibfnamefont
  {M.}~\bibnamefont {Fertl}}, \bibinfo {author} {\bibfnamefont {J.~A.}\
  \bibnamefont {Formaggio}}, \bibinfo {author} {\bibfnamefont {J.~K.}\
  \bibnamefont {Gaison}}, \bibinfo {author} {\bibfnamefont {L.}~\bibnamefont
  {Gladstone}}, \bibinfo {author} {\bibfnamefont {M.}~\bibnamefont {Guigue}},
  \bibinfo {author} {\bibfnamefont {J.}~\bibnamefont {Hartse}}, \bibinfo
  {author} {\bibfnamefont {K.~M.}\ \bibnamefont {Heeger}}, \bibinfo {author}
  {\bibfnamefont {X.}~\bibnamefont {Huyan}}, \bibinfo {author} {\bibfnamefont
  {A.~M.}\ \bibnamefont {Jones}}, \bibinfo {author} {\bibfnamefont
  {K.}~\bibnamefont {Kazkaz}}, \bibinfo {author} {\bibfnamefont {B.~H.}\
  \bibnamefont {LaRoque}}, \bibinfo {author} {\bibfnamefont {M.}~\bibnamefont
  {Li}}, \bibinfo {author} {\bibfnamefont {A.}~\bibnamefont {Lindman}},
  \bibinfo {author} {\bibfnamefont {E.}~\bibnamefont {Machado}}, \bibinfo
  {author} {\bibfnamefont {A.}~\bibnamefont {Marsteller}}, \bibinfo {author}
  {\bibfnamefont {C.}~\bibnamefont {Matth\'e}}, \bibinfo {author}
  {\bibfnamefont {R.}~\bibnamefont {Mohiuddin}}, \bibinfo {author}
  {\bibfnamefont {B.}~\bibnamefont {Monreal}}, \bibinfo {author} {\bibfnamefont
  {R.}~\bibnamefont {Mueller}}, \bibinfo {author} {\bibfnamefont {J.~A.}\
  \bibnamefont {Nikkel}}, \bibinfo {author} {\bibfnamefont {E.}~\bibnamefont
  {Novitski}}, \bibinfo {author} {\bibfnamefont {N.~S.}\ \bibnamefont
  {Oblath}}, \bibinfo {author} {\bibfnamefont {J.~I.}\ \bibnamefont {Pe\~na}},
  \bibinfo {author} {\bibfnamefont {W.}~\bibnamefont {Pettus}}, \bibinfo
  {author} {\bibfnamefont {R.}~\bibnamefont {Reimann}}, \bibinfo {author}
  {\bibfnamefont {R.~G.~H.}\ \bibnamefont {Robertson}}, \bibinfo {author}
  {\bibfnamefont {D.}~\bibnamefont {Rosa De~Jes\'us}}, \bibinfo {author}
  {\bibfnamefont {G.}~\bibnamefont {Rybka}}, \bibinfo {author} {\bibfnamefont
  {L.}~\bibnamefont {Salda\~na}}, \bibinfo {author} {\bibfnamefont
  {M.}~\bibnamefont {Schram}}, \bibinfo {author} {\bibfnamefont {P.~L.}\
  \bibnamefont {Slocum}}, \bibinfo {author} {\bibfnamefont {J.}~\bibnamefont
  {Stachurska}}, \bibinfo {author} {\bibfnamefont {Y.-H.}\ \bibnamefont {Sun}},
  \bibinfo {author} {\bibfnamefont {P.~T.}\ \bibnamefont {Surukuchi}}, \bibinfo
  {author} {\bibfnamefont {J.~R.}\ \bibnamefont {Tedeschi}}, \bibinfo {author}
  {\bibfnamefont {A.~B.}\ \bibnamefont {Telles}}, \bibinfo {author}
  {\bibfnamefont {F.}~\bibnamefont {Thomas}}, \bibinfo {author} {\bibfnamefont
  {M.}~\bibnamefont {Thomas}}, \bibinfo {author} {\bibfnamefont {L.~A.}\
  \bibnamefont {Thorne}}, \bibinfo {author} {\bibfnamefont {T.}~\bibnamefont
  {Th\"ummler}}, \bibinfo {author} {\bibfnamefont {L.}~\bibnamefont
  {Tvrznikova}}, \bibinfo {author} {\bibfnamefont {W.}~\bibnamefont {Van
  De~Pontseele}}, \bibinfo {author} {\bibfnamefont {B.~A.}\ \bibnamefont
  {VanDevender}}, \bibinfo {author} {\bibfnamefont {J.}~\bibnamefont
  {Weintroub}}, \bibinfo {author} {\bibfnamefont {T.~E.}\ \bibnamefont
  {Weiss}}, \bibinfo {author} {\bibfnamefont {T.}~\bibnamefont {Wendler}},
  \bibinfo {author} {\bibfnamefont {A.}~\bibnamefont {Young}}, \bibinfo
  {author} {\bibfnamefont {E.}~\bibnamefont {Zayas}},\ and\ \bibinfo {author}
  {\bibfnamefont {A.}~\bibnamefont {Ziegler}} (\bibinfo {collaboration}
  {Project 8 Collaboration}),\ }\href
  {https://doi.org/10.1103/PhysRevC.109.035503} {\bibfield  {journal} {\bibinfo
   {journal} {Phys. Rev. C}\ }\textbf {\bibinfo {volume} {109}},\ \bibinfo
  {pages} {035503} (\bibinfo {year} {2024})}\BibitemShut {NoStop}%
\bibitem [{\citenamefont {Schreckenbach}\ \emph {et~al.}(1981)\citenamefont
  {Schreckenbach}, \citenamefont {Faust}, \citenamefont {{von Feilitzsch}},
  \citenamefont {Hahn}, \citenamefont {Hawerkamp},\ and\ \citenamefont
  {Vuilleumier}}]{SCHRECKENBACH1}%
  \BibitemOpen
  \bibfield  {author} {\bibinfo {author} {\bibfnamefont {K.}~\bibnamefont
  {Schreckenbach}}, \bibinfo {author} {\bibfnamefont {H.}~\bibnamefont
  {Faust}}, \bibinfo {author} {\bibfnamefont {F.}~\bibnamefont {{von
  Feilitzsch}}}, \bibinfo {author} {\bibfnamefont {A.}~\bibnamefont {Hahn}},
  \bibinfo {author} {\bibfnamefont {K.}~\bibnamefont {Hawerkamp}},\ and\
  \bibinfo {author} {\bibfnamefont {J.}~\bibnamefont {Vuilleumier}},\ }\href
  {https://doi.org/https://doi.org/10.1016/0370-2693(81)91120-5} {\bibfield
  {journal} {\bibinfo  {journal} {Phys. Lett. B}\ }\textbf {\bibinfo {volume}
  {99}},\ \bibinfo {pages} {251} (\bibinfo {year} {1981})}\BibitemShut
  {NoStop}%
\bibitem [{\citenamefont {Schreckenbach}\ \emph {et~al.}(1985)\citenamefont
  {Schreckenbach}, \citenamefont {Colvin}, \citenamefont {Gelletly},\ and\
  \citenamefont {{Von Feilitzsch}}}]{SCHRECKENBACH2}%
  \BibitemOpen
  \bibfield  {author} {\bibinfo {author} {\bibfnamefont {K.}~\bibnamefont
  {Schreckenbach}}, \bibinfo {author} {\bibfnamefont {G.}~\bibnamefont
  {Colvin}}, \bibinfo {author} {\bibfnamefont {W.}~\bibnamefont {Gelletly}},\
  and\ \bibinfo {author} {\bibfnamefont {F.}~\bibnamefont {{Von Feilitzsch}}},\
  }\href {https://doi.org/https://doi.org/10.1016/0370-2693(85)91337-1}
  {\bibfield  {journal} {\bibinfo  {journal} {Phys. Lett. B}\ }\textbf
  {\bibinfo {volume} {160}},\ \bibinfo {pages} {325} (\bibinfo {year}
  {1985})}\BibitemShut {NoStop}%
\bibitem [{\citenamefont {Hahn}\ \emph {et~al.}(1989)\citenamefont {Hahn},
  \citenamefont {Schreckenbach}, \citenamefont {Gelletly}, \citenamefont {{von
  Feilitzsch}}, \citenamefont {Colvin},\ and\ \citenamefont
  {Krusche}}]{SCHRECKENBACH3}%
  \BibitemOpen
  \bibfield  {author} {\bibinfo {author} {\bibfnamefont {A.}~\bibnamefont
  {Hahn}}, \bibinfo {author} {\bibfnamefont {K.}~\bibnamefont {Schreckenbach}},
  \bibinfo {author} {\bibfnamefont {W.}~\bibnamefont {Gelletly}}, \bibinfo
  {author} {\bibfnamefont {F.}~\bibnamefont {{von Feilitzsch}}}, \bibinfo
  {author} {\bibfnamefont {G.}~\bibnamefont {Colvin}},\ and\ \bibinfo {author}
  {\bibfnamefont {B.}~\bibnamefont {Krusche}},\ }\href
  {https://doi.org/https://doi.org/10.1016/0370-2693(89)91598-0} {\bibfield
  {journal} {\bibinfo  {journal} {Phys. Lett. B}\ }\textbf {\bibinfo {volume}
  {218}},\ \bibinfo {pages} {365} (\bibinfo {year} {1989})}\BibitemShut
  {NoStop}%
\bibitem [{\citenamefont {Haag}\ \emph {et~al.}(2014)\citenamefont {Haag},
  \citenamefont {G\"utlein}, \citenamefont {Hofmann}, \citenamefont {Oberauer},
  \citenamefont {Potzel}, \citenamefont {Schreckenbach},\ and\ \citenamefont
  {Wagner}}]{HAAG}%
  \BibitemOpen
  \bibfield  {author} {\bibinfo {author} {\bibfnamefont {N.}~\bibnamefont
  {Haag}}, \bibinfo {author} {\bibfnamefont {A.}~\bibnamefont {G\"utlein}},
  \bibinfo {author} {\bibfnamefont {M.}~\bibnamefont {Hofmann}}, \bibinfo
  {author} {\bibfnamefont {L.}~\bibnamefont {Oberauer}}, \bibinfo {author}
  {\bibfnamefont {W.}~\bibnamefont {Potzel}}, \bibinfo {author} {\bibfnamefont
  {K.}~\bibnamefont {Schreckenbach}},\ and\ \bibinfo {author} {\bibfnamefont
  {F.~M.}\ \bibnamefont {Wagner}},\ }\href
  {https://doi.org/10.1103/PhysRevLett.112.122501} {\bibfield  {journal}
  {\bibinfo  {journal} {Phys. Rev. Lett.}\ }\textbf {\bibinfo {volume} {112}},\
  \bibinfo {pages} {122501} (\bibinfo {year} {2014})}\BibitemShut {NoStop}%
\bibitem [{\citenamefont {Huber}(2011)}]{HUBER}%
  \BibitemOpen
  \bibfield  {author} {\bibinfo {author} {\bibfnamefont {P.}~\bibnamefont
  {Huber}},\ }\href {https://doi.org/10.1103/PhysRevC.84.024617} {\bibfield
  {journal} {\bibinfo  {journal} {Phys. Rev. C}\ }\textbf {\bibinfo {volume}
  {84}},\ \bibinfo {pages} {024617} (\bibinfo {year} {2011})}\BibitemShut
  {NoStop}%
\bibitem [{\citenamefont {Mueller}\ \emph {et~al.}(2011)\citenamefont
  {Mueller}, \citenamefont {Lhuillier}, \citenamefont {Fallot}, \citenamefont
  {Letourneau}, \citenamefont {Cormon}, \citenamefont {Fechner}, \citenamefont
  {Giot}, \citenamefont {Lasserre}, \citenamefont {Martino}, \citenamefont
  {Mention}, \citenamefont {Porta},\ and\ \citenamefont {Yermia}}]{MUELLER}%
  \BibitemOpen
  \bibfield  {author} {\bibinfo {author} {\bibfnamefont {T.~A.}\ \bibnamefont
  {Mueller}}, \bibinfo {author} {\bibfnamefont {D.}~\bibnamefont {Lhuillier}},
  \bibinfo {author} {\bibfnamefont {M.}~\bibnamefont {Fallot}}, \bibinfo
  {author} {\bibfnamefont {A.}~\bibnamefont {Letourneau}}, \bibinfo {author}
  {\bibfnamefont {S.}~\bibnamefont {Cormon}}, \bibinfo {author} {\bibfnamefont
  {M.}~\bibnamefont {Fechner}}, \bibinfo {author} {\bibfnamefont
  {L.}~\bibnamefont {Giot}}, \bibinfo {author} {\bibfnamefont {T.}~\bibnamefont
  {Lasserre}}, \bibinfo {author} {\bibfnamefont {J.}~\bibnamefont {Martino}},
  \bibinfo {author} {\bibfnamefont {G.}~\bibnamefont {Mention}}, \bibinfo
  {author} {\bibfnamefont {A.}~\bibnamefont {Porta}},\ and\ \bibinfo {author}
  {\bibfnamefont {F.}~\bibnamefont {Yermia}},\ }\href
  {https://doi.org/10.1103/PhysRevC.83.054615} {\bibfield  {journal} {\bibinfo
  {journal} {Phys. Rev. C}\ }\textbf {\bibinfo {volume} {83}},\ \bibinfo
  {pages} {054615} (\bibinfo {year} {2011})}\BibitemShut {NoStop}%
\bibitem [{\citenamefont {Fallot}\ \emph {et~al.}(2017)\citenamefont {Fallot},
  \citenamefont {Porta}, \citenamefont {Meur}, \citenamefont {Briz},
  \citenamefont {Zakari-Issoufou}, \citenamefont {Guadilla}, \citenamefont
  {Algora}, \citenamefont {Taìn}, \citenamefont {Valencia}, \citenamefont
  {Rice}, \citenamefont {Bui}, \citenamefont {Cormon}, \citenamefont
  {Estienne}, \citenamefont {Agramunt}, \citenamefont {Äystö}, \citenamefont
  {Batist}, \citenamefont {Bowry}, \citenamefont {Caballero-Folch},
  \citenamefont {Cano-Ott}, \citenamefont {Cucoanes}, \citenamefont {Elomaa},
  \citenamefont {Eronen}, \citenamefont {Estévez}, \citenamefont {Farrelly},
  \citenamefont {Fraile}, \citenamefont {Fleming}, \citenamefont {Ganogliu},
  \citenamefont {Garcia}, \citenamefont {Gelletly}, \citenamefont
  {Gomez-Hornillos}, \citenamefont {Gorelov}, \citenamefont {Gorlychev},
  \citenamefont {Hakala}, \citenamefont {Jokinen}, \citenamefont {Jordan},
  \citenamefont {Kankainen}, \citenamefont {Karvonen}, \citenamefont
  {Kolhinen}, \citenamefont {Kondev}, \citenamefont {Koponen}, \citenamefont
  {Lebois}, \citenamefont {Martinez}, \citenamefont {Mason}, \citenamefont
  {Mendoza}, \citenamefont {Molina}, \citenamefont {Monserrate}, \citenamefont
  {Montaner-Pizá}, \citenamefont {Moore}, \citenamefont {Nácher},
  \citenamefont {Orrigo}, \citenamefont {Penttilä}, \citenamefont {Perez},
  \citenamefont {Podolyák}, \citenamefont {Pohjalainen}, \citenamefont
  {Regan}, \citenamefont {Reinikainen}, \citenamefont {Reponen}, \citenamefont
  {Rinta-Antila}, \citenamefont {Rissanen}, \citenamefont {Rubio},
  \citenamefont {Shiba}, \citenamefont {Sonnenschein}, \citenamefont
  {Sonzogni}, \citenamefont {Sublet}, \citenamefont {Vedia}, \citenamefont
  {Voss}, \citenamefont {Weber},\ and\ \citenamefont {Wilson}}]{FALLOT}%
  \BibitemOpen
  \bibfield  {author} {\bibinfo {author} {\bibfnamefont {M.}~\bibnamefont
  {Fallot}}, \bibinfo {author} {\bibfnamefont {A.}~\bibnamefont {Porta}},
  \bibinfo {author} {\bibfnamefont {L.~L.}\ \bibnamefont {Meur}}, \bibinfo
  {author} {\bibfnamefont {J.~A.}\ \bibnamefont {Briz}}, \bibinfo {author}
  {\bibfnamefont {A.-A.}\ \bibnamefont {Zakari-Issoufou}}, \bibinfo {author}
  {\bibfnamefont {V.}~\bibnamefont {Guadilla}}, \bibinfo {author}
  {\bibfnamefont {A.}~\bibnamefont {Algora}}, \bibinfo {author} {\bibfnamefont
  {J.-L.}\ \bibnamefont {Taìn}}, \bibinfo {author} {\bibfnamefont
  {E.}~\bibnamefont {Valencia}}, \bibinfo {author} {\bibfnamefont
  {S.}~\bibnamefont {Rice}}, \bibinfo {author} {\bibfnamefont {V.}~\bibnamefont
  {Bui}}, \bibinfo {author} {\bibfnamefont {S.}~\bibnamefont {Cormon}},
  \bibinfo {author} {\bibfnamefont {M.}~\bibnamefont {Estienne}}, \bibinfo
  {author} {\bibfnamefont {J.}~\bibnamefont {Agramunt}}, \bibinfo {author}
  {\bibfnamefont {J.}~\bibnamefont {Äystö}}, \bibinfo {author} {\bibfnamefont
  {L.}~\bibnamefont {Batist}}, \bibinfo {author} {\bibfnamefont
  {M.}~\bibnamefont {Bowry}}, \bibinfo {author} {\bibfnamefont
  {R.}~\bibnamefont {Caballero-Folch}}, \bibinfo {author} {\bibfnamefont
  {D.}~\bibnamefont {Cano-Ott}}, \bibinfo {author} {\bibfnamefont
  {A.}~\bibnamefont {Cucoanes}}, \bibinfo {author} {\bibfnamefont {V.-V.}\
  \bibnamefont {Elomaa}}, \bibinfo {author} {\bibfnamefont {T.}~\bibnamefont
  {Eronen}}, \bibinfo {author} {\bibfnamefont {E.}~\bibnamefont {Estévez}},
  \bibinfo {author} {\bibfnamefont {G.}~\bibnamefont {Farrelly}}, \bibinfo
  {author} {\bibfnamefont {L.}~\bibnamefont {Fraile}}, \bibinfo {author}
  {\bibfnamefont {M.}~\bibnamefont {Fleming}}, \bibinfo {author} {\bibfnamefont
  {E.}~\bibnamefont {Ganogliu}}, \bibinfo {author} {\bibfnamefont
  {A.}~\bibnamefont {Garcia}}, \bibinfo {author} {\bibfnamefont
  {W.}~\bibnamefont {Gelletly}}, \bibinfo {author} {\bibfnamefont
  {M.}~\bibnamefont {Gomez-Hornillos}}, \bibinfo {author} {\bibfnamefont
  {D.}~\bibnamefont {Gorelov}}, \bibinfo {author} {\bibfnamefont
  {V.}~\bibnamefont {Gorlychev}}, \bibinfo {author} {\bibfnamefont
  {J.}~\bibnamefont {Hakala}}, \bibinfo {author} {\bibfnamefont
  {A.}~\bibnamefont {Jokinen}}, \bibinfo {author} {\bibfnamefont
  {M.}~\bibnamefont {Jordan}}, \bibinfo {author} {\bibfnamefont
  {A.}~\bibnamefont {Kankainen}}, \bibinfo {author} {\bibfnamefont
  {P.}~\bibnamefont {Karvonen}}, \bibinfo {author} {\bibfnamefont
  {V.}~\bibnamefont {Kolhinen}}, \bibinfo {author} {\bibfnamefont
  {F.}~\bibnamefont {Kondev}}, \bibinfo {author} {\bibfnamefont
  {J.}~\bibnamefont {Koponen}}, \bibinfo {author} {\bibfnamefont
  {M.}~\bibnamefont {Lebois}}, \bibinfo {author} {\bibfnamefont
  {T.}~\bibnamefont {Martinez}}, \bibinfo {author} {\bibfnamefont
  {P.}~\bibnamefont {Mason}}, \bibinfo {author} {\bibfnamefont
  {E.}~\bibnamefont {Mendoza}}, \bibinfo {author} {\bibfnamefont
  {F.}~\bibnamefont {Molina}}, \bibinfo {author} {\bibfnamefont
  {M.}~\bibnamefont {Monserrate}}, \bibinfo {author} {\bibfnamefont
  {A.}~\bibnamefont {Montaner-Pizá}}, \bibinfo {author} {\bibfnamefont
  {I.}~\bibnamefont {Moore}}, \bibinfo {author} {\bibfnamefont
  {E.}~\bibnamefont {Nácher}}, \bibinfo {author} {\bibfnamefont
  {S.}~\bibnamefont {Orrigo}}, \bibinfo {author} {\bibfnamefont
  {H.}~\bibnamefont {Penttilä}}, \bibinfo {author} {\bibfnamefont
  {A.}~\bibnamefont {Perez}}, \bibinfo {author} {\bibfnamefont
  {Z.}~\bibnamefont {Podolyák}}, \bibinfo {author} {\bibfnamefont
  {I.}~\bibnamefont {Pohjalainen}}, \bibinfo {author} {\bibfnamefont
  {P.}~\bibnamefont {Regan}}, \bibinfo {author} {\bibfnamefont
  {J.}~\bibnamefont {Reinikainen}}, \bibinfo {author} {\bibfnamefont
  {M.}~\bibnamefont {Reponen}}, \bibinfo {author} {\bibfnamefont
  {S.}~\bibnamefont {Rinta-Antila}}, \bibinfo {author} {\bibfnamefont
  {J.}~\bibnamefont {Rissanen}}, \bibinfo {author} {\bibfnamefont
  {B.}~\bibnamefont {Rubio}}, \bibinfo {author} {\bibfnamefont
  {T.}~\bibnamefont {Shiba}}, \bibinfo {author} {\bibfnamefont
  {V.}~\bibnamefont {Sonnenschein}}, \bibinfo {author} {\bibfnamefont
  {A.}~\bibnamefont {Sonzogni}}, \bibinfo {author} {\bibfnamefont {J.-C.}\
  \bibnamefont {Sublet}}, \bibinfo {author} {\bibfnamefont {V.}~\bibnamefont
  {Vedia}}, \bibinfo {author} {\bibfnamefont {A.}~\bibnamefont {Voss}},
  \bibinfo {author} {\bibfnamefont {C.}~\bibnamefont {Weber}},\ and\ \bibinfo
  {author} {\bibfnamefont {J.}~\bibnamefont {Wilson}},\ }\href
  {https://doi.org/10.1051/epjconf/201714610002} {\bibfield  {journal}
  {\bibinfo  {journal} {EPJ Web Conf.}\ }\textbf {\bibinfo {volume} {146}},\
  \bibinfo {pages} {10002} (\bibinfo {year} {2017})}\BibitemShut {NoStop}%
\bibitem [{\citenamefont {Estienne}\ \emph {et~al.}(2019)\citenamefont
  {Estienne}, \citenamefont {Fallot}, \citenamefont {Algora}, \citenamefont
  {Briz-Monago}, \citenamefont {Bui}, \citenamefont {Cormon}, \citenamefont
  {Gelletly}, \citenamefont {Giot}, \citenamefont {Guadilla}, \citenamefont
  {Jordan}, \citenamefont {Le~Meur}, \citenamefont {Porta}, \citenamefont
  {Rice}, \citenamefont {Rubio}, \citenamefont {Ta\'{\i}n}, \citenamefont
  {Valencia},\ and\ \citenamefont {Zakari-Issoufou}}]{ESTIENNE}%
  \BibitemOpen
  \bibfield  {author} {\bibinfo {author} {\bibfnamefont {M.}~\bibnamefont
  {Estienne}}, \bibinfo {author} {\bibfnamefont {M.}~\bibnamefont {Fallot}},
  \bibinfo {author} {\bibfnamefont {A.}~\bibnamefont {Algora}}, \bibinfo
  {author} {\bibfnamefont {J.}~\bibnamefont {Briz-Monago}}, \bibinfo {author}
  {\bibfnamefont {V.~M.}\ \bibnamefont {Bui}}, \bibinfo {author} {\bibfnamefont
  {S.}~\bibnamefont {Cormon}}, \bibinfo {author} {\bibfnamefont
  {W.}~\bibnamefont {Gelletly}}, \bibinfo {author} {\bibfnamefont
  {L.}~\bibnamefont {Giot}}, \bibinfo {author} {\bibfnamefont {V.}~\bibnamefont
  {Guadilla}}, \bibinfo {author} {\bibfnamefont {D.}~\bibnamefont {Jordan}},
  \bibinfo {author} {\bibfnamefont {L.}~\bibnamefont {Le~Meur}}, \bibinfo
  {author} {\bibfnamefont {A.}~\bibnamefont {Porta}}, \bibinfo {author}
  {\bibfnamefont {S.}~\bibnamefont {Rice}}, \bibinfo {author} {\bibfnamefont
  {B.}~\bibnamefont {Rubio}}, \bibinfo {author} {\bibfnamefont {J.~L.}\
  \bibnamefont {Ta\'{\i}n}}, \bibinfo {author} {\bibfnamefont {E.}~\bibnamefont
  {Valencia}},\ and\ \bibinfo {author} {\bibfnamefont {A.-A.}\ \bibnamefont
  {Zakari-Issoufou}},\ }\href {https://doi.org/10.1103/PhysRevLett.123.022502}
  {\bibfield  {journal} {\bibinfo  {journal} {Phys. Rev. Lett.}\ }\textbf
  {\bibinfo {volume} {123}},\ \bibinfo {pages} {022502} (\bibinfo {year}
  {2019})}\BibitemShut {NoStop}%
\bibitem [{\citenamefont {Sonzogni}\ \emph {et~al.}(2015)\citenamefont
  {Sonzogni}, \citenamefont {Johnson},\ and\ \citenamefont
  {McCutchan}}]{SONZOGNI2}%
  \BibitemOpen
  \bibfield  {author} {\bibinfo {author} {\bibfnamefont {A.~A.}\ \bibnamefont
  {Sonzogni}}, \bibinfo {author} {\bibfnamefont {T.~D.}\ \bibnamefont
  {Johnson}},\ and\ \bibinfo {author} {\bibfnamefont {E.~A.}\ \bibnamefont
  {McCutchan}},\ }\href {https://doi.org/10.1103/PhysRevC.91.011301} {\bibfield
   {journal} {\bibinfo  {journal} {Phys. Rev. C}\ }\textbf {\bibinfo {volume}
  {91}},\ \bibinfo {pages} {011301} (\bibinfo {year} {2015})}\BibitemShut
  {NoStop}%
\bibitem [{\citenamefont {P\'eriss\'e}\ \emph {et~al.}(2023)\citenamefont
  {P\'eriss\'e}, \citenamefont {Onillon}, \citenamefont {Mougeot},
  \citenamefont {Vivier}, \citenamefont {Lasserre}, \citenamefont {Letourneau},
  \citenamefont {Lhuillier},\ and\ \citenamefont {Mention}}]{PERISSE}%
  \BibitemOpen
  \bibfield  {author} {\bibinfo {author} {\bibfnamefont {L.}~\bibnamefont
  {P\'eriss\'e}}, \bibinfo {author} {\bibfnamefont {A.}~\bibnamefont
  {Onillon}}, \bibinfo {author} {\bibfnamefont {X.}~\bibnamefont {Mougeot}},
  \bibinfo {author} {\bibfnamefont {M.}~\bibnamefont {Vivier}}, \bibinfo
  {author} {\bibfnamefont {T.}~\bibnamefont {Lasserre}}, \bibinfo {author}
  {\bibfnamefont {A.}~\bibnamefont {Letourneau}}, \bibinfo {author}
  {\bibfnamefont {D.}~\bibnamefont {Lhuillier}},\ and\ \bibinfo {author}
  {\bibfnamefont {G.}~\bibnamefont {Mention}},\ }\href
  {https://doi.org/10.1103/PhysRevC.108.055501} {\bibfield  {journal} {\bibinfo
   {journal} {Phys. Rev. C}\ }\textbf {\bibinfo {volume} {108}},\ \bibinfo
  {pages} {055501} (\bibinfo {year} {2023})}\BibitemShut {NoStop}%
\bibitem [{\citenamefont {Sonzogni}\ \emph {et~al.}(2017)\citenamefont
  {Sonzogni}, \citenamefont {McCutchan},\ and\ \citenamefont
  {Hayes}}]{SONZOGNI1}%
  \BibitemOpen
  \bibfield  {author} {\bibinfo {author} {\bibfnamefont {A.~A.}\ \bibnamefont
  {Sonzogni}}, \bibinfo {author} {\bibfnamefont {E.~A.}\ \bibnamefont
  {McCutchan}},\ and\ \bibinfo {author} {\bibfnamefont {A.~C.}\ \bibnamefont
  {Hayes}},\ }\href {https://doi.org/10.1103/PhysRevLett.119.112501} {\bibfield
   {journal} {\bibinfo  {journal} {Phys. Rev. Lett.}\ }\textbf {\bibinfo
  {volume} {119}},\ \bibinfo {pages} {112501} (\bibinfo {year}
  {2017})}\BibitemShut {NoStop}%
\bibitem [{\citenamefont {Hardy}\ \emph {et~al.}(1977)\citenamefont {Hardy},
  \citenamefont {Carraz}, \citenamefont {Jonson},\ and\ \citenamefont
  {Hansen}}]{HARDY}%
  \BibitemOpen
  \bibfield  {author} {\bibinfo {author} {\bibfnamefont {J.}~\bibnamefont
  {Hardy}}, \bibinfo {author} {\bibfnamefont {L.}~\bibnamefont {Carraz}},
  \bibinfo {author} {\bibfnamefont {B.}~\bibnamefont {Jonson}},\ and\ \bibinfo
  {author} {\bibfnamefont {P.}~\bibnamefont {Hansen}},\ }\href
  {https://doi.org/https://doi.org/10.1016/0370-2693(77)90223-4} {\bibfield
  {journal} {\bibinfo  {journal} {Phys. Lett. B}\ }\textbf {\bibinfo {volume}
  {71}},\ \bibinfo {pages} {307} (\bibinfo {year} {1977})}\BibitemShut
  {NoStop}%
\bibitem [{\citenamefont {{G. Mention, M. Fechner, Th. Lasserre, Th. A.
  Mueller, D. Lhuillier, M. Cribier, and A. Letourneau}}(2011)}]{MENTION}%
  \BibitemOpen
  \bibfield  {author} {\bibinfo {author} {\bibnamefont {{G. Mention, M.
  Fechner, Th. Lasserre, Th. A. Mueller, D. Lhuillier, M. Cribier, and A.
  Letourneau}}},\ }\href {https://doi.org/10.1103/PhysRevD.83.073006}
  {\bibfield  {journal} {\bibinfo  {journal} {Phys. Rev. D}\ }\textbf {\bibinfo
  {volume} {83}},\ \bibinfo {pages} {073006} (\bibinfo {year}
  {2011})}\BibitemShut {NoStop}%
\bibitem [{\citenamefont {Apollonio}\ \emph {et~al.}(1999)\citenamefont
  {Apollonio}, \citenamefont {Baldini}, \citenamefont {Bemporad}, \citenamefont
  {Caffau}, \citenamefont {Cei}, \citenamefont {Déclais}, \citenamefont {{de
  Kerret}}, \citenamefont {Dieterle}, \citenamefont {Etenko}, \citenamefont
  {George}, \citenamefont {Giannini}, \citenamefont {Grassi}, \citenamefont
  {Kozlov}, \citenamefont {Kropp}, \citenamefont {Kryn}, \citenamefont
  {Laiman}, \citenamefont {Lane}, \citenamefont {Lefièvre}, \citenamefont
  {Machulin}, \citenamefont {Martemyanov}, \citenamefont {Martemyanov},
  \citenamefont {Mikaelyan}, \citenamefont {Nicolò}, \citenamefont
  {Obolensky}, \citenamefont {Pazzi}, \citenamefont {Pieri}, \citenamefont
  {Price}, \citenamefont {Riley}, \citenamefont {Reeder}, \citenamefont
  {Sabelnikov}, \citenamefont {Santin}, \citenamefont {Skorokhvatov},
  \citenamefont {Sobel}, \citenamefont {Steele}, \citenamefont {Steinberg},
  \citenamefont {Sukhotin}, \citenamefont {Tomshaw}, \citenamefont {Veron},\
  and\ \citenamefont {Vyrodov}}]{DCHOOZ1}%
  \BibitemOpen
  \bibfield  {author} {\bibinfo {author} {\bibfnamefont {M.}~\bibnamefont
  {Apollonio}}, \bibinfo {author} {\bibfnamefont {A.}~\bibnamefont {Baldini}},
  \bibinfo {author} {\bibfnamefont {C.}~\bibnamefont {Bemporad}}, \bibinfo
  {author} {\bibfnamefont {E.}~\bibnamefont {Caffau}}, \bibinfo {author}
  {\bibfnamefont {F.}~\bibnamefont {Cei}}, \bibinfo {author} {\bibfnamefont
  {Y.}~\bibnamefont {Déclais}}, \bibinfo {author} {\bibfnamefont
  {H.}~\bibnamefont {{de Kerret}}}, \bibinfo {author} {\bibfnamefont
  {B.}~\bibnamefont {Dieterle}}, \bibinfo {author} {\bibfnamefont
  {A.}~\bibnamefont {Etenko}}, \bibinfo {author} {\bibfnamefont
  {J.}~\bibnamefont {George}}, \bibinfo {author} {\bibfnamefont
  {G.}~\bibnamefont {Giannini}}, \bibinfo {author} {\bibfnamefont
  {M.}~\bibnamefont {Grassi}}, \bibinfo {author} {\bibfnamefont
  {Y.}~\bibnamefont {Kozlov}}, \bibinfo {author} {\bibfnamefont
  {W.}~\bibnamefont {Kropp}}, \bibinfo {author} {\bibfnamefont
  {D.}~\bibnamefont {Kryn}}, \bibinfo {author} {\bibfnamefont {M.}~\bibnamefont
  {Laiman}}, \bibinfo {author} {\bibfnamefont {C.}~\bibnamefont {Lane}},
  \bibinfo {author} {\bibfnamefont {B.}~\bibnamefont {Lefièvre}}, \bibinfo
  {author} {\bibfnamefont {I.}~\bibnamefont {Machulin}}, \bibinfo {author}
  {\bibfnamefont {A.}~\bibnamefont {Martemyanov}}, \bibinfo {author}
  {\bibfnamefont {V.}~\bibnamefont {Martemyanov}}, \bibinfo {author}
  {\bibfnamefont {L.}~\bibnamefont {Mikaelyan}}, \bibinfo {author}
  {\bibfnamefont {D.}~\bibnamefont {Nicolò}}, \bibinfo {author} {\bibfnamefont
  {M.}~\bibnamefont {Obolensky}}, \bibinfo {author} {\bibfnamefont
  {R.}~\bibnamefont {Pazzi}}, \bibinfo {author} {\bibfnamefont
  {G.}~\bibnamefont {Pieri}}, \bibinfo {author} {\bibfnamefont
  {L.}~\bibnamefont {Price}}, \bibinfo {author} {\bibfnamefont
  {S.}~\bibnamefont {Riley}}, \bibinfo {author} {\bibfnamefont
  {R.}~\bibnamefont {Reeder}}, \bibinfo {author} {\bibfnamefont
  {A.}~\bibnamefont {Sabelnikov}}, \bibinfo {author} {\bibfnamefont
  {G.}~\bibnamefont {Santin}}, \bibinfo {author} {\bibfnamefont
  {M.}~\bibnamefont {Skorokhvatov}}, \bibinfo {author} {\bibfnamefont
  {H.}~\bibnamefont {Sobel}}, \bibinfo {author} {\bibfnamefont
  {J.}~\bibnamefont {Steele}}, \bibinfo {author} {\bibfnamefont
  {R.}~\bibnamefont {Steinberg}}, \bibinfo {author} {\bibfnamefont
  {S.}~\bibnamefont {Sukhotin}}, \bibinfo {author} {\bibfnamefont
  {S.}~\bibnamefont {Tomshaw}}, \bibinfo {author} {\bibfnamefont
  {D.}~\bibnamefont {Veron}},\ and\ \bibinfo {author} {\bibfnamefont
  {V.}~\bibnamefont {Vyrodov}} (\bibinfo {collaboration} {Double Chooz
  Collaboration}),\ }\href
  {https://doi.org/https://doi.org/10.1016/S0370-2693(99)01072-2} {\bibfield
  {journal} {\bibinfo  {journal} {Phys. Lett. B}\ }\textbf {\bibinfo {volume}
  {466}},\ \bibinfo {pages} {415} (\bibinfo {year} {1999})}\BibitemShut
  {NoStop}%
\bibitem [{\citenamefont {Apollonio}\ \emph {et~al.}(2003)\citenamefont
  {Apollonio}, \citenamefont {Baldini}, \citenamefont {Bemporad}, \citenamefont
  {Caffau}, \citenamefont {Cei}, \citenamefont {Déclais}, \citenamefont
  {de~Kerret}, \citenamefont {Dieterle}, \citenamefont {Etenko}, \citenamefont
  {Foresti}, \citenamefont {George}, \citenamefont {Giannini}, \citenamefont
  {Grassi}, \citenamefont {Kozlov}, \citenamefont {Kropp}, \citenamefont
  {Kryn}, \citenamefont {Laiman}, \citenamefont {Lane}, \citenamefont
  {Lefièvre}, \citenamefont {Machulin}, \citenamefont {Martemyanov},
  \citenamefont {Martemyanov}, \citenamefont {Mikaelyan}, \citenamefont
  {Nicolò}, \citenamefont {Obolensky}, \citenamefont {Pazzi}, \citenamefont
  {Pieri}, \citenamefont {Price}, \citenamefont {Riley}, \citenamefont
  {Reeder}, \citenamefont {Sabelnikov}, \citenamefont {Santin}, \citenamefont
  {Skorokhvatov}, \citenamefont {Sobel}, \citenamefont {Steele}, \citenamefont
  {Steinberg}, \citenamefont {Sukhotin}, \citenamefont {Tomshaw}, \citenamefont
  {Veron},\ and\ \citenamefont {Vyrodov}}]{DCHOOZ2}%
  \BibitemOpen
  \bibfield  {author} {\bibinfo {author} {\bibfnamefont {M.}~\bibnamefont
  {Apollonio}}, \bibinfo {author} {\bibfnamefont {A.}~\bibnamefont {Baldini}},
  \bibinfo {author} {\bibfnamefont {C.}~\bibnamefont {Bemporad}}, \bibinfo
  {author} {\bibfnamefont {E.}~\bibnamefont {Caffau}}, \bibinfo {author}
  {\bibfnamefont {F.}~\bibnamefont {Cei}}, \bibinfo {author} {\bibfnamefont
  {Y.}~\bibnamefont {Déclais}}, \bibinfo {author} {\bibfnamefont
  {H.}~\bibnamefont {de~Kerret}}, \bibinfo {author} {\bibfnamefont
  {B.}~\bibnamefont {Dieterle}}, \bibinfo {author} {\bibfnamefont
  {A.}~\bibnamefont {Etenko}}, \bibinfo {author} {\bibfnamefont
  {L.}~\bibnamefont {Foresti}}, \bibinfo {author} {\bibfnamefont
  {J.}~\bibnamefont {George}}, \bibinfo {author} {\bibfnamefont
  {G.}~\bibnamefont {Giannini}}, \bibinfo {author} {\bibfnamefont
  {M.}~\bibnamefont {Grassi}}, \bibinfo {author} {\bibfnamefont
  {Y.}~\bibnamefont {Kozlov}}, \bibinfo {author} {\bibfnamefont
  {W.}~\bibnamefont {Kropp}}, \bibinfo {author} {\bibfnamefont
  {D.}~\bibnamefont {Kryn}}, \bibinfo {author} {\bibfnamefont {M.}~\bibnamefont
  {Laiman}}, \bibinfo {author} {\bibfnamefont {C.~E.}\ \bibnamefont {Lane}},
  \bibinfo {author} {\bibfnamefont {B.}~\bibnamefont {Lefièvre}}, \bibinfo
  {author} {\bibfnamefont {I.}~\bibnamefont {Machulin}}, \bibinfo {author}
  {\bibfnamefont {A.}~\bibnamefont {Martemyanov}}, \bibinfo {author}
  {\bibfnamefont {V.}~\bibnamefont {Martemyanov}}, \bibinfo {author}
  {\bibfnamefont {L.}~\bibnamefont {Mikaelyan}}, \bibinfo {author}
  {\bibfnamefont {D.}~\bibnamefont {Nicolò}}, \bibinfo {author} {\bibfnamefont
  {M.}~\bibnamefont {Obolensky}}, \bibinfo {author} {\bibfnamefont
  {R.}~\bibnamefont {Pazzi}}, \bibinfo {author} {\bibfnamefont
  {G.}~\bibnamefont {Pieri}}, \bibinfo {author} {\bibfnamefont
  {L.}~\bibnamefont {Price}}, \bibinfo {author} {\bibfnamefont
  {S.}~\bibnamefont {Riley}}, \bibinfo {author} {\bibfnamefont
  {R.}~\bibnamefont {Reeder}}, \bibinfo {author} {\bibfnamefont
  {A.}~\bibnamefont {Sabelnikov}}, \bibinfo {author} {\bibfnamefont
  {G.}~\bibnamefont {Santin}}, \bibinfo {author} {\bibfnamefont
  {M.}~\bibnamefont {Skorokhvatov}}, \bibinfo {author} {\bibfnamefont
  {H.}~\bibnamefont {Sobel}}, \bibinfo {author} {\bibfnamefont
  {J.}~\bibnamefont {Steele}}, \bibinfo {author} {\bibfnamefont
  {R.}~\bibnamefont {Steinberg}}, \bibinfo {author} {\bibfnamefont
  {S.}~\bibnamefont {Sukhotin}}, \bibinfo {author} {\bibfnamefont
  {S.}~\bibnamefont {Tomshaw}}, \bibinfo {author} {\bibfnamefont
  {D.}~\bibnamefont {Veron}},\ and\ \bibinfo {author} {\bibfnamefont
  {V.}~\bibnamefont {Vyrodov}} (\bibinfo {collaboration} {Double Chooz
  Collaboration}),\ }\href {https://doi.org/10.1140/epjc/s2002-01127-9}
  {\bibfield  {journal} {\bibinfo  {journal} {Eur. Phys. J. C}\ }\textbf
  {\bibinfo {volume} {27}},\ \bibinfo {pages} {331} (\bibinfo {year}
  {2003})}\BibitemShut {NoStop}%
\bibitem [{\citenamefont {Choi}\ \emph {et~al.}(2016)\citenamefont {Choi},
  \citenamefont {Choi}, \citenamefont {Choi}, \citenamefont {Jang},
  \citenamefont {Jang}, \citenamefont {Jeon}, \citenamefont {Joo},
  \citenamefont {Kim}, \citenamefont {Kim}, \citenamefont {Kim}, \citenamefont
  {Kim}, \citenamefont {Kim}, \citenamefont {Kim}, \citenamefont {Kim},
  \citenamefont {Ko}, \citenamefont {Lee}, \citenamefont {Lim}, \citenamefont
  {Pac}, \citenamefont {Park}, \citenamefont {Park}, \citenamefont {Park},
  \citenamefont {Seo}, \citenamefont {Seo}, \citenamefont {Seon}, \citenamefont
  {Shin}, \citenamefont {Siyeon}, \citenamefont {Yang}, \citenamefont {Yeo},\
  and\ \citenamefont {Yu}}]{RENO}%
  \BibitemOpen
  \bibfield  {author} {\bibinfo {author} {\bibfnamefont {J.~H.}\ \bibnamefont
  {Choi}}, \bibinfo {author} {\bibfnamefont {W.~Q.}\ \bibnamefont {Choi}},
  \bibinfo {author} {\bibfnamefont {Y.}~\bibnamefont {Choi}}, \bibinfo {author}
  {\bibfnamefont {H.~I.}\ \bibnamefont {Jang}}, \bibinfo {author}
  {\bibfnamefont {J.~S.}\ \bibnamefont {Jang}}, \bibinfo {author}
  {\bibfnamefont {E.~J.}\ \bibnamefont {Jeon}}, \bibinfo {author}
  {\bibfnamefont {K.~K.}\ \bibnamefont {Joo}}, \bibinfo {author} {\bibfnamefont
  {B.~R.}\ \bibnamefont {Kim}}, \bibinfo {author} {\bibfnamefont {H.~S.}\
  \bibnamefont {Kim}}, \bibinfo {author} {\bibfnamefont {J.~Y.}\ \bibnamefont
  {Kim}}, \bibinfo {author} {\bibfnamefont {S.~B.}\ \bibnamefont {Kim}},
  \bibinfo {author} {\bibfnamefont {S.~Y.}\ \bibnamefont {Kim}}, \bibinfo
  {author} {\bibfnamefont {W.}~\bibnamefont {Kim}}, \bibinfo {author}
  {\bibfnamefont {Y.~D.}\ \bibnamefont {Kim}}, \bibinfo {author} {\bibfnamefont
  {Y.}~\bibnamefont {Ko}}, \bibinfo {author} {\bibfnamefont {D.~H.}\
  \bibnamefont {Lee}}, \bibinfo {author} {\bibfnamefont {I.~T.}\ \bibnamefont
  {Lim}}, \bibinfo {author} {\bibfnamefont {M.~Y.}\ \bibnamefont {Pac}},
  \bibinfo {author} {\bibfnamefont {I.~G.}\ \bibnamefont {Park}}, \bibinfo
  {author} {\bibfnamefont {J.~S.}\ \bibnamefont {Park}}, \bibinfo {author}
  {\bibfnamefont {R.~G.}\ \bibnamefont {Park}}, \bibinfo {author}
  {\bibfnamefont {H.}~\bibnamefont {Seo}}, \bibinfo {author} {\bibfnamefont
  {S.~H.}\ \bibnamefont {Seo}}, \bibinfo {author} {\bibfnamefont {Y.~G.}\
  \bibnamefont {Seon}}, \bibinfo {author} {\bibfnamefont {C.~D.}\ \bibnamefont
  {Shin}}, \bibinfo {author} {\bibfnamefont {K.}~\bibnamefont {Siyeon}},
  \bibinfo {author} {\bibfnamefont {J.~H.}\ \bibnamefont {Yang}}, \bibinfo
  {author} {\bibfnamefont {I.~S.}\ \bibnamefont {Yeo}},\ and\ \bibinfo {author}
  {\bibfnamefont {I.}~\bibnamefont {Yu}} (\bibinfo {collaboration} {RENO
  Collaboration}),\ }\href {https://doi.org/10.1103/PhysRevLett.116.211801}
  {\bibfield  {journal} {\bibinfo  {journal} {Phys. Rev. Lett.}\ }\textbf
  {\bibinfo {volume} {116}},\ \bibinfo {pages} {211801} (\bibinfo {year}
  {2016})}\BibitemShut {NoStop}%
\bibitem [{\citenamefont {An}\ \emph {et~al.}(2016)\citenamefont {An},
  \citenamefont {Balantekin}, \citenamefont {Band}, \citenamefont {Bishai},
  \citenamefont {Blyth}, \citenamefont {Butorov}, \citenamefont {Cao},
  \citenamefont {Cao}, \citenamefont {Cao}, \citenamefont {Cen}, \citenamefont
  {Chan}, \citenamefont {Chang}, \citenamefont {Chang}, \citenamefont {Chang},
  \citenamefont {Chen}, \citenamefont {Chen}, \citenamefont {Chen},
  \citenamefont {Chen}, \citenamefont {Chen}, \citenamefont {Cheng},
  \citenamefont {Cheng}, \citenamefont {Cheng}, \citenamefont {Cherwinka},
  \citenamefont {Chu}, \citenamefont {Cummings}, \citenamefont {de~Arcos},
  \citenamefont {Deng}, \citenamefont {Ding}, \citenamefont {Ding},
  \citenamefont {Diwan}, \citenamefont {Dove}, \citenamefont {Draeger},
  \citenamefont {Dwyer}, \citenamefont {Edwards}, \citenamefont {Ely},
  \citenamefont {Gill}, \citenamefont {Gonchar}, \citenamefont {Gong},
  \citenamefont {Gong}, \citenamefont {Grassi}, \citenamefont {Gu},
  \citenamefont {Guan}, \citenamefont {Guo}, \citenamefont {Guo}, \citenamefont
  {Hackenburg}, \citenamefont {Han}, \citenamefont {Hans}, \citenamefont {He},
  \citenamefont {Heeger}, \citenamefont {Heng}, \citenamefont {Higuera},
  \citenamefont {Hor}, \citenamefont {Hsiung}, \citenamefont {Hu},
  \citenamefont {Hu}, \citenamefont {Hu}, \citenamefont {Hu}, \citenamefont
  {Hu}, \citenamefont {Huang}, \citenamefont {Huang}, \citenamefont {Huang},
  \citenamefont {Huber}, \citenamefont {Hussain}, \citenamefont {Jaffe},
  \citenamefont {Jaffke}, \citenamefont {Jen}, \citenamefont {Jetter},
  \citenamefont {Ji}, \citenamefont {Ji}, \citenamefont {Jiao}, \citenamefont
  {Johnson}, \citenamefont {Kang}, \citenamefont {Kettell}, \citenamefont
  {Kohn}, \citenamefont {Kramer}, \citenamefont {Kwan}, \citenamefont {Kwok},
  \citenamefont {Kwok}, \citenamefont {Langford}, \citenamefont {Lau},
  \citenamefont {Lebanowski}, \citenamefont {Lee}, \citenamefont {Lei},
  \citenamefont {Leitner}, \citenamefont {Leung}, \citenamefont {Leung},
  \citenamefont {Lewis}, \citenamefont {Li}, \citenamefont {Li}, \citenamefont
  {Li}, \citenamefont {Li}, \citenamefont {Li}, \citenamefont {Li},
  \citenamefont {Li}, \citenamefont {Li}, \citenamefont {Li}, \citenamefont
  {Li}, \citenamefont {Liang}, \citenamefont {Lin}, \citenamefont {Lin},
  \citenamefont {Lin}, \citenamefont {Lin}, \citenamefont {Ling}, \citenamefont
  {Link}, \citenamefont {Littenberg}, \citenamefont {Littlejohn}, \citenamefont
  {Liu}, \citenamefont {Liu}, \citenamefont {Liu}, \citenamefont {Liu},
  \citenamefont {Liu}, \citenamefont {Lu}, \citenamefont {Lu}, \citenamefont
  {Lu}, \citenamefont {Luk}, \citenamefont {Ma}, \citenamefont {Ma},
  \citenamefont {Ma}, \citenamefont {Ma}, \citenamefont {Martinez~Caicedo},
  \citenamefont {McDonald}, \citenamefont {McKeown}, \citenamefont {Meng},
  \citenamefont {Mitchell}, \citenamefont {Monari~Kebwaro}, \citenamefont
  {Nakajima}, \citenamefont {Napolitano}, \citenamefont {Naumov}, \citenamefont
  {Naumova}, \citenamefont {Ngai}, \citenamefont {Ning}, \citenamefont
  {Ochoa-Ricoux}, \citenamefont {Olshevski}, \citenamefont {Pan}, \citenamefont
  {Park}, \citenamefont {Patton}, \citenamefont {Pec}, \citenamefont {Peng},
  \citenamefont {Piilonen}, \citenamefont {Pinsky}, \citenamefont {Pun},
  \citenamefont {Qi}, \citenamefont {Qi}, \citenamefont {Qian}, \citenamefont
  {Raper}, \citenamefont {Ren}, \citenamefont {Ren}, \citenamefont {Rosero},
  \citenamefont {Roskovec}, \citenamefont {Ruan}, \citenamefont {Shao},
  \citenamefont {Steiner}, \citenamefont {Sun}, \citenamefont {Sun},
  \citenamefont {Tang}, \citenamefont {Taychenachev}, \citenamefont {Tsang},
  \citenamefont {Tull}, \citenamefont {Tung}, \citenamefont {Viaux},
  \citenamefont {Viren}, \citenamefont {Vorobel}, \citenamefont {Wang},
  \citenamefont {Wang}, \citenamefont {Wang}, \citenamefont {Wang},
  \citenamefont {Wang}, \citenamefont {Wang}, \citenamefont {Wang},
  \citenamefont {Wang}, \citenamefont {Wang}, \citenamefont {Wang},
  \citenamefont {Wang}, \citenamefont {Wei}, \citenamefont {Wen}, \citenamefont
  {Whisnant}, \citenamefont {White}, \citenamefont {Whitehead}, \citenamefont
  {Wise}, \citenamefont {Wong}, \citenamefont {Wong}, \citenamefont
  {Worcester}, \citenamefont {Wu}, \citenamefont {Xia}, \citenamefont {Xia},
  \citenamefont {Xia}, \citenamefont {Xing}, \citenamefont {Xu}, \citenamefont
  {Xu}, \citenamefont {Xu}, \citenamefont {Xu}, \citenamefont {Xue},
  \citenamefont {Yan}, \citenamefont {Yang}, \citenamefont {Yang},
  \citenamefont {Yang}, \citenamefont {Yang}, \citenamefont {Ye}, \citenamefont
  {Yeh}, \citenamefont {Young}, \citenamefont {Yu}, \citenamefont {Yu},
  \citenamefont {Zang}, \citenamefont {Zhan}, \citenamefont {Zhang},
  \citenamefont {Zhang}, \citenamefont {Zhang}, \citenamefont {Zhang},
  \citenamefont {Zhang}, \citenamefont {Zhang}, \citenamefont {Zhang},
  \citenamefont {Zhang}, \citenamefont {Zhang}, \citenamefont {Zhang},
  \citenamefont {Zhao}, \citenamefont {Zhao}, \citenamefont {Zhao},
  \citenamefont {Zhao}, \citenamefont {Zheng}, \citenamefont {Zhong},
  \citenamefont {Zhou}, \citenamefont {Zhou}, \citenamefont {Zhuang},\ and\
  \citenamefont {Zou}}]{DAYABAY}%
  \BibitemOpen
  \bibfield  {author} {\bibinfo {author} {\bibfnamefont {F.~P.}\ \bibnamefont
  {An}}, \bibinfo {author} {\bibfnamefont {A.~B.}\ \bibnamefont {Balantekin}},
  \bibinfo {author} {\bibfnamefont {H.~R.}\ \bibnamefont {Band}}, \bibinfo
  {author} {\bibfnamefont {M.}~\bibnamefont {Bishai}}, \bibinfo {author}
  {\bibfnamefont {S.}~\bibnamefont {Blyth}}, \bibinfo {author} {\bibfnamefont
  {I.}~\bibnamefont {Butorov}}, \bibinfo {author} {\bibfnamefont
  {D.}~\bibnamefont {Cao}}, \bibinfo {author} {\bibfnamefont {G.~F.}\
  \bibnamefont {Cao}}, \bibinfo {author} {\bibfnamefont {J.}~\bibnamefont
  {Cao}}, \bibinfo {author} {\bibfnamefont {W.~R.}\ \bibnamefont {Cen}},
  \bibinfo {author} {\bibfnamefont {Y.~L.}\ \bibnamefont {Chan}}, \bibinfo
  {author} {\bibfnamefont {J.~F.}\ \bibnamefont {Chang}}, \bibinfo {author}
  {\bibfnamefont {L.~C.}\ \bibnamefont {Chang}}, \bibinfo {author}
  {\bibfnamefont {Y.}~\bibnamefont {Chang}}, \bibinfo {author} {\bibfnamefont
  {H.~S.}\ \bibnamefont {Chen}}, \bibinfo {author} {\bibfnamefont {Q.~Y.}\
  \bibnamefont {Chen}}, \bibinfo {author} {\bibfnamefont {S.~M.}\ \bibnamefont
  {Chen}}, \bibinfo {author} {\bibfnamefont {Y.~X.}\ \bibnamefont {Chen}},
  \bibinfo {author} {\bibfnamefont {Y.}~\bibnamefont {Chen}}, \bibinfo {author}
  {\bibfnamefont {J.~H.}\ \bibnamefont {Cheng}}, \bibinfo {author}
  {\bibfnamefont {J.}~\bibnamefont {Cheng}}, \bibinfo {author} {\bibfnamefont
  {Y.~P.}\ \bibnamefont {Cheng}}, \bibinfo {author} {\bibfnamefont {J.~J.}\
  \bibnamefont {Cherwinka}}, \bibinfo {author} {\bibfnamefont {M.~C.}\
  \bibnamefont {Chu}}, \bibinfo {author} {\bibfnamefont {J.~P.}\ \bibnamefont
  {Cummings}}, \bibinfo {author} {\bibfnamefont {J.}~\bibnamefont {de~Arcos}},
  \bibinfo {author} {\bibfnamefont {Z.~Y.}\ \bibnamefont {Deng}}, \bibinfo
  {author} {\bibfnamefont {X.~F.}\ \bibnamefont {Ding}}, \bibinfo {author}
  {\bibfnamefont {Y.~Y.}\ \bibnamefont {Ding}}, \bibinfo {author}
  {\bibfnamefont {M.~V.}\ \bibnamefont {Diwan}}, \bibinfo {author}
  {\bibfnamefont {J.}~\bibnamefont {Dove}}, \bibinfo {author} {\bibfnamefont
  {E.}~\bibnamefont {Draeger}}, \bibinfo {author} {\bibfnamefont {D.~A.}\
  \bibnamefont {Dwyer}}, \bibinfo {author} {\bibfnamefont {W.~R.}\ \bibnamefont
  {Edwards}}, \bibinfo {author} {\bibfnamefont {S.~R.}\ \bibnamefont {Ely}},
  \bibinfo {author} {\bibfnamefont {R.}~\bibnamefont {Gill}}, \bibinfo {author}
  {\bibfnamefont {M.}~\bibnamefont {Gonchar}}, \bibinfo {author} {\bibfnamefont
  {G.~H.}\ \bibnamefont {Gong}}, \bibinfo {author} {\bibfnamefont
  {H.}~\bibnamefont {Gong}}, \bibinfo {author} {\bibfnamefont {M.}~\bibnamefont
  {Grassi}}, \bibinfo {author} {\bibfnamefont {W.~Q.}\ \bibnamefont {Gu}},
  \bibinfo {author} {\bibfnamefont {M.~Y.}\ \bibnamefont {Guan}}, \bibinfo
  {author} {\bibfnamefont {L.}~\bibnamefont {Guo}}, \bibinfo {author}
  {\bibfnamefont {X.~H.}\ \bibnamefont {Guo}}, \bibinfo {author} {\bibfnamefont
  {R.~W.}\ \bibnamefont {Hackenburg}}, \bibinfo {author} {\bibfnamefont
  {R.}~\bibnamefont {Han}}, \bibinfo {author} {\bibfnamefont {S.}~\bibnamefont
  {Hans}}, \bibinfo {author} {\bibfnamefont {M.}~\bibnamefont {He}}, \bibinfo
  {author} {\bibfnamefont {K.~M.}\ \bibnamefont {Heeger}}, \bibinfo {author}
  {\bibfnamefont {Y.~K.}\ \bibnamefont {Heng}}, \bibinfo {author}
  {\bibfnamefont {A.}~\bibnamefont {Higuera}}, \bibinfo {author} {\bibfnamefont
  {Y.~K.}\ \bibnamefont {Hor}}, \bibinfo {author} {\bibfnamefont {Y.~B.}\
  \bibnamefont {Hsiung}}, \bibinfo {author} {\bibfnamefont {B.~Z.}\
  \bibnamefont {Hu}}, \bibinfo {author} {\bibfnamefont {L.~M.}\ \bibnamefont
  {Hu}}, \bibinfo {author} {\bibfnamefont {L.~J.}\ \bibnamefont {Hu}}, \bibinfo
  {author} {\bibfnamefont {T.}~\bibnamefont {Hu}}, \bibinfo {author}
  {\bibfnamefont {W.}~\bibnamefont {Hu}}, \bibinfo {author} {\bibfnamefont
  {E.~C.}\ \bibnamefont {Huang}}, \bibinfo {author} {\bibfnamefont {H.~X.}\
  \bibnamefont {Huang}}, \bibinfo {author} {\bibfnamefont {X.~T.}\ \bibnamefont
  {Huang}}, \bibinfo {author} {\bibfnamefont {P.}~\bibnamefont {Huber}},
  \bibinfo {author} {\bibfnamefont {G.}~\bibnamefont {Hussain}}, \bibinfo
  {author} {\bibfnamefont {D.~E.}\ \bibnamefont {Jaffe}}, \bibinfo {author}
  {\bibfnamefont {P.}~\bibnamefont {Jaffke}}, \bibinfo {author} {\bibfnamefont
  {K.~L.}\ \bibnamefont {Jen}}, \bibinfo {author} {\bibfnamefont
  {S.}~\bibnamefont {Jetter}}, \bibinfo {author} {\bibfnamefont {X.~P.}\
  \bibnamefont {Ji}}, \bibinfo {author} {\bibfnamefont {X.~L.}\ \bibnamefont
  {Ji}}, \bibinfo {author} {\bibfnamefont {J.~B.}\ \bibnamefont {Jiao}},
  \bibinfo {author} {\bibfnamefont {R.~A.}\ \bibnamefont {Johnson}}, \bibinfo
  {author} {\bibfnamefont {L.}~\bibnamefont {Kang}}, \bibinfo {author}
  {\bibfnamefont {S.~H.}\ \bibnamefont {Kettell}}, \bibinfo {author}
  {\bibfnamefont {S.}~\bibnamefont {Kohn}}, \bibinfo {author} {\bibfnamefont
  {M.}~\bibnamefont {Kramer}}, \bibinfo {author} {\bibfnamefont {K.~K.}\
  \bibnamefont {Kwan}}, \bibinfo {author} {\bibfnamefont {M.~W.}\ \bibnamefont
  {Kwok}}, \bibinfo {author} {\bibfnamefont {T.}~\bibnamefont {Kwok}}, \bibinfo
  {author} {\bibfnamefont {T.~J.}\ \bibnamefont {Langford}}, \bibinfo {author}
  {\bibfnamefont {K.}~\bibnamefont {Lau}}, \bibinfo {author} {\bibfnamefont
  {L.}~\bibnamefont {Lebanowski}}, \bibinfo {author} {\bibfnamefont
  {J.}~\bibnamefont {Lee}}, \bibinfo {author} {\bibfnamefont {R.~T.}\
  \bibnamefont {Lei}}, \bibinfo {author} {\bibfnamefont {R.}~\bibnamefont
  {Leitner}}, \bibinfo {author} {\bibfnamefont {K.~Y.}\ \bibnamefont {Leung}},
  \bibinfo {author} {\bibfnamefont {J.~K.~C.}\ \bibnamefont {Leung}}, \bibinfo
  {author} {\bibfnamefont {C.~A.}\ \bibnamefont {Lewis}}, \bibinfo {author}
  {\bibfnamefont {D.~J.}\ \bibnamefont {Li}}, \bibinfo {author} {\bibfnamefont
  {F.}~\bibnamefont {Li}}, \bibinfo {author} {\bibfnamefont {G.~S.}\
  \bibnamefont {Li}}, \bibinfo {author} {\bibfnamefont {Q.~J.}\ \bibnamefont
  {Li}}, \bibinfo {author} {\bibfnamefont {S.~C.}\ \bibnamefont {Li}}, \bibinfo
  {author} {\bibfnamefont {W.~D.}\ \bibnamefont {Li}}, \bibinfo {author}
  {\bibfnamefont {X.~N.}\ \bibnamefont {Li}}, \bibinfo {author} {\bibfnamefont
  {X.~Q.}\ \bibnamefont {Li}}, \bibinfo {author} {\bibfnamefont {Y.~F.}\
  \bibnamefont {Li}}, \bibinfo {author} {\bibfnamefont {Z.~B.}\ \bibnamefont
  {Li}}, \bibinfo {author} {\bibfnamefont {H.}~\bibnamefont {Liang}}, \bibinfo
  {author} {\bibfnamefont {C.~J.}\ \bibnamefont {Lin}}, \bibinfo {author}
  {\bibfnamefont {G.~L.}\ \bibnamefont {Lin}}, \bibinfo {author} {\bibfnamefont
  {P.~Y.}\ \bibnamefont {Lin}}, \bibinfo {author} {\bibfnamefont {S.~K.}\
  \bibnamefont {Lin}}, \bibinfo {author} {\bibfnamefont {J.~J.}\ \bibnamefont
  {Ling}}, \bibinfo {author} {\bibfnamefont {J.~M.}\ \bibnamefont {Link}},
  \bibinfo {author} {\bibfnamefont {L.}~\bibnamefont {Littenberg}}, \bibinfo
  {author} {\bibfnamefont {B.~R.}\ \bibnamefont {Littlejohn}}, \bibinfo
  {author} {\bibfnamefont {D.~W.}\ \bibnamefont {Liu}}, \bibinfo {author}
  {\bibfnamefont {H.}~\bibnamefont {Liu}}, \bibinfo {author} {\bibfnamefont
  {J.~L.}\ \bibnamefont {Liu}}, \bibinfo {author} {\bibfnamefont {J.~C.}\
  \bibnamefont {Liu}}, \bibinfo {author} {\bibfnamefont {S.~S.}\ \bibnamefont
  {Liu}}, \bibinfo {author} {\bibfnamefont {C.}~\bibnamefont {Lu}}, \bibinfo
  {author} {\bibfnamefont {H.~Q.}\ \bibnamefont {Lu}}, \bibinfo {author}
  {\bibfnamefont {J.~S.}\ \bibnamefont {Lu}}, \bibinfo {author} {\bibfnamefont
  {K.~B.}\ \bibnamefont {Luk}}, \bibinfo {author} {\bibfnamefont {Q.~M.}\
  \bibnamefont {Ma}}, \bibinfo {author} {\bibfnamefont {X.~Y.}\ \bibnamefont
  {Ma}}, \bibinfo {author} {\bibfnamefont {X.~B.}\ \bibnamefont {Ma}}, \bibinfo
  {author} {\bibfnamefont {Y.~Q.}\ \bibnamefont {Ma}}, \bibinfo {author}
  {\bibfnamefont {D.~A.}\ \bibnamefont {Martinez~Caicedo}}, \bibinfo {author}
  {\bibfnamefont {K.~T.}\ \bibnamefont {McDonald}}, \bibinfo {author}
  {\bibfnamefont {R.~D.}\ \bibnamefont {McKeown}}, \bibinfo {author}
  {\bibfnamefont {Y.}~\bibnamefont {Meng}}, \bibinfo {author} {\bibfnamefont
  {I.}~\bibnamefont {Mitchell}}, \bibinfo {author} {\bibfnamefont
  {J.}~\bibnamefont {Monari~Kebwaro}}, \bibinfo {author} {\bibfnamefont
  {Y.}~\bibnamefont {Nakajima}}, \bibinfo {author} {\bibfnamefont
  {J.}~\bibnamefont {Napolitano}}, \bibinfo {author} {\bibfnamefont
  {D.}~\bibnamefont {Naumov}}, \bibinfo {author} {\bibfnamefont
  {E.}~\bibnamefont {Naumova}}, \bibinfo {author} {\bibfnamefont {H.~Y.}\
  \bibnamefont {Ngai}}, \bibinfo {author} {\bibfnamefont {Z.}~\bibnamefont
  {Ning}}, \bibinfo {author} {\bibfnamefont {J.~P.}\ \bibnamefont
  {Ochoa-Ricoux}}, \bibinfo {author} {\bibfnamefont {A.}~\bibnamefont
  {Olshevski}}, \bibinfo {author} {\bibfnamefont {H.-R.}\ \bibnamefont {Pan}},
  \bibinfo {author} {\bibfnamefont {J.}~\bibnamefont {Park}}, \bibinfo {author}
  {\bibfnamefont {S.}~\bibnamefont {Patton}}, \bibinfo {author} {\bibfnamefont
  {V.}~\bibnamefont {Pec}}, \bibinfo {author} {\bibfnamefont {J.~C.}\
  \bibnamefont {Peng}}, \bibinfo {author} {\bibfnamefont {L.~E.}\ \bibnamefont
  {Piilonen}}, \bibinfo {author} {\bibfnamefont {L.}~\bibnamefont {Pinsky}},
  \bibinfo {author} {\bibfnamefont {C.~S.~J.}\ \bibnamefont {Pun}}, \bibinfo
  {author} {\bibfnamefont {F.~Z.}\ \bibnamefont {Qi}}, \bibinfo {author}
  {\bibfnamefont {M.}~\bibnamefont {Qi}}, \bibinfo {author} {\bibfnamefont
  {X.}~\bibnamefont {Qian}}, \bibinfo {author} {\bibfnamefont {N.}~\bibnamefont
  {Raper}}, \bibinfo {author} {\bibfnamefont {B.}~\bibnamefont {Ren}}, \bibinfo
  {author} {\bibfnamefont {J.}~\bibnamefont {Ren}}, \bibinfo {author}
  {\bibfnamefont {R.}~\bibnamefont {Rosero}}, \bibinfo {author} {\bibfnamefont
  {B.}~\bibnamefont {Roskovec}}, \bibinfo {author} {\bibfnamefont {X.~C.}\
  \bibnamefont {Ruan}}, \bibinfo {author} {\bibfnamefont {B.~B.}\ \bibnamefont
  {Shao}}, \bibinfo {author} {\bibfnamefont {H.}~\bibnamefont {Steiner}},
  \bibinfo {author} {\bibfnamefont {G.~X.}\ \bibnamefont {Sun}}, \bibinfo
  {author} {\bibfnamefont {J.~L.}\ \bibnamefont {Sun}}, \bibinfo {author}
  {\bibfnamefont {W.}~\bibnamefont {Tang}}, \bibinfo {author} {\bibfnamefont
  {D.}~\bibnamefont {Taychenachev}}, \bibinfo {author} {\bibfnamefont {K.~V.}\
  \bibnamefont {Tsang}}, \bibinfo {author} {\bibfnamefont {C.~E.}\ \bibnamefont
  {Tull}}, \bibinfo {author} {\bibfnamefont {Y.~C.}\ \bibnamefont {Tung}},
  \bibinfo {author} {\bibfnamefont {N.}~\bibnamefont {Viaux}}, \bibinfo
  {author} {\bibfnamefont {B.}~\bibnamefont {Viren}}, \bibinfo {author}
  {\bibfnamefont {V.}~\bibnamefont {Vorobel}}, \bibinfo {author} {\bibfnamefont
  {C.~H.}\ \bibnamefont {Wang}}, \bibinfo {author} {\bibfnamefont
  {M.}~\bibnamefont {Wang}}, \bibinfo {author} {\bibfnamefont {N.~Y.}\
  \bibnamefont {Wang}}, \bibinfo {author} {\bibfnamefont {R.~G.}\ \bibnamefont
  {Wang}}, \bibinfo {author} {\bibfnamefont {W.}~\bibnamefont {Wang}}, \bibinfo
  {author} {\bibfnamefont {W.~W.}\ \bibnamefont {Wang}}, \bibinfo {author}
  {\bibfnamefont {X.}~\bibnamefont {Wang}}, \bibinfo {author} {\bibfnamefont
  {Y.~F.}\ \bibnamefont {Wang}}, \bibinfo {author} {\bibfnamefont
  {Z.}~\bibnamefont {Wang}}, \bibinfo {author} {\bibfnamefont {Z.}~\bibnamefont
  {Wang}}, \bibinfo {author} {\bibfnamefont {Z.~M.}\ \bibnamefont {Wang}},
  \bibinfo {author} {\bibfnamefont {H.~Y.}\ \bibnamefont {Wei}}, \bibinfo
  {author} {\bibfnamefont {L.~J.}\ \bibnamefont {Wen}}, \bibinfo {author}
  {\bibfnamefont {K.}~\bibnamefont {Whisnant}}, \bibinfo {author}
  {\bibfnamefont {C.~G.}\ \bibnamefont {White}}, \bibinfo {author}
  {\bibfnamefont {L.}~\bibnamefont {Whitehead}}, \bibinfo {author}
  {\bibfnamefont {T.}~\bibnamefont {Wise}}, \bibinfo {author} {\bibfnamefont
  {H.~L.~H.}\ \bibnamefont {Wong}}, \bibinfo {author} {\bibfnamefont
  {S.~C.~F.}\ \bibnamefont {Wong}}, \bibinfo {author} {\bibfnamefont
  {E.}~\bibnamefont {Worcester}}, \bibinfo {author} {\bibfnamefont
  {Q.}~\bibnamefont {Wu}}, \bibinfo {author} {\bibfnamefont {D.~M.}\
  \bibnamefont {Xia}}, \bibinfo {author} {\bibfnamefont {J.~K.}\ \bibnamefont
  {Xia}}, \bibinfo {author} {\bibfnamefont {X.}~\bibnamefont {Xia}}, \bibinfo
  {author} {\bibfnamefont {Z.~Z.}\ \bibnamefont {Xing}}, \bibinfo {author}
  {\bibfnamefont {J.~Y.}\ \bibnamefont {Xu}}, \bibinfo {author} {\bibfnamefont
  {J.~L.}\ \bibnamefont {Xu}}, \bibinfo {author} {\bibfnamefont
  {J.}~\bibnamefont {Xu}}, \bibinfo {author} {\bibfnamefont {Y.}~\bibnamefont
  {Xu}}, \bibinfo {author} {\bibfnamefont {T.}~\bibnamefont {Xue}}, \bibinfo
  {author} {\bibfnamefont {J.}~\bibnamefont {Yan}}, \bibinfo {author}
  {\bibfnamefont {C.~G.}\ \bibnamefont {Yang}}, \bibinfo {author}
  {\bibfnamefont {L.}~\bibnamefont {Yang}}, \bibinfo {author} {\bibfnamefont
  {M.~S.}\ \bibnamefont {Yang}}, \bibinfo {author} {\bibfnamefont {M.~T.}\
  \bibnamefont {Yang}}, \bibinfo {author} {\bibfnamefont {M.}~\bibnamefont
  {Ye}}, \bibinfo {author} {\bibfnamefont {M.}~\bibnamefont {Yeh}}, \bibinfo
  {author} {\bibfnamefont {B.~L.}\ \bibnamefont {Young}}, \bibinfo {author}
  {\bibfnamefont {G.~Y.}\ \bibnamefont {Yu}}, \bibinfo {author} {\bibfnamefont
  {Z.~Y.}\ \bibnamefont {Yu}}, \bibinfo {author} {\bibfnamefont {S.~L.}\
  \bibnamefont {Zang}}, \bibinfo {author} {\bibfnamefont {L.}~\bibnamefont
  {Zhan}}, \bibinfo {author} {\bibfnamefont {C.}~\bibnamefont {Zhang}},
  \bibinfo {author} {\bibfnamefont {H.~H.}\ \bibnamefont {Zhang}}, \bibinfo
  {author} {\bibfnamefont {J.~W.}\ \bibnamefont {Zhang}}, \bibinfo {author}
  {\bibfnamefont {Q.~M.}\ \bibnamefont {Zhang}}, \bibinfo {author}
  {\bibfnamefont {Y.~M.}\ \bibnamefont {Zhang}}, \bibinfo {author}
  {\bibfnamefont {Y.~X.}\ \bibnamefont {Zhang}}, \bibinfo {author}
  {\bibfnamefont {Y.~M.}\ \bibnamefont {Zhang}}, \bibinfo {author}
  {\bibfnamefont {Z.~J.}\ \bibnamefont {Zhang}}, \bibinfo {author}
  {\bibfnamefont {Z.~Y.}\ \bibnamefont {Zhang}}, \bibinfo {author}
  {\bibfnamefont {Z.~P.}\ \bibnamefont {Zhang}}, \bibinfo {author}
  {\bibfnamefont {J.}~\bibnamefont {Zhao}}, \bibinfo {author} {\bibfnamefont
  {Q.~W.}\ \bibnamefont {Zhao}}, \bibinfo {author} {\bibfnamefont {Y.~F.}\
  \bibnamefont {Zhao}}, \bibinfo {author} {\bibfnamefont {Y.~B.}\ \bibnamefont
  {Zhao}}, \bibinfo {author} {\bibfnamefont {L.}~\bibnamefont {Zheng}},
  \bibinfo {author} {\bibfnamefont {W.~L.}\ \bibnamefont {Zhong}}, \bibinfo
  {author} {\bibfnamefont {L.}~\bibnamefont {Zhou}}, \bibinfo {author}
  {\bibfnamefont {N.}~\bibnamefont {Zhou}}, \bibinfo {author} {\bibfnamefont
  {H.~L.}\ \bibnamefont {Zhuang}},\ and\ \bibinfo {author} {\bibfnamefont
  {J.~H.}\ \bibnamefont {Zou}} (\bibinfo {collaboration} {Daya Bay
  Collaboration}),\ }\href {https://doi.org/10.1103/PhysRevLett.116.061801}
  {\bibfield  {journal} {\bibinfo  {journal} {Phys. Rev. Lett.}\ }\textbf
  {\bibinfo {volume} {116}},\ \bibinfo {pages} {061801} (\bibinfo {year}
  {2016})}\BibitemShut {NoStop}%
\bibitem [{\citenamefont {Hayes}\ \emph {et~al.}(2014)\citenamefont {Hayes},
  \citenamefont {Friar}, \citenamefont {Garvey}, \citenamefont {Jungman},\ and\
  \citenamefont {Jonkmans}}]{HAYES2}%
  \BibitemOpen
  \bibfield  {author} {\bibinfo {author} {\bibfnamefont {A.~C.}\ \bibnamefont
  {Hayes}}, \bibinfo {author} {\bibfnamefont {J.~L.}\ \bibnamefont {Friar}},
  \bibinfo {author} {\bibfnamefont {G.~T.}\ \bibnamefont {Garvey}}, \bibinfo
  {author} {\bibfnamefont {G.}~\bibnamefont {Jungman}},\ and\ \bibinfo {author}
  {\bibfnamefont {G.}~\bibnamefont {Jonkmans}},\ }\href
  {https://doi.org/10.1103/PhysRevLett.112.202501} {\bibfield  {journal}
  {\bibinfo  {journal} {Phys. Rev. Lett.}\ }\textbf {\bibinfo {volume} {112}},\
  \bibinfo {pages} {202501} (\bibinfo {year} {2014})}\BibitemShut {NoStop}%
\bibitem [{\citenamefont {Hayen}\ \emph {et~al.}(2019)\citenamefont {Hayen},
  \citenamefont {Kostensalo}, \citenamefont {Severijns},\ and\ \citenamefont
  {Suhonen}}]{HAYEN_1F}%
  \BibitemOpen
  \bibfield  {author} {\bibinfo {author} {\bibfnamefont {L.}~\bibnamefont
  {Hayen}}, \bibinfo {author} {\bibfnamefont {J.}~\bibnamefont {Kostensalo}},
  \bibinfo {author} {\bibfnamefont {N.}~\bibnamefont {Severijns}},\ and\
  \bibinfo {author} {\bibfnamefont {J.}~\bibnamefont {Suhonen}},\ }\href
  {https://doi.org/10.1103/PhysRevC.100.054323} {\bibfield  {journal} {\bibinfo
   {journal} {Phys. Rev. C}\ }\textbf {\bibinfo {volume} {100}},\ \bibinfo
  {pages} {054323} (\bibinfo {year} {2019})}\BibitemShut {NoStop}%
\bibitem [{\citenamefont {Kopeikin}\ \emph {et~al.}(2021)\citenamefont
  {Kopeikin}, \citenamefont {Skorokhvatov},\ and\ \citenamefont
  {Titov}}]{KOPEIKIN}%
  \BibitemOpen
  \bibfield  {author} {\bibinfo {author} {\bibfnamefont {V.}~\bibnamefont
  {Kopeikin}}, \bibinfo {author} {\bibfnamefont {M.}~\bibnamefont
  {Skorokhvatov}},\ and\ \bibinfo {author} {\bibfnamefont {O.}~\bibnamefont
  {Titov}},\ }\href {https://doi.org/10.1103/PhysRevD.104.L071301} {\bibfield
  {journal} {\bibinfo  {journal} {Phys. Rev. D}\ }\textbf {\bibinfo {volume}
  {104}},\ \bibinfo {pages} {L071301} (\bibinfo {year} {2021})}\BibitemShut
  {NoStop}%
\bibitem [{\citenamefont {Abe}\ \emph {et~al.}(2014)\citenamefont {Abe},
  \citenamefont {dos Anjos}, \citenamefont {Barriere}, \citenamefont {Baussan},
  \citenamefont {Bekman}, \citenamefont {Bergevin}, \citenamefont {Bezerra},
  \citenamefont {Bezrukov}, \citenamefont {Blucher}, \citenamefont {Buck},
  \citenamefont {Busenitz}, \citenamefont {Cabrera}, \citenamefont {Caden},
  \citenamefont {Camilleri}, \citenamefont {Carr}, \citenamefont {Cerrada},
  \citenamefont {Chang}, \citenamefont {Chauveau}, \citenamefont {Chimenti},
  \citenamefont {Collin}, \citenamefont {Conover}, \citenamefont {Conrad},
  \citenamefont {Crespo-Anad{\'o}n}, \citenamefont {Crum}, \citenamefont
  {Cucoanes}, \citenamefont {Damon}, \citenamefont {Dawson}, \citenamefont
  {Dhooghe}, \citenamefont {Dietrich}, \citenamefont {Djurcic}, \citenamefont
  {Dracos}, \citenamefont {Elnimr}, \citenamefont {Etenko}, \citenamefont
  {Fallot}, \citenamefont {von Feilitzsch}, \citenamefont {Felde},
  \citenamefont {Fernandes}, \citenamefont {Fischer}, \citenamefont {Franco},
  \citenamefont {Franke}, \citenamefont {Furuta}, \citenamefont {Gil-Botella},
  \citenamefont {Giot}, \citenamefont {G{\"o}ger-Neff}, \citenamefont
  {Gonzalez}, \citenamefont {Goodenough}, \citenamefont {Goodman},
  \citenamefont {Grant}, \citenamefont {Haag}, \citenamefont {Hara},
  \citenamefont {Haser}, \citenamefont {Hofmann}, \citenamefont {Horton-Smith},
  \citenamefont {Hourlier}, \citenamefont {Ishitsuka}, \citenamefont {Jochum},
  \citenamefont {Jollet}, \citenamefont {Kaether}, \citenamefont {Kalousis},
  \citenamefont {Kamyshkov}, \citenamefont {Kaplan}, \citenamefont {Kawasaki},
  \citenamefont {Kemp}, \citenamefont {de~Kerret}, \citenamefont {Kryn},
  \citenamefont {Kuze}, \citenamefont {Lachenmaier}, \citenamefont {Lane},
  \citenamefont {Lasserre}, \citenamefont {Letourneau}, \citenamefont
  {Lhuillier}, \citenamefont {Lima}, \citenamefont {Lindner}, \citenamefont
  {L{\'o}pez-Casta{\~n}o}, \citenamefont {LoSecco}, \citenamefont
  {Lubsandorzhiev}, \citenamefont {Lucht}, \citenamefont {Maeda}, \citenamefont
  {Mariani}, \citenamefont {Maricic}, \citenamefont {Martino}, \citenamefont
  {Matsubara}, \citenamefont {Mention}, \citenamefont {Meregaglia},
  \citenamefont {Miletic}, \citenamefont {Milincic}, \citenamefont {Minotti},
  \citenamefont {Nagasaka}, \citenamefont {Nikitenko}, \citenamefont {Novella},
  \citenamefont {Oberauer}, \citenamefont {Obolensky}, \citenamefont {Onillon},
  \citenamefont {Osborn}, \citenamefont {Palomares}, \citenamefont {Pepe},
  \citenamefont {Perasso}, \citenamefont {Pfahler}, \citenamefont {Porta},
  \citenamefont {Pronost}, \citenamefont {Reichenbacher}, \citenamefont
  {Reinhold}, \citenamefont {R{\"o}hling}, \citenamefont {Roncin},
  \citenamefont {Roth}, \citenamefont {Rybolt}, \citenamefont {Sakamoto},
  \citenamefont {Santorelli}, \citenamefont {Schilithz}, \citenamefont
  {Sch{\"o}nert}, \citenamefont {Schoppmann}, \citenamefont {Shaevitz},
  \citenamefont {Sharankova}, \citenamefont {Shimojima}, \citenamefont
  {Shrestha}, \citenamefont {Sibille}, \citenamefont {Sinev}, \citenamefont
  {Skorokhvatov}, \citenamefont {Smith}, \citenamefont {Spitz}, \citenamefont
  {Stahl}, \citenamefont {Stancu}, \citenamefont {Stokes}, \citenamefont
  {Strait}, \citenamefont {St{\"u}ken}, \citenamefont {Suekane}, \citenamefont
  {Sukhotin}, \citenamefont {Sumiyoshi}, \citenamefont {Sun}, \citenamefont
  {Svoboda}, \citenamefont {Terao}, \citenamefont {Tonazzo}, \citenamefont
  {Thi}, \citenamefont {Valdiviesso}, \citenamefont {Vassilopoulos},
  \citenamefont {Veyssiere}, \citenamefont {Vivier}, \citenamefont {Wagner},
  \citenamefont {Walsh}, \citenamefont {Watanabe}, \citenamefont {Wiebusch},
  \citenamefont {Winslow}, \citenamefont {Wurm}, \citenamefont {Yang},
  \citenamefont {Yermia},\ and\ \citenamefont {Zimmer}}]{DOUBLECHOOZ_bump}%
  \BibitemOpen
  \bibfield  {author} {\bibinfo {author} {\bibfnamefont {Y.}~\bibnamefont
  {Abe}}, \bibinfo {author} {\bibfnamefont {J.~C.}\ \bibnamefont {dos Anjos}},
  \bibinfo {author} {\bibfnamefont {J.~C.}\ \bibnamefont {Barriere}}, \bibinfo
  {author} {\bibfnamefont {E.}~\bibnamefont {Baussan}}, \bibinfo {author}
  {\bibfnamefont {I.}~\bibnamefont {Bekman}}, \bibinfo {author} {\bibfnamefont
  {M.}~\bibnamefont {Bergevin}}, \bibinfo {author} {\bibfnamefont {T.~J.~C.}\
  \bibnamefont {Bezerra}}, \bibinfo {author} {\bibfnamefont {L.}~\bibnamefont
  {Bezrukov}}, \bibinfo {author} {\bibfnamefont {E.}~\bibnamefont {Blucher}},
  \bibinfo {author} {\bibfnamefont {C.}~\bibnamefont {Buck}}, \bibinfo {author}
  {\bibfnamefont {J.}~\bibnamefont {Busenitz}}, \bibinfo {author}
  {\bibfnamefont {A.}~\bibnamefont {Cabrera}}, \bibinfo {author} {\bibfnamefont
  {E.}~\bibnamefont {Caden}}, \bibinfo {author} {\bibfnamefont
  {L.}~\bibnamefont {Camilleri}}, \bibinfo {author} {\bibfnamefont
  {R.}~\bibnamefont {Carr}}, \bibinfo {author} {\bibfnamefont {M.}~\bibnamefont
  {Cerrada}}, \bibinfo {author} {\bibfnamefont {P.~J.}\ \bibnamefont {Chang}},
  \bibinfo {author} {\bibfnamefont {E.}~\bibnamefont {Chauveau}}, \bibinfo
  {author} {\bibfnamefont {P.}~\bibnamefont {Chimenti}}, \bibinfo {author}
  {\bibfnamefont {A.~P.}\ \bibnamefont {Collin}}, \bibinfo {author}
  {\bibfnamefont {E.}~\bibnamefont {Conover}}, \bibinfo {author} {\bibfnamefont
  {J.~M.}\ \bibnamefont {Conrad}}, \bibinfo {author} {\bibfnamefont {J.~I.}\
  \bibnamefont {Crespo-Anad{\'o}n}}, \bibinfo {author} {\bibfnamefont
  {K.}~\bibnamefont {Crum}}, \bibinfo {author} {\bibfnamefont {A.~S.}\
  \bibnamefont {Cucoanes}}, \bibinfo {author} {\bibfnamefont {E.}~\bibnamefont
  {Damon}}, \bibinfo {author} {\bibfnamefont {J.~V.}\ \bibnamefont {Dawson}},
  \bibinfo {author} {\bibfnamefont {J.}~\bibnamefont {Dhooghe}}, \bibinfo
  {author} {\bibfnamefont {D.}~\bibnamefont {Dietrich}}, \bibinfo {author}
  {\bibfnamefont {Z.}~\bibnamefont {Djurcic}}, \bibinfo {author} {\bibfnamefont
  {M.}~\bibnamefont {Dracos}}, \bibinfo {author} {\bibfnamefont
  {M.}~\bibnamefont {Elnimr}}, \bibinfo {author} {\bibfnamefont
  {A.}~\bibnamefont {Etenko}}, \bibinfo {author} {\bibfnamefont
  {M.}~\bibnamefont {Fallot}}, \bibinfo {author} {\bibfnamefont
  {F.}~\bibnamefont {von Feilitzsch}}, \bibinfo {author} {\bibfnamefont
  {J.}~\bibnamefont {Felde}}, \bibinfo {author} {\bibfnamefont {S.~M.}\
  \bibnamefont {Fernandes}}, \bibinfo {author} {\bibfnamefont {V.}~\bibnamefont
  {Fischer}}, \bibinfo {author} {\bibfnamefont {D.}~\bibnamefont {Franco}},
  \bibinfo {author} {\bibfnamefont {M.}~\bibnamefont {Franke}}, \bibinfo
  {author} {\bibfnamefont {H.}~\bibnamefont {Furuta}}, \bibinfo {author}
  {\bibfnamefont {I.}~\bibnamefont {Gil-Botella}}, \bibinfo {author}
  {\bibfnamefont {L.}~\bibnamefont {Giot}}, \bibinfo {author} {\bibfnamefont
  {M.}~\bibnamefont {G{\"o}ger-Neff}}, \bibinfo {author} {\bibfnamefont
  {L.~F.~G.}\ \bibnamefont {Gonzalez}}, \bibinfo {author} {\bibfnamefont
  {L.}~\bibnamefont {Goodenough}}, \bibinfo {author} {\bibfnamefont {M.~C.}\
  \bibnamefont {Goodman}}, \bibinfo {author} {\bibfnamefont {C.}~\bibnamefont
  {Grant}}, \bibinfo {author} {\bibfnamefont {N.}~\bibnamefont {Haag}},
  \bibinfo {author} {\bibfnamefont {T.}~\bibnamefont {Hara}}, \bibinfo {author}
  {\bibfnamefont {J.}~\bibnamefont {Haser}}, \bibinfo {author} {\bibfnamefont
  {M.}~\bibnamefont {Hofmann}}, \bibinfo {author} {\bibfnamefont {G.~A.}\
  \bibnamefont {Horton-Smith}}, \bibinfo {author} {\bibfnamefont
  {A.}~\bibnamefont {Hourlier}}, \bibinfo {author} {\bibfnamefont
  {M.}~\bibnamefont {Ishitsuka}}, \bibinfo {author} {\bibfnamefont
  {J.}~\bibnamefont {Jochum}}, \bibinfo {author} {\bibfnamefont
  {C.}~\bibnamefont {Jollet}}, \bibinfo {author} {\bibfnamefont
  {F.}~\bibnamefont {Kaether}}, \bibinfo {author} {\bibfnamefont {L.~N.}\
  \bibnamefont {Kalousis}}, \bibinfo {author} {\bibfnamefont {Y.}~\bibnamefont
  {Kamyshkov}}, \bibinfo {author} {\bibfnamefont {D.~M.}\ \bibnamefont
  {Kaplan}}, \bibinfo {author} {\bibfnamefont {T.}~\bibnamefont {Kawasaki}},
  \bibinfo {author} {\bibfnamefont {E.}~\bibnamefont {Kemp}}, \bibinfo {author}
  {\bibfnamefont {H.}~\bibnamefont {de~Kerret}}, \bibinfo {author}
  {\bibfnamefont {D.}~\bibnamefont {Kryn}}, \bibinfo {author} {\bibfnamefont
  {M.}~\bibnamefont {Kuze}}, \bibinfo {author} {\bibfnamefont {T.}~\bibnamefont
  {Lachenmaier}}, \bibinfo {author} {\bibfnamefont {C.~E.}\ \bibnamefont
  {Lane}}, \bibinfo {author} {\bibfnamefont {T.}~\bibnamefont {Lasserre}},
  \bibinfo {author} {\bibfnamefont {A.}~\bibnamefont {Letourneau}}, \bibinfo
  {author} {\bibfnamefont {D.}~\bibnamefont {Lhuillier}}, \bibinfo {author}
  {\bibfnamefont {H.~P.}\ \bibnamefont {Lima}}, \bibinfo {author}
  {\bibfnamefont {M.}~\bibnamefont {Lindner}}, \bibinfo {author} {\bibfnamefont
  {J.~M.}\ \bibnamefont {L{\'o}pez-Casta{\~n}o}}, \bibinfo {author}
  {\bibfnamefont {J.~M.}\ \bibnamefont {LoSecco}}, \bibinfo {author}
  {\bibfnamefont {B.}~\bibnamefont {Lubsandorzhiev}}, \bibinfo {author}
  {\bibfnamefont {S.}~\bibnamefont {Lucht}}, \bibinfo {author} {\bibfnamefont
  {J.}~\bibnamefont {Maeda}}, \bibinfo {author} {\bibfnamefont
  {C.}~\bibnamefont {Mariani}}, \bibinfo {author} {\bibfnamefont
  {J.}~\bibnamefont {Maricic}}, \bibinfo {author} {\bibfnamefont
  {J.}~\bibnamefont {Martino}}, \bibinfo {author} {\bibfnamefont
  {T.}~\bibnamefont {Matsubara}}, \bibinfo {author} {\bibfnamefont
  {G.}~\bibnamefont {Mention}}, \bibinfo {author} {\bibfnamefont
  {A.}~\bibnamefont {Meregaglia}}, \bibinfo {author} {\bibfnamefont
  {T.}~\bibnamefont {Miletic}}, \bibinfo {author} {\bibfnamefont
  {R.}~\bibnamefont {Milincic}}, \bibinfo {author} {\bibfnamefont
  {A.}~\bibnamefont {Minotti}}, \bibinfo {author} {\bibfnamefont
  {Y.}~\bibnamefont {Nagasaka}}, \bibinfo {author} {\bibfnamefont
  {Y.}~\bibnamefont {Nikitenko}}, \bibinfo {author} {\bibfnamefont
  {P.}~\bibnamefont {Novella}}, \bibinfo {author} {\bibfnamefont
  {L.}~\bibnamefont {Oberauer}}, \bibinfo {author} {\bibfnamefont
  {M.}~\bibnamefont {Obolensky}}, \bibinfo {author} {\bibfnamefont
  {A.}~\bibnamefont {Onillon}}, \bibinfo {author} {\bibfnamefont
  {A.}~\bibnamefont {Osborn}}, \bibinfo {author} {\bibfnamefont
  {C.}~\bibnamefont {Palomares}}, \bibinfo {author} {\bibfnamefont {I.~M.}\
  \bibnamefont {Pepe}}, \bibinfo {author} {\bibfnamefont {S.}~\bibnamefont
  {Perasso}}, \bibinfo {author} {\bibfnamefont {P.}~\bibnamefont {Pfahler}},
  \bibinfo {author} {\bibfnamefont {A.}~\bibnamefont {Porta}}, \bibinfo
  {author} {\bibfnamefont {G.}~\bibnamefont {Pronost}}, \bibinfo {author}
  {\bibfnamefont {J.}~\bibnamefont {Reichenbacher}}, \bibinfo {author}
  {\bibfnamefont {B.}~\bibnamefont {Reinhold}}, \bibinfo {author}
  {\bibfnamefont {M.}~\bibnamefont {R{\"o}hling}}, \bibinfo {author}
  {\bibfnamefont {R.}~\bibnamefont {Roncin}}, \bibinfo {author} {\bibfnamefont
  {S.}~\bibnamefont {Roth}}, \bibinfo {author} {\bibfnamefont {B.}~\bibnamefont
  {Rybolt}}, \bibinfo {author} {\bibfnamefont {Y.}~\bibnamefont {Sakamoto}},
  \bibinfo {author} {\bibfnamefont {R.}~\bibnamefont {Santorelli}}, \bibinfo
  {author} {\bibfnamefont {A.~C.}\ \bibnamefont {Schilithz}}, \bibinfo {author}
  {\bibfnamefont {S.}~\bibnamefont {Sch{\"o}nert}}, \bibinfo {author}
  {\bibfnamefont {S.}~\bibnamefont {Schoppmann}}, \bibinfo {author}
  {\bibfnamefont {M.~H.}\ \bibnamefont {Shaevitz}}, \bibinfo {author}
  {\bibfnamefont {R.}~\bibnamefont {Sharankova}}, \bibinfo {author}
  {\bibfnamefont {S.}~\bibnamefont {Shimojima}}, \bibinfo {author}
  {\bibfnamefont {D.}~\bibnamefont {Shrestha}}, \bibinfo {author}
  {\bibfnamefont {V.}~\bibnamefont {Sibille}}, \bibinfo {author} {\bibfnamefont
  {V.}~\bibnamefont {Sinev}}, \bibinfo {author} {\bibfnamefont
  {M.}~\bibnamefont {Skorokhvatov}}, \bibinfo {author} {\bibfnamefont
  {E.}~\bibnamefont {Smith}}, \bibinfo {author} {\bibfnamefont
  {J.}~\bibnamefont {Spitz}}, \bibinfo {author} {\bibfnamefont
  {A.}~\bibnamefont {Stahl}}, \bibinfo {author} {\bibfnamefont
  {I.}~\bibnamefont {Stancu}}, \bibinfo {author} {\bibfnamefont {L.~F.~F.}\
  \bibnamefont {Stokes}}, \bibinfo {author} {\bibfnamefont {M.}~\bibnamefont
  {Strait}}, \bibinfo {author} {\bibfnamefont {A.}~\bibnamefont {St{\"u}ken}},
  \bibinfo {author} {\bibfnamefont {F.}~\bibnamefont {Suekane}}, \bibinfo
  {author} {\bibfnamefont {S.}~\bibnamefont {Sukhotin}}, \bibinfo {author}
  {\bibfnamefont {T.}~\bibnamefont {Sumiyoshi}}, \bibinfo {author}
  {\bibfnamefont {Y.}~\bibnamefont {Sun}}, \bibinfo {author} {\bibfnamefont
  {R.}~\bibnamefont {Svoboda}}, \bibinfo {author} {\bibfnamefont
  {K.}~\bibnamefont {Terao}}, \bibinfo {author} {\bibfnamefont
  {A.}~\bibnamefont {Tonazzo}}, \bibinfo {author} {\bibfnamefont {H.~H.~T.}\
  \bibnamefont {Thi}}, \bibinfo {author} {\bibfnamefont {G.}~\bibnamefont
  {Valdiviesso}}, \bibinfo {author} {\bibfnamefont {N.}~\bibnamefont
  {Vassilopoulos}}, \bibinfo {author} {\bibfnamefont {C.}~\bibnamefont
  {Veyssiere}}, \bibinfo {author} {\bibfnamefont {M.}~\bibnamefont {Vivier}},
  \bibinfo {author} {\bibfnamefont {S.}~\bibnamefont {Wagner}}, \bibinfo
  {author} {\bibfnamefont {N.}~\bibnamefont {Walsh}}, \bibinfo {author}
  {\bibfnamefont {H.}~\bibnamefont {Watanabe}}, \bibinfo {author}
  {\bibfnamefont {C.}~\bibnamefont {Wiebusch}}, \bibinfo {author}
  {\bibfnamefont {L.}~\bibnamefont {Winslow}}, \bibinfo {author} {\bibfnamefont
  {M.}~\bibnamefont {Wurm}}, \bibinfo {author} {\bibfnamefont {G.}~\bibnamefont
  {Yang}}, \bibinfo {author} {\bibfnamefont {F.}~\bibnamefont {Yermia}},\ and\
  \bibinfo {author} {\bibfnamefont {V.}~\bibnamefont {Zimmer}} (\bibinfo
  {collaboration} {The Double Chooz collaboration}),\ }\href
  {https://doi.org/10.1007/JHEP10(2014)086} {\bibfield  {journal} {\bibinfo
  {journal} {J. High Energ. Phys.}\ }\textbf {\bibinfo {volume} {2014}}\bibinfo
   {number} { (10)},\ \bibinfo {pages} {86}}\BibitemShut {NoStop}%
\bibitem [{MEE(2019)}]{MEETING1}%
  \BibitemOpen
\bibfield  {number} {  }\href@noop {} {\bibinfo {title} {{Technical Meeting on
  Antineutrino Spectra and Applications, International Atomic Energy Agency
  (IAEA), 2019), INDC-NDS-0786 Report}}} (\bibinfo {year} {2019})\BibitemShut
  {NoStop}%
\bibitem [{MEE(2023)}]{MEETING2}%
  \BibitemOpen
  \href@noop {} {\bibinfo {title} {{2nd IAEA Technical Meeting on Nuclear Data
  Needs for Antineutrino Spectra Applications, IAEA}}} (\bibinfo {year}
  {2023})\BibitemShut {NoStop}%
\bibitem [{MEE(2020)}]{MEETING3}%
  \BibitemOpen
  \href@noop {} {\bibinfo {title} {{Prospect Collaboration Workshop, on-line}}}
  (\bibinfo {year} {2020})\BibitemShut {NoStop}%
\bibitem [{\citenamefont {Al~Kharusi}\ \emph {et~al.}(2020)\citenamefont
  {Al~Kharusi}, \citenamefont {Anton}, \citenamefont {Badhrees}, \citenamefont
  {Barbeau}, \citenamefont {Beck}, \citenamefont {Belov}, \citenamefont
  {Bhatta}, \citenamefont {Breidenbach}, \citenamefont {Brunner}, \citenamefont
  {Cao}, \citenamefont {Cen}, \citenamefont {Chambers}, \citenamefont
  {Cleveland}, \citenamefont {Coon}, \citenamefont {Craycraft}, \citenamefont
  {Daniels}, \citenamefont {Darroch}, \citenamefont {Daugherty}, \citenamefont
  {Davis}, \citenamefont {Delaquis}, \citenamefont {Der Mesrobian-Kabakian},
  \citenamefont {DeVoe}, \citenamefont {Dilling}, \citenamefont {Dolgolenko},
  \citenamefont {Dolinski}, \citenamefont {Echevers}, \citenamefont {Fairbank},
  \citenamefont {Fairbank}, \citenamefont {Farine}, \citenamefont {Feyzbakhsh},
  \citenamefont {Fierlinger}, \citenamefont {Fudenberg}, \citenamefont
  {Gautam}, \citenamefont {Gornea}, \citenamefont {Gratta}, \citenamefont
  {Hall}, \citenamefont {Hansen}, \citenamefont {Hoessl}, \citenamefont
  {Hufschmidt}, \citenamefont {Hughes}, \citenamefont {Iverson}, \citenamefont
  {Jamil}, \citenamefont {Jessiman}, \citenamefont {Jewell}, \citenamefont
  {Johnson}, \citenamefont {Karelin}, \citenamefont {Kaufman}, \citenamefont
  {Koffas}, \citenamefont {Kostensalo}, \citenamefont {Kr\"ucken},
  \citenamefont {Kuchenkov}, \citenamefont {Kumar}, \citenamefont {Lan},
  \citenamefont {Larson}, \citenamefont {Lenardo}, \citenamefont {Leonard},
  \citenamefont {Li}, \citenamefont {Li}, \citenamefont {Li}, \citenamefont
  {Licciardi}, \citenamefont {Lin}, \citenamefont {MacLellan}, \citenamefont
  {McElroy}, \citenamefont {Michel}, \citenamefont {Mong}, \citenamefont
  {Moore}, \citenamefont {Murray}, \citenamefont {Nakarmi}, \citenamefont
  {Njoya}, \citenamefont {Nusair}, \citenamefont {Odian}, \citenamefont
  {Ostrovskiy}, \citenamefont {Piepke}, \citenamefont {Pocar}, \citenamefont
  {Reti\`ere}, \citenamefont {Robinson}, \citenamefont {Rowson}, \citenamefont
  {Ruddell}, \citenamefont {Runge}, \citenamefont {Schmidt}, \citenamefont
  {Sinclair}, \citenamefont {Skarpaas}, \citenamefont {Soma}, \citenamefont
  {Stekhanov}, \citenamefont {Suhonen}, \citenamefont {Tarka}, \citenamefont
  {Thibado}, \citenamefont {Todd}, \citenamefont {Tolba}, \citenamefont
  {Totev}, \citenamefont {Tsang}, \citenamefont {Veenstra}, \citenamefont
  {Veeraraghavan}, \citenamefont {Vogel}, \citenamefont {Vuilleumier},
  \citenamefont {Wagenpfeil}, \citenamefont {Watkins}, \citenamefont {Weber},
  \citenamefont {Wen}, \citenamefont {Wichoski}, \citenamefont {Wrede},
  \citenamefont {Wu}, \citenamefont {Xia}, \citenamefont {Yahne}, \citenamefont
  {Yang}, \citenamefont {Yen}, \citenamefont {Zeldovich},\ and\ \citenamefont
  {Ziegler}}]{EXO200}%
  \BibitemOpen
  \bibfield  {author} {\bibinfo {author} {\bibfnamefont {S.}~\bibnamefont
  {Al~Kharusi}}, \bibinfo {author} {\bibfnamefont {G.}~\bibnamefont {Anton}},
  \bibinfo {author} {\bibfnamefont {I.}~\bibnamefont {Badhrees}}, \bibinfo
  {author} {\bibfnamefont {P.~S.}\ \bibnamefont {Barbeau}}, \bibinfo {author}
  {\bibfnamefont {D.}~\bibnamefont {Beck}}, \bibinfo {author} {\bibfnamefont
  {V.}~\bibnamefont {Belov}}, \bibinfo {author} {\bibfnamefont
  {T.}~\bibnamefont {Bhatta}}, \bibinfo {author} {\bibfnamefont
  {M.}~\bibnamefont {Breidenbach}}, \bibinfo {author} {\bibfnamefont
  {T.}~\bibnamefont {Brunner}}, \bibinfo {author} {\bibfnamefont {G.~F.}\
  \bibnamefont {Cao}}, \bibinfo {author} {\bibfnamefont {W.~R.}\ \bibnamefont
  {Cen}}, \bibinfo {author} {\bibfnamefont {C.}~\bibnamefont {Chambers}},
  \bibinfo {author} {\bibfnamefont {B.}~\bibnamefont {Cleveland}}, \bibinfo
  {author} {\bibfnamefont {M.}~\bibnamefont {Coon}}, \bibinfo {author}
  {\bibfnamefont {A.}~\bibnamefont {Craycraft}}, \bibinfo {author}
  {\bibfnamefont {T.}~\bibnamefont {Daniels}}, \bibinfo {author} {\bibfnamefont
  {L.}~\bibnamefont {Darroch}}, \bibinfo {author} {\bibfnamefont {S.~J.}\
  \bibnamefont {Daugherty}}, \bibinfo {author} {\bibfnamefont {J.}~\bibnamefont
  {Davis}}, \bibinfo {author} {\bibfnamefont {S.}~\bibnamefont {Delaquis}},
  \bibinfo {author} {\bibfnamefont {A.}~\bibnamefont {Der Mesrobian-Kabakian}},
  \bibinfo {author} {\bibfnamefont {R.}~\bibnamefont {DeVoe}}, \bibinfo
  {author} {\bibfnamefont {J.}~\bibnamefont {Dilling}}, \bibinfo {author}
  {\bibfnamefont {A.}~\bibnamefont {Dolgolenko}}, \bibinfo {author}
  {\bibfnamefont {M.~J.}\ \bibnamefont {Dolinski}}, \bibinfo {author}
  {\bibfnamefont {J.}~\bibnamefont {Echevers}}, \bibinfo {author}
  {\bibfnamefont {W.}~\bibnamefont {Fairbank}}, \bibinfo {author}
  {\bibfnamefont {D.}~\bibnamefont {Fairbank}}, \bibinfo {author}
  {\bibfnamefont {J.}~\bibnamefont {Farine}}, \bibinfo {author} {\bibfnamefont
  {S.}~\bibnamefont {Feyzbakhsh}}, \bibinfo {author} {\bibfnamefont
  {P.}~\bibnamefont {Fierlinger}}, \bibinfo {author} {\bibfnamefont
  {D.}~\bibnamefont {Fudenberg}}, \bibinfo {author} {\bibfnamefont
  {P.}~\bibnamefont {Gautam}}, \bibinfo {author} {\bibfnamefont
  {R.}~\bibnamefont {Gornea}}, \bibinfo {author} {\bibfnamefont
  {G.}~\bibnamefont {Gratta}}, \bibinfo {author} {\bibfnamefont
  {C.}~\bibnamefont {Hall}}, \bibinfo {author} {\bibfnamefont {E.~V.}\
  \bibnamefont {Hansen}}, \bibinfo {author} {\bibfnamefont {J.}~\bibnamefont
  {Hoessl}}, \bibinfo {author} {\bibfnamefont {P.}~\bibnamefont {Hufschmidt}},
  \bibinfo {author} {\bibfnamefont {M.}~\bibnamefont {Hughes}}, \bibinfo
  {author} {\bibfnamefont {A.}~\bibnamefont {Iverson}}, \bibinfo {author}
  {\bibfnamefont {A.}~\bibnamefont {Jamil}}, \bibinfo {author} {\bibfnamefont
  {C.}~\bibnamefont {Jessiman}}, \bibinfo {author} {\bibfnamefont {M.~J.}\
  \bibnamefont {Jewell}}, \bibinfo {author} {\bibfnamefont {A.}~\bibnamefont
  {Johnson}}, \bibinfo {author} {\bibfnamefont {A.}~\bibnamefont {Karelin}},
  \bibinfo {author} {\bibfnamefont {L.~J.}\ \bibnamefont {Kaufman}}, \bibinfo
  {author} {\bibfnamefont {T.}~\bibnamefont {Koffas}}, \bibinfo {author}
  {\bibfnamefont {J.}~\bibnamefont {Kostensalo}}, \bibinfo {author}
  {\bibfnamefont {R.}~\bibnamefont {Kr\"ucken}}, \bibinfo {author}
  {\bibfnamefont {A.}~\bibnamefont {Kuchenkov}}, \bibinfo {author}
  {\bibfnamefont {K.~S.}\ \bibnamefont {Kumar}}, \bibinfo {author}
  {\bibfnamefont {Y.}~\bibnamefont {Lan}}, \bibinfo {author} {\bibfnamefont
  {A.}~\bibnamefont {Larson}}, \bibinfo {author} {\bibfnamefont {B.~G.}\
  \bibnamefont {Lenardo}}, \bibinfo {author} {\bibfnamefont {D.~S.}\
  \bibnamefont {Leonard}}, \bibinfo {author} {\bibfnamefont {G.~S.}\
  \bibnamefont {Li}}, \bibinfo {author} {\bibfnamefont {S.}~\bibnamefont {Li}},
  \bibinfo {author} {\bibfnamefont {Z.}~\bibnamefont {Li}}, \bibinfo {author}
  {\bibfnamefont {C.}~\bibnamefont {Licciardi}}, \bibinfo {author}
  {\bibfnamefont {Y.~H.}\ \bibnamefont {Lin}}, \bibinfo {author} {\bibfnamefont
  {R.}~\bibnamefont {MacLellan}}, \bibinfo {author} {\bibfnamefont
  {T.}~\bibnamefont {McElroy}}, \bibinfo {author} {\bibfnamefont
  {T.}~\bibnamefont {Michel}}, \bibinfo {author} {\bibfnamefont
  {B.}~\bibnamefont {Mong}}, \bibinfo {author} {\bibfnamefont {D.~C.}\
  \bibnamefont {Moore}}, \bibinfo {author} {\bibfnamefont {K.}~\bibnamefont
  {Murray}}, \bibinfo {author} {\bibfnamefont {P.}~\bibnamefont {Nakarmi}},
  \bibinfo {author} {\bibfnamefont {O.}~\bibnamefont {Njoya}}, \bibinfo
  {author} {\bibfnamefont {O.}~\bibnamefont {Nusair}}, \bibinfo {author}
  {\bibfnamefont {A.}~\bibnamefont {Odian}}, \bibinfo {author} {\bibfnamefont
  {I.}~\bibnamefont {Ostrovskiy}}, \bibinfo {author} {\bibfnamefont
  {A.}~\bibnamefont {Piepke}}, \bibinfo {author} {\bibfnamefont
  {A.}~\bibnamefont {Pocar}}, \bibinfo {author} {\bibfnamefont
  {F.}~\bibnamefont {Reti\`ere}}, \bibinfo {author} {\bibfnamefont {A.~L.}\
  \bibnamefont {Robinson}}, \bibinfo {author} {\bibfnamefont {P.~C.}\
  \bibnamefont {Rowson}}, \bibinfo {author} {\bibfnamefont {D.}~\bibnamefont
  {Ruddell}}, \bibinfo {author} {\bibfnamefont {J.}~\bibnamefont {Runge}},
  \bibinfo {author} {\bibfnamefont {S.}~\bibnamefont {Schmidt}}, \bibinfo
  {author} {\bibfnamefont {D.}~\bibnamefont {Sinclair}}, \bibinfo {author}
  {\bibfnamefont {K.}~\bibnamefont {Skarpaas}}, \bibinfo {author}
  {\bibfnamefont {A.~K.}\ \bibnamefont {Soma}}, \bibinfo {author}
  {\bibfnamefont {V.}~\bibnamefont {Stekhanov}}, \bibinfo {author}
  {\bibfnamefont {J.}~\bibnamefont {Suhonen}}, \bibinfo {author} {\bibfnamefont
  {M.}~\bibnamefont {Tarka}}, \bibinfo {author} {\bibfnamefont
  {S.}~\bibnamefont {Thibado}}, \bibinfo {author} {\bibfnamefont
  {J.}~\bibnamefont {Todd}}, \bibinfo {author} {\bibfnamefont {T.}~\bibnamefont
  {Tolba}}, \bibinfo {author} {\bibfnamefont {T.~I.}\ \bibnamefont {Totev}},
  \bibinfo {author} {\bibfnamefont {R.}~\bibnamefont {Tsang}}, \bibinfo
  {author} {\bibfnamefont {B.}~\bibnamefont {Veenstra}}, \bibinfo {author}
  {\bibfnamefont {V.}~\bibnamefont {Veeraraghavan}}, \bibinfo {author}
  {\bibfnamefont {P.}~\bibnamefont {Vogel}}, \bibinfo {author} {\bibfnamefont
  {J.-L.}\ \bibnamefont {Vuilleumier}}, \bibinfo {author} {\bibfnamefont
  {M.}~\bibnamefont {Wagenpfeil}}, \bibinfo {author} {\bibfnamefont
  {J.}~\bibnamefont {Watkins}}, \bibinfo {author} {\bibfnamefont
  {M.}~\bibnamefont {Weber}}, \bibinfo {author} {\bibfnamefont {L.~J.}\
  \bibnamefont {Wen}}, \bibinfo {author} {\bibfnamefont {U.}~\bibnamefont
  {Wichoski}}, \bibinfo {author} {\bibfnamefont {G.}~\bibnamefont {Wrede}},
  \bibinfo {author} {\bibfnamefont {S.~X.}\ \bibnamefont {Wu}}, \bibinfo
  {author} {\bibfnamefont {Q.}~\bibnamefont {Xia}}, \bibinfo {author}
  {\bibfnamefont {D.~R.}\ \bibnamefont {Yahne}}, \bibinfo {author}
  {\bibfnamefont {L.}~\bibnamefont {Yang}}, \bibinfo {author} {\bibfnamefont
  {Y.-R.}\ \bibnamefont {Yen}}, \bibinfo {author} {\bibfnamefont {O.~Y.}\
  \bibnamefont {Zeldovich}},\ and\ \bibinfo {author} {\bibfnamefont
  {T.}~\bibnamefont {Ziegler}} (\bibinfo {collaboration} {EXO-200
  Collaboration}),\ }\href {https://doi.org/10.1103/PhysRevLett.124.232502}
  {\bibfield  {journal} {\bibinfo  {journal} {Phys. Rev. Lett.}\ }\textbf
  {\bibinfo {volume} {124}},\ \bibinfo {pages} {232502} (\bibinfo {year}
  {2020})}\BibitemShut {NoStop}%
\bibitem [{\citenamefont {Tain}\ and\ \citenamefont {Cano-Ott}(2007)}]{TAIN}%
  \BibitemOpen
  \bibfield  {author} {\bibinfo {author} {\bibfnamefont {J.}~\bibnamefont
  {Tain}}\ and\ \bibinfo {author} {\bibfnamefont {D.}~\bibnamefont
  {Cano-Ott}},\ }\href
  {https://doi.org/https://doi.org/10.1016/j.nima.2006.10.098} {\bibfield
  {journal} {\bibinfo  {journal} {Nucl. Instrum. Meth. A}\ }\textbf {\bibinfo
  {volume} {571}},\ \bibinfo {pages} {728} (\bibinfo {year}
  {2007})}\BibitemShut {NoStop}%
\bibitem [{\citenamefont {Moore}\ \emph {et~al.}(2014)\citenamefont {Moore},
  \citenamefont {Dendooven},\ and\ \citenamefont {Ärje}}]{MOORE}%
  \BibitemOpen
  \bibfield  {author} {\bibinfo {author} {\bibfnamefont {I.~D.}\ \bibnamefont
  {Moore}}, \bibinfo {author} {\bibfnamefont {P.}~\bibnamefont {Dendooven}},\
  and\ \bibinfo {author} {\bibfnamefont {J.}~\bibnamefont {Ärje}},\ }\href
  {https://doi.org/10.1007/s10751-013-0871-0} {\bibfield  {journal} {\bibinfo
  {journal} {Hyperfine Interact.}\ }\textbf {\bibinfo {volume} {223}},\
  \bibinfo {pages} {17–62} (\bibinfo {year} {2014})}\BibitemShut {NoStop}%
\bibitem [{\citenamefont {Äystö}(2001)}]{AYSTO}%
  \BibitemOpen
  \bibfield  {author} {\bibinfo {author} {\bibfnamefont {J.}~\bibnamefont
  {Äystö}},\ }\href
  {https://doi.org/https://doi.org/10.1016/S0375-9474(01)00923-X} {\bibfield
  {journal} {\bibinfo  {journal} {Nucl. Phys. A}\ }\textbf {\bibinfo {volume}
  {693}},\ \bibinfo {pages} {477} (\bibinfo {year} {2001})}\BibitemShut
  {NoStop}%
\bibitem [{\citenamefont {Eronen}\ \emph {et~al.}(2012)\citenamefont {Eronen},
  \citenamefont {Kolhinen}, \citenamefont {Elomaa}, \citenamefont {Gorelov},
  \citenamefont {Hager}, \citenamefont {Hakala}, \citenamefont {Jokinen},
  \citenamefont {Kankainen}, \citenamefont {Karvonen}, \citenamefont {Kopecky},
  \citenamefont {Moore}, \citenamefont {Penttilä}, \citenamefont {Rahaman},
  \citenamefont {Rinta-Antila}, \citenamefont {Rissanen}, \citenamefont
  {Saastamoinen}, \citenamefont {Szerypo}, \citenamefont {Weber},\ and\
  \citenamefont {Äystö}}]{ERONEN}%
  \BibitemOpen
  \bibfield  {author} {\bibinfo {author} {\bibfnamefont {T.}~\bibnamefont
  {Eronen}}, \bibinfo {author} {\bibfnamefont {V.~S.}\ \bibnamefont
  {Kolhinen}}, \bibinfo {author} {\bibfnamefont {V.~V.}\ \bibnamefont
  {Elomaa}}, \bibinfo {author} {\bibfnamefont {D.}~\bibnamefont {Gorelov}},
  \bibinfo {author} {\bibfnamefont {U.}~\bibnamefont {Hager}}, \bibinfo
  {author} {\bibfnamefont {J.}~\bibnamefont {Hakala}}, \bibinfo {author}
  {\bibfnamefont {A.}~\bibnamefont {Jokinen}}, \bibinfo {author} {\bibfnamefont
  {A.}~\bibnamefont {Kankainen}}, \bibinfo {author} {\bibfnamefont
  {P.}~\bibnamefont {Karvonen}}, \bibinfo {author} {\bibfnamefont
  {S.}~\bibnamefont {Kopecky}}, \bibinfo {author} {\bibfnamefont {I.~D.}\
  \bibnamefont {Moore}}, \bibinfo {author} {\bibfnamefont {H.}~\bibnamefont
  {Penttilä}}, \bibinfo {author} {\bibfnamefont {S.}~\bibnamefont {Rahaman}},
  \bibinfo {author} {\bibfnamefont {S.}~\bibnamefont {Rinta-Antila}}, \bibinfo
  {author} {\bibfnamefont {J.}~\bibnamefont {Rissanen}}, \bibinfo {author}
  {\bibfnamefont {A.}~\bibnamefont {Saastamoinen}}, \bibinfo {author}
  {\bibfnamefont {J.}~\bibnamefont {Szerypo}}, \bibinfo {author} {\bibfnamefont
  {C.}~\bibnamefont {Weber}},\ and\ \bibinfo {author} {\bibfnamefont
  {J.}~\bibnamefont {Äystö}},\ }\href
  {https://doi.org/10.1140/epja/i2012-12046-1} {\bibfield  {journal} {\bibinfo
  {journal} {Eur. Phys. J. A}\ }\textbf {\bibinfo {volume} {48}},\ \bibinfo
  {pages} {46} (\bibinfo {year} {2012})}\BibitemShut {NoStop}%
\bibitem [{\citenamefont {Guadilla}\ \emph {et~al.}(2024)\citenamefont
  {Guadilla}, \citenamefont {Algora}, \citenamefont {Estienne}, \citenamefont
  {Fallot}, \citenamefont {Gelletly}, \citenamefont {Porta}, \citenamefont
  {Rigalleau},\ and\ \citenamefont {Stutzmann}}]{GUADILLA}%
  \BibitemOpen
  \bibfield  {author} {\bibinfo {author} {\bibfnamefont {V.}~\bibnamefont
  {Guadilla}}, \bibinfo {author} {\bibfnamefont {A.}~\bibnamefont {Algora}},
  \bibinfo {author} {\bibfnamefont {M.}~\bibnamefont {Estienne}}, \bibinfo
  {author} {\bibfnamefont {M.}~\bibnamefont {Fallot}}, \bibinfo {author}
  {\bibfnamefont {W.}~\bibnamefont {Gelletly}}, \bibinfo {author}
  {\bibfnamefont {A.}~\bibnamefont {Porta}}, \bibinfo {author} {\bibfnamefont
  {L.-M.}\ \bibnamefont {Rigalleau}},\ and\ \bibinfo {author} {\bibfnamefont
  {J.-S.}\ \bibnamefont {Stutzmann}},\ }\href
  {https://doi.org/10.1088/1748-0221/19/02/P02027} {\bibfield  {journal}
  {\bibinfo  {journal} {JINST}\ }\textbf {\bibinfo {volume} {19}},\ \bibinfo
  {pages} {P02027}}\BibitemShut {NoStop}%
\bibitem [{\citenamefont {{D. Etasse et al.}}()}]{FASTER}%
  \BibitemOpen
  \bibfield  {author} {\bibinfo {author} {\bibnamefont {{D. Etasse et al.}}},\
  }\href@noop {} {\bibinfo {title} {{Fast Acquisition SysTem for nuclEar
  Research (FASTER)}}},\ \bibinfo {note} {\url{https://faster.in2p3.fr/}, last
  accessed: January 2025}\BibitemShut {NoStop}%
\bibitem [{\citenamefont {Blachot}(2012)}]{ENSDF_114Ag}%
  \BibitemOpen
  \bibfield  {author} {\bibinfo {author} {\bibfnamefont {J.}~\bibnamefont
  {Blachot}},\ }\href
  {https://doi.org/https://doi.org/10.1016/j.nds.2012.02.002} {\bibfield
  {journal} {\bibinfo  {journal} {Nuclear Data Sheets}\ }\textbf {\bibinfo
  {volume} {113}},\ \bibinfo {pages} {515} (\bibinfo {year}
  {2012})}\BibitemShut {NoStop}%
\bibitem [{\citenamefont {Poole}\ \emph
  {et~al.}(2012{\natexlab{a}})\citenamefont {Poole}, \citenamefont {Cornelius},
  \citenamefont {Trapp},\ and\ \citenamefont {Langton}}]{CADMESH1}%
  \BibitemOpen
  \bibfield  {author} {\bibinfo {author} {\bibfnamefont {C.~M.}\ \bibnamefont
  {Poole}}, \bibinfo {author} {\bibfnamefont {I.}~\bibnamefont {Cornelius}},
  \bibinfo {author} {\bibfnamefont {J.~V.}\ \bibnamefont {Trapp}},\ and\
  \bibinfo {author} {\bibfnamefont {C.~M.}\ \bibnamefont {Langton}},\ }\href
  {https://doi.org/10.1007/s13246-012-0159-8} {\bibfield  {journal} {\bibinfo
  {journal} {Australasian Phys. Eng. Sci. Med.}\ }\textbf {\bibinfo {volume}
  {35}},\ \bibinfo {pages} {329} (\bibinfo {year}
  {2012}{\natexlab{a}})}\BibitemShut {NoStop}%
\bibitem [{\citenamefont {Poole}\ \emph
  {et~al.}(2012{\natexlab{b}})\citenamefont {Poole}, \citenamefont {Cornelius},
  \citenamefont {Trapp},\ and\ \citenamefont {Langton}}]{CADMESH2}%
  \BibitemOpen
  \bibfield  {author} {\bibinfo {author} {\bibfnamefont {C.~M.}\ \bibnamefont
  {Poole}}, \bibinfo {author} {\bibfnamefont {I.}~\bibnamefont {Cornelius}},
  \bibinfo {author} {\bibfnamefont {J.~V.}\ \bibnamefont {Trapp}},\ and\
  \bibinfo {author} {\bibfnamefont {C.~M.}\ \bibnamefont {Langton}},\ }\href
  {https://doi.org/10.1109/TNS.2012.2197415} {\bibfield  {journal} {\bibinfo
  {journal} {IEEE Trans. Nucl. Sci.}\ }\textbf {\bibinfo {volume} {59}},\
  \bibinfo {pages} {1695} (\bibinfo {year} {2012}{\natexlab{b}})}\BibitemShut
  {NoStop}%
\bibitem [{\citenamefont {Hayen}\ \emph {et~al.}(2018)\citenamefont {Hayen},
  \citenamefont {Severijns}, \citenamefont {Bodek}, \citenamefont {Rozpedzik},\
  and\ \citenamefont {Mougeot}}]{HAYEN_Allowed}%
  \BibitemOpen
  \bibfield  {author} {\bibinfo {author} {\bibfnamefont {L.}~\bibnamefont
  {Hayen}}, \bibinfo {author} {\bibfnamefont {N.}~\bibnamefont {Severijns}},
  \bibinfo {author} {\bibfnamefont {K.}~\bibnamefont {Bodek}}, \bibinfo
  {author} {\bibfnamefont {D.}~\bibnamefont {Rozpedzik}},\ and\ \bibinfo
  {author} {\bibfnamefont {X.}~\bibnamefont {Mougeot}},\ }\href
  {https://doi.org/10.1103/RevModPhys.90.015008} {\bibfield  {journal}
  {\bibinfo  {journal} {Rev. Mod. Phys.}\ }\textbf {\bibinfo {volume} {90}},\
  \bibinfo {pages} {015008} (\bibinfo {year} {2018})}\BibitemShut {NoStop}%
\bibitem [{\citenamefont {Alcalá}\ \emph {et~al.}()\citenamefont {Alcalá}
  \emph {et~al.}}]{ALCALA2}%
  \BibitemOpen
  \bibfield  {author} {\bibinfo {author} {\bibfnamefont {G.~A.}\ \bibnamefont
  {Alcalá}} \emph {et~al.},\ }\href@noop {} {}\bibinfo {note} {Manuscript in
  preparation}\BibitemShut {NoStop}%
\bibitem [{\citenamefont {Zakari-Issoufou}\ \emph {et~al.}(2015)\citenamefont
  {Zakari-Issoufou}, \citenamefont {Fallot}, \citenamefont {Porta},
  \citenamefont {Algora}, \citenamefont {Tain}, \citenamefont {Valencia},
  \citenamefont {Rice}, \citenamefont {Bui}, \citenamefont {Cormon},
  \citenamefont {Estienne}, \citenamefont {Agramunt}, \citenamefont
  {\"Ayst\"o}, \citenamefont {Bowry}, \citenamefont {Briz}, \citenamefont
  {Caballero-Folch}, \citenamefont {Cano-Ott}, \citenamefont {Cucoanes},
  \citenamefont {Elomaa}, \citenamefont {Eronen}, \citenamefont {Est\'evez},
  \citenamefont {Farrelly}, \citenamefont {Garcia}, \citenamefont {Gelletly},
  \citenamefont {Gomez-Hornillos}, \citenamefont {Gorlychev}, \citenamefont
  {Hakala}, \citenamefont {Jokinen}, \citenamefont {Jordan}, \citenamefont
  {Kankainen}, \citenamefont {Karvonen}, \citenamefont {Kolhinen},
  \citenamefont {Kondev}, \citenamefont {Martinez}, \citenamefont {Mendoza},
  \citenamefont {Molina}, \citenamefont {Moore}, \citenamefont
  {Perez-Cerd\'an}, \citenamefont {Podoly\'ak}, \citenamefont {Penttil\"a},
  \citenamefont {Regan}, \citenamefont {Reponen}, \citenamefont {Rissanen},
  \citenamefont {Rubio}, \citenamefont {Shiba}, \citenamefont {Sonzogni},\ and\
  \citenamefont {Weber}}]{ZAKARI}%
  \BibitemOpen
  \bibfield  {author} {\bibinfo {author} {\bibfnamefont {A.-A.}\ \bibnamefont
  {Zakari-Issoufou}}, \bibinfo {author} {\bibfnamefont {M.}~\bibnamefont
  {Fallot}}, \bibinfo {author} {\bibfnamefont {A.}~\bibnamefont {Porta}},
  \bibinfo {author} {\bibfnamefont {A.}~\bibnamefont {Algora}}, \bibinfo
  {author} {\bibfnamefont {J.~L.}\ \bibnamefont {Tain}}, \bibinfo {author}
  {\bibfnamefont {E.}~\bibnamefont {Valencia}}, \bibinfo {author}
  {\bibfnamefont {S.}~\bibnamefont {Rice}}, \bibinfo {author} {\bibfnamefont
  {V.~M.}\ \bibnamefont {Bui}}, \bibinfo {author} {\bibfnamefont
  {S.}~\bibnamefont {Cormon}}, \bibinfo {author} {\bibfnamefont
  {M.}~\bibnamefont {Estienne}}, \bibinfo {author} {\bibfnamefont
  {J.}~\bibnamefont {Agramunt}}, \bibinfo {author} {\bibfnamefont
  {J.}~\bibnamefont {\"Ayst\"o}}, \bibinfo {author} {\bibfnamefont
  {M.}~\bibnamefont {Bowry}}, \bibinfo {author} {\bibfnamefont {J.~A.}\
  \bibnamefont {Briz}}, \bibinfo {author} {\bibfnamefont {R.}~\bibnamefont
  {Caballero-Folch}}, \bibinfo {author} {\bibfnamefont {D.}~\bibnamefont
  {Cano-Ott}}, \bibinfo {author} {\bibfnamefont {A.}~\bibnamefont {Cucoanes}},
  \bibinfo {author} {\bibfnamefont {V.-V.}\ \bibnamefont {Elomaa}}, \bibinfo
  {author} {\bibfnamefont {T.}~\bibnamefont {Eronen}}, \bibinfo {author}
  {\bibfnamefont {E.}~\bibnamefont {Est\'evez}}, \bibinfo {author}
  {\bibfnamefont {G.~F.}\ \bibnamefont {Farrelly}}, \bibinfo {author}
  {\bibfnamefont {A.~R.}\ \bibnamefont {Garcia}}, \bibinfo {author}
  {\bibfnamefont {W.}~\bibnamefont {Gelletly}}, \bibinfo {author}
  {\bibfnamefont {M.~B.}\ \bibnamefont {Gomez-Hornillos}}, \bibinfo {author}
  {\bibfnamefont {V.}~\bibnamefont {Gorlychev}}, \bibinfo {author}
  {\bibfnamefont {J.}~\bibnamefont {Hakala}}, \bibinfo {author} {\bibfnamefont
  {A.}~\bibnamefont {Jokinen}}, \bibinfo {author} {\bibfnamefont {M.~D.}\
  \bibnamefont {Jordan}}, \bibinfo {author} {\bibfnamefont {A.}~\bibnamefont
  {Kankainen}}, \bibinfo {author} {\bibfnamefont {P.}~\bibnamefont {Karvonen}},
  \bibinfo {author} {\bibfnamefont {V.~S.}\ \bibnamefont {Kolhinen}}, \bibinfo
  {author} {\bibfnamefont {F.~G.}\ \bibnamefont {Kondev}}, \bibinfo {author}
  {\bibfnamefont {T.}~\bibnamefont {Martinez}}, \bibinfo {author}
  {\bibfnamefont {E.}~\bibnamefont {Mendoza}}, \bibinfo {author} {\bibfnamefont
  {F.}~\bibnamefont {Molina}}, \bibinfo {author} {\bibfnamefont
  {I.}~\bibnamefont {Moore}}, \bibinfo {author} {\bibfnamefont {A.~B.}\
  \bibnamefont {Perez-Cerd\'an}}, \bibinfo {author} {\bibfnamefont
  {Z.}~\bibnamefont {Podoly\'ak}}, \bibinfo {author} {\bibfnamefont
  {H.}~\bibnamefont {Penttil\"a}}, \bibinfo {author} {\bibfnamefont {P.~H.}\
  \bibnamefont {Regan}}, \bibinfo {author} {\bibfnamefont {M.}~\bibnamefont
  {Reponen}}, \bibinfo {author} {\bibfnamefont {J.}~\bibnamefont {Rissanen}},
  \bibinfo {author} {\bibfnamefont {B.}~\bibnamefont {Rubio}}, \bibinfo
  {author} {\bibfnamefont {T.}~\bibnamefont {Shiba}}, \bibinfo {author}
  {\bibfnamefont {A.~A.}\ \bibnamefont {Sonzogni}},\ and\ \bibinfo {author}
  {\bibfnamefont {C.}~\bibnamefont {Weber}} (\bibinfo {collaboration} {IGISOL
  collaboration}),\ }\href {https://doi.org/10.1103/PhysRevLett.115.102503}
  {\bibfield  {journal} {\bibinfo  {journal} {Phys. Rev. Lett.}\ }\textbf
  {\bibinfo {volume} {115}},\ \bibinfo {pages} {102503} (\bibinfo {year}
  {2015})}\BibitemShut {NoStop}%
\bibitem [{\citenamefont {Baglin}(2012)}]{ENSDF_92Rb}%
  \BibitemOpen
  \bibfield  {author} {\bibinfo {author} {\bibfnamefont {C.~M.}\ \bibnamefont
  {Baglin}},\ }\href
  {https://doi.org/https://doi.org/10.1016/j.nds.2012.10.001} {\bibfield
  {journal} {\bibinfo  {journal} {Nucl. Data Sheets}\ }\textbf {\bibinfo
  {volume} {113}},\ \bibinfo {pages} {2187} (\bibinfo {year}
  {2012})}\BibitemShut {NoStop}%
\bibitem [{\citenamefont {Rasco}\ \emph {et~al.}(2016)\citenamefont {Rasco},
  \citenamefont {Woli\ifmmode\acute{n}\else\'{n}\fi{}ska-Cichocka},
  \citenamefont {Fija\l{}kowska}, \citenamefont {Rykaczewski}, \citenamefont
  {Karny}, \citenamefont {Grzywacz}, \citenamefont {Goetz}, \citenamefont
  {Gross}, \citenamefont {Stracener}, \citenamefont {Zganjar}, \citenamefont
  {Batchelder}, \citenamefont {Blackmon}, \citenamefont {Brewer}, \citenamefont
  {Go}, \citenamefont {Heffron}, \citenamefont {King}, \citenamefont {Matta},
  \citenamefont {Miernik}, \citenamefont {Nesaraja}, \citenamefont
  {Paulauskas}, \citenamefont {Rajabali}, \citenamefont {Wang}, \citenamefont
  {Winger}, \citenamefont {Xiao},\ and\ \citenamefont {Zachary}}]{RASCO}%
  \BibitemOpen
  \bibfield  {author} {\bibinfo {author} {\bibfnamefont {B.~C.}\ \bibnamefont
  {Rasco}}, \bibinfo {author} {\bibfnamefont {M.}~\bibnamefont
  {Woli\ifmmode\acute{n}\else\'{n}\fi{}ska-Cichocka}}, \bibinfo {author}
  {\bibfnamefont {A.}~\bibnamefont {Fija\l{}kowska}}, \bibinfo {author}
  {\bibfnamefont {K.~P.}\ \bibnamefont {Rykaczewski}}, \bibinfo {author}
  {\bibfnamefont {M.}~\bibnamefont {Karny}}, \bibinfo {author} {\bibfnamefont
  {R.~K.}\ \bibnamefont {Grzywacz}}, \bibinfo {author} {\bibfnamefont {K.~C.}\
  \bibnamefont {Goetz}}, \bibinfo {author} {\bibfnamefont {C.~J.}\ \bibnamefont
  {Gross}}, \bibinfo {author} {\bibfnamefont {D.~W.}\ \bibnamefont
  {Stracener}}, \bibinfo {author} {\bibfnamefont {E.~F.}\ \bibnamefont
  {Zganjar}}, \bibinfo {author} {\bibfnamefont {J.~C.}\ \bibnamefont
  {Batchelder}}, \bibinfo {author} {\bibfnamefont {J.~C.}\ \bibnamefont
  {Blackmon}}, \bibinfo {author} {\bibfnamefont {N.~T.}\ \bibnamefont
  {Brewer}}, \bibinfo {author} {\bibfnamefont {S.}~\bibnamefont {Go}}, \bibinfo
  {author} {\bibfnamefont {B.}~\bibnamefont {Heffron}}, \bibinfo {author}
  {\bibfnamefont {T.}~\bibnamefont {King}}, \bibinfo {author} {\bibfnamefont
  {J.~T.}\ \bibnamefont {Matta}}, \bibinfo {author} {\bibfnamefont
  {K.}~\bibnamefont {Miernik}}, \bibinfo {author} {\bibfnamefont {C.~D.}\
  \bibnamefont {Nesaraja}}, \bibinfo {author} {\bibfnamefont {S.~V.}\
  \bibnamefont {Paulauskas}}, \bibinfo {author} {\bibfnamefont {M.~M.}\
  \bibnamefont {Rajabali}}, \bibinfo {author} {\bibfnamefont {E.~H.}\
  \bibnamefont {Wang}}, \bibinfo {author} {\bibfnamefont {J.~A.}\ \bibnamefont
  {Winger}}, \bibinfo {author} {\bibfnamefont {Y.}~\bibnamefont {Xiao}},\ and\
  \bibinfo {author} {\bibfnamefont {C.~J.}\ \bibnamefont {Zachary}},\ }\href
  {https://doi.org/10.1103/PhysRevLett.117.092501} {\bibfield  {journal}
  {\bibinfo  {journal} {Phys. Rev. Lett.}\ }\textbf {\bibinfo {volume} {117}},\
  \bibinfo {pages} {092501} (\bibinfo {year} {2016})}\BibitemShut {NoStop}%
\bibitem [{\citenamefont {Johnson}\ \emph {et~al.}(2011)\citenamefont
  {Johnson}, \citenamefont {Symochko}, \citenamefont {Fadil},\ and\
  \citenamefont {Tuli}}]{ENSDF_142Cs}%
  \BibitemOpen
  \bibfield  {author} {\bibinfo {author} {\bibfnamefont {T.}~\bibnamefont
  {Johnson}}, \bibinfo {author} {\bibfnamefont {D.}~\bibnamefont {Symochko}},
  \bibinfo {author} {\bibfnamefont {M.}~\bibnamefont {Fadil}},\ and\ \bibinfo
  {author} {\bibfnamefont {J.}~\bibnamefont {Tuli}},\ }\href
  {https://doi.org/https://doi.org/10.1016/j.nds.2011.08.002} {\bibfield
  {journal} {\bibinfo  {journal} {Nucl. Data Sheets}\ }\textbf {\bibinfo
  {volume} {112}},\ \bibinfo {pages} {1949} (\bibinfo {year}
  {2011})}\BibitemShut {NoStop}%
\bibitem [{\citenamefont {Woli\ifmmode\acute{n}\else\'{n}\fi{}ska-Cichocka}\
  \emph {et~al.}(2023)\citenamefont
  {Woli\ifmmode\acute{n}\else\'{n}\fi{}ska-Cichocka}, \citenamefont {Rasco},
  \citenamefont {Rykaczewski}, \citenamefont {Brewer}, \citenamefont
  {Fija\l{}kowska}, \citenamefont {Karny}, \citenamefont {Grzywacz},
  \citenamefont {Goetz}, \citenamefont {Gross}, \citenamefont {Stracener},
  \citenamefont {Zganjar}, \citenamefont {Batchelder}, \citenamefont
  {Blackmon}, \citenamefont {Go}, \citenamefont {Heffron}, \citenamefont
  {Johnson}, \citenamefont {King}, \citenamefont {Matta}, \citenamefont
  {Miernik}, \citenamefont {Madurga}, \citenamefont {McCutchan}, \citenamefont
  {Miller}, \citenamefont {Nesaraja}, \citenamefont {Paulauskas}, \citenamefont
  {Rajabali}, \citenamefont {Taylor}, \citenamefont {Sonzogni}, \citenamefont
  {Wang}, \citenamefont {Winger}, \citenamefont {Xiao},\ and\ \citenamefont
  {Zachary}}]{WOLINSKA}%
  \BibitemOpen
  \bibfield  {author} {\bibinfo {author} {\bibfnamefont {M.}~\bibnamefont
  {Woli\ifmmode\acute{n}\else\'{n}\fi{}ska-Cichocka}}, \bibinfo {author}
  {\bibfnamefont {B.~C.}\ \bibnamefont {Rasco}}, \bibinfo {author}
  {\bibfnamefont {K.~P.}\ \bibnamefont {Rykaczewski}}, \bibinfo {author}
  {\bibfnamefont {N.~T.}\ \bibnamefont {Brewer}}, \bibinfo {author}
  {\bibfnamefont {A.}~\bibnamefont {Fija\l{}kowska}}, \bibinfo {author}
  {\bibfnamefont {M.}~\bibnamefont {Karny}}, \bibinfo {author} {\bibfnamefont
  {R.~K.}\ \bibnamefont {Grzywacz}}, \bibinfo {author} {\bibfnamefont {K.~C.}\
  \bibnamefont {Goetz}}, \bibinfo {author} {\bibfnamefont {C.~J.}\ \bibnamefont
  {Gross}}, \bibinfo {author} {\bibfnamefont {D.~W.}\ \bibnamefont
  {Stracener}}, \bibinfo {author} {\bibfnamefont {E.~F.}\ \bibnamefont
  {Zganjar}}, \bibinfo {author} {\bibfnamefont {J.~C.}\ \bibnamefont
  {Batchelder}}, \bibinfo {author} {\bibfnamefont {J.~C.}\ \bibnamefont
  {Blackmon}}, \bibinfo {author} {\bibfnamefont {S.}~\bibnamefont {Go}},
  \bibinfo {author} {\bibfnamefont {B.}~\bibnamefont {Heffron}}, \bibinfo
  {author} {\bibfnamefont {J.}~\bibnamefont {Johnson}}, \bibinfo {author}
  {\bibfnamefont {T.~T.}\ \bibnamefont {King}}, \bibinfo {author}
  {\bibfnamefont {J.~T.}\ \bibnamefont {Matta}}, \bibinfo {author}
  {\bibfnamefont {K.}~\bibnamefont {Miernik}}, \bibinfo {author} {\bibfnamefont
  {M.}~\bibnamefont {Madurga}}, \bibinfo {author} {\bibfnamefont {E.~A.}\
  \bibnamefont {McCutchan}}, \bibinfo {author} {\bibfnamefont {D.}~\bibnamefont
  {Miller}}, \bibinfo {author} {\bibfnamefont {C.~D.}\ \bibnamefont
  {Nesaraja}}, \bibinfo {author} {\bibfnamefont {S.~V.}\ \bibnamefont
  {Paulauskas}}, \bibinfo {author} {\bibfnamefont {M.~M.}\ \bibnamefont
  {Rajabali}}, \bibinfo {author} {\bibfnamefont {S.}~\bibnamefont {Taylor}},
  \bibinfo {author} {\bibfnamefont {A.~A.}\ \bibnamefont {Sonzogni}}, \bibinfo
  {author} {\bibfnamefont {E.~H.}\ \bibnamefont {Wang}}, \bibinfo {author}
  {\bibfnamefont {J.~A.}\ \bibnamefont {Winger}}, \bibinfo {author}
  {\bibfnamefont {Y.}~\bibnamefont {Xiao}},\ and\ \bibinfo {author}
  {\bibfnamefont {C.~J.}\ \bibnamefont {Zachary}},\ }\href
  {https://doi.org/10.1103/PhysRevC.107.034303} {\bibfield  {journal} {\bibinfo
   {journal} {Phys. Rev. C}\ }\textbf {\bibinfo {volume} {107}},\ \bibinfo
  {pages} {034303} (\bibinfo {year} {2023})}\BibitemShut {NoStop}%
\bibitem [{\citenamefont {Hayes}\ \emph {et~al.}(2015)\citenamefont {Hayes},
  \citenamefont {Friar}, \citenamefont {Garvey}, \citenamefont {Ibeling},
  \citenamefont {Jungman}, \citenamefont {Kawano},\ and\ \citenamefont
  {Mills}}]{HAYES1}%
  \BibitemOpen
  \bibfield  {author} {\bibinfo {author} {\bibfnamefont {A.~C.}\ \bibnamefont
  {Hayes}}, \bibinfo {author} {\bibfnamefont {J.~L.}\ \bibnamefont {Friar}},
  \bibinfo {author} {\bibfnamefont {G.~T.}\ \bibnamefont {Garvey}}, \bibinfo
  {author} {\bibfnamefont {D.}~\bibnamefont {Ibeling}}, \bibinfo {author}
  {\bibfnamefont {G.}~\bibnamefont {Jungman}}, \bibinfo {author} {\bibfnamefont
  {T.}~\bibnamefont {Kawano}},\ and\ \bibinfo {author} {\bibfnamefont {R.~W.}\
  \bibnamefont {Mills}},\ }\href {https://doi.org/10.1103/PhysRevD.92.033015}
  {\bibfield  {journal} {\bibinfo  {journal} {Phys. Rev. D}\ }\textbf {\bibinfo
  {volume} {92}},\ \bibinfo {pages} {033015} (\bibinfo {year}
  {2015})}\BibitemShut {NoStop}%
\bibitem [{\citenamefont {Guadilla}\ \emph {et~al.}(2020)\citenamefont
  {Guadilla}, \citenamefont {Tain}, \citenamefont {Algora}, \citenamefont
  {Agramunt}, \citenamefont {Jordan}, \citenamefont {Monserrate}, \citenamefont
  {Montaner-Piz\'a}, \citenamefont {Orrigo}, \citenamefont {Rubio},
  \citenamefont {Valencia}, \citenamefont {Briz}, \citenamefont {Cucoanes},
  \citenamefont {Estienne}, \citenamefont {Fallot}, \citenamefont {Le~Meur},
  \citenamefont {Porta}, \citenamefont {Shiba}, \citenamefont
  {Zakari-Issoufou}, \citenamefont {\"Ayst\"o}, \citenamefont {Eronen},
  \citenamefont {Gorelov}, \citenamefont {Hakala}, \citenamefont {Jokinen},
  \citenamefont {Kankainen}, \citenamefont {Kolhinen}, \citenamefont {Koponen},
  \citenamefont {Moore}, \citenamefont {Penttil\"a}, \citenamefont
  {Pohjalainen}, \citenamefont {Reinikainen}, \citenamefont {Reponen},
  \citenamefont {Rinta-Antila}, \citenamefont {Rytk\"onen}, \citenamefont
  {Sonnenschein}, \citenamefont {Voss}, \citenamefont {Fraile}, \citenamefont
  {Vedia}, \citenamefont {Ganio\ifmmode~\breve{g}\else \u{g}\fi{}lu},
  \citenamefont {Gelletly}, \citenamefont {Lebois}, \citenamefont {Wilson},
  \citenamefont {Martinez}, \citenamefont {N\'acher},\ and\ \citenamefont
  {Sonzogni}}]{GUADILLA3}%
  \BibitemOpen
  \bibfield  {author} {\bibinfo {author} {\bibfnamefont {V.}~\bibnamefont
  {Guadilla}}, \bibinfo {author} {\bibfnamefont {J.~L.}\ \bibnamefont {Tain}},
  \bibinfo {author} {\bibfnamefont {A.}~\bibnamefont {Algora}}, \bibinfo
  {author} {\bibfnamefont {J.}~\bibnamefont {Agramunt}}, \bibinfo {author}
  {\bibfnamefont {D.}~\bibnamefont {Jordan}}, \bibinfo {author} {\bibfnamefont
  {M.}~\bibnamefont {Monserrate}}, \bibinfo {author} {\bibfnamefont
  {A.}~\bibnamefont {Montaner-Piz\'a}}, \bibinfo {author} {\bibfnamefont
  {S.~E.~A.}\ \bibnamefont {Orrigo}}, \bibinfo {author} {\bibfnamefont
  {B.}~\bibnamefont {Rubio}}, \bibinfo {author} {\bibfnamefont
  {E.}~\bibnamefont {Valencia}}, \bibinfo {author} {\bibfnamefont {J.~A.}\
  \bibnamefont {Briz}}, \bibinfo {author} {\bibfnamefont {A.}~\bibnamefont
  {Cucoanes}}, \bibinfo {author} {\bibfnamefont {M.}~\bibnamefont {Estienne}},
  \bibinfo {author} {\bibfnamefont {M.}~\bibnamefont {Fallot}}, \bibinfo
  {author} {\bibfnamefont {L.}~\bibnamefont {Le~Meur}}, \bibinfo {author}
  {\bibfnamefont {A.}~\bibnamefont {Porta}}, \bibinfo {author} {\bibfnamefont
  {T.}~\bibnamefont {Shiba}}, \bibinfo {author} {\bibfnamefont {A.-A.}\
  \bibnamefont {Zakari-Issoufou}}, \bibinfo {author} {\bibfnamefont
  {J.}~\bibnamefont {\"Ayst\"o}}, \bibinfo {author} {\bibfnamefont
  {T.}~\bibnamefont {Eronen}}, \bibinfo {author} {\bibfnamefont
  {D.}~\bibnamefont {Gorelov}}, \bibinfo {author} {\bibfnamefont
  {J.}~\bibnamefont {Hakala}}, \bibinfo {author} {\bibfnamefont
  {A.}~\bibnamefont {Jokinen}}, \bibinfo {author} {\bibfnamefont
  {A.}~\bibnamefont {Kankainen}}, \bibinfo {author} {\bibfnamefont {V.~S.}\
  \bibnamefont {Kolhinen}}, \bibinfo {author} {\bibfnamefont {J.}~\bibnamefont
  {Koponen}}, \bibinfo {author} {\bibfnamefont {I.~D.}\ \bibnamefont {Moore}},
  \bibinfo {author} {\bibfnamefont {H.}~\bibnamefont {Penttil\"a}}, \bibinfo
  {author} {\bibfnamefont {I.}~\bibnamefont {Pohjalainen}}, \bibinfo {author}
  {\bibfnamefont {J.}~\bibnamefont {Reinikainen}}, \bibinfo {author}
  {\bibfnamefont {M.}~\bibnamefont {Reponen}}, \bibinfo {author} {\bibfnamefont
  {S.}~\bibnamefont {Rinta-Antila}}, \bibinfo {author} {\bibfnamefont
  {K.}~\bibnamefont {Rytk\"onen}}, \bibinfo {author} {\bibfnamefont
  {V.}~\bibnamefont {Sonnenschein}}, \bibinfo {author} {\bibfnamefont
  {A.}~\bibnamefont {Voss}}, \bibinfo {author} {\bibfnamefont {L.~M.}\
  \bibnamefont {Fraile}}, \bibinfo {author} {\bibfnamefont {V.}~\bibnamefont
  {Vedia}}, \bibinfo {author} {\bibfnamefont {E.}~\bibnamefont
  {Ganio\ifmmode~\breve{g}\else \u{g}\fi{}lu}}, \bibinfo {author}
  {\bibfnamefont {W.}~\bibnamefont {Gelletly}}, \bibinfo {author}
  {\bibfnamefont {M.}~\bibnamefont {Lebois}}, \bibinfo {author} {\bibfnamefont
  {J.~N.}\ \bibnamefont {Wilson}}, \bibinfo {author} {\bibfnamefont
  {T.}~\bibnamefont {Martinez}}, \bibinfo {author} {\bibfnamefont
  {E.}~\bibnamefont {N\'acher}},\ and\ \bibinfo {author} {\bibfnamefont
  {A.~A.}\ \bibnamefont {Sonzogni}},\ }\href
  {https://doi.org/10.1103/PhysRevC.102.064304} {\bibfield  {journal} {\bibinfo
   {journal} {Phys. Rev. C}\ }\textbf {\bibinfo {volume} {102}},\ \bibinfo
  {pages} {064304} (\bibinfo {year} {2020})}\BibitemShut {NoStop}%
\bibitem [{\citenamefont {Guadilla}(2024)}]{GUADILLA_GS}%
  \BibitemOpen
  \bibfield  {author} {\bibinfo {author} {\bibfnamefont {V.}~\bibnamefont
  {Guadilla}},\ }\bibfield  {journal} {\bibinfo  {journal} {Front. Phys.}\
  }\textbf {\bibinfo {volume} {12}},\ \href
  {https://doi.org/10.3389/fphy.2024.1452988} {10.3389/fphy.2024.1452988}
  (\bibinfo {year} {2024})\BibitemShut {NoStop}%
\end{thebibliography}%
  \bibliographystyle{apsrev4-2}

  \newpage
  \vskip -1 cm
  \section{End Matter} \label{sec:EndM}
  \renewcommand\thefigure{EM.\arabic{figure}}
  \setcounter{figure}{0}
  \renewcommand\theequation{EM.\arabic{equation}}
  \setcounter{equation}{0}

  Here we briefly summarize the shape correction factors employed in our comparisons with Hayen {\it et al.} \cite{HAYEN_Allowed} and Huber \cite{HUBER} models, in particular for the allowed transitions. For the first-forbidden transitions, we refer the reader to \cite{HAYEN_1F}.

  Correction factors are multiplied to the standard shape of the beta decay spectrum to improve the description of the energy distribution of the emitted electrons. The corrected beta spectrum is written in natural units as
  \begin{equation}
    \label{eq_as_eq3}
    S_\beta(W) \propto \eta W (W_0 - W)^2\,F(W,Z_d,A)\,C(Z_d, W)\,K(Z_d, W) \;,
  \end{equation}
  {\noindent}where $\eta W (W_0 - W)^2$ is the statistical factor obtained in Fermi theory, being $\eta$, $W$ and $W_0$ the electron momentum, energy and maximum available energy in the decay, respectively. $F(W,Z_d,A)$ is the Fermi function, needed to take into account at first order the Coulomb interaction of the emitted electron with the daughter nucleus characterized by $A$ and $Z_d$. $C(Z_d, W)$ is the ``shape correction factor'', which accounts for the dependence of the beta transition on nuclear structure information . For allowed transitions, the shape factor is taken as 1. For forbidden transitions, the shape correction factors depends on the emitted electrons energy, electron momentum \cite{HAYEN_1F} and matrix elements of various operators (see Table I of Hayes {\it et al.} \cite{HAYES2}). $K(Z_d, W)$ includes the remaining higher order allowed transition corrections. Those are related to atomic and nuclear properties of the daughter nucleus and to the weak interaction itself \cite{HAYEN_Allowed,HUBER}.

  Hayen {\it et al.} \cite{HAYEN_Allowed} proposed several terms for the improvement of the description of allowed transitions that can be tested with our measurements. From those proposed in \cite{HAYEN_Allowed}, only the most relevant factors for the energy range inspected here were selected. The implemented corrections in our model were: finite size of the nucleus, radiative correction with relativistic correction, atomic exchange, atomic mismatch, atomic screening, recoiling nucleus, and distorted Coulomb potential due to recoil. Huber \cite{HUBER} weak magnetism term was also included.

  In our comparisons with the Huber model, all correction terms of \cite{HUBER} for the shape of allowed beta spectra were employed since all considered factors are relevant in the explored energy range. In this case the implemented corrections were: finite size of the nucleus, weak interaction finite size of the nucleus, atomic screening, radiative correction, and the weak magnetism term.

\begin{figure}[htb]
    \resizebox{1.0\columnwidth}{!}{\rotatebox{0}{\includegraphics[clip=]{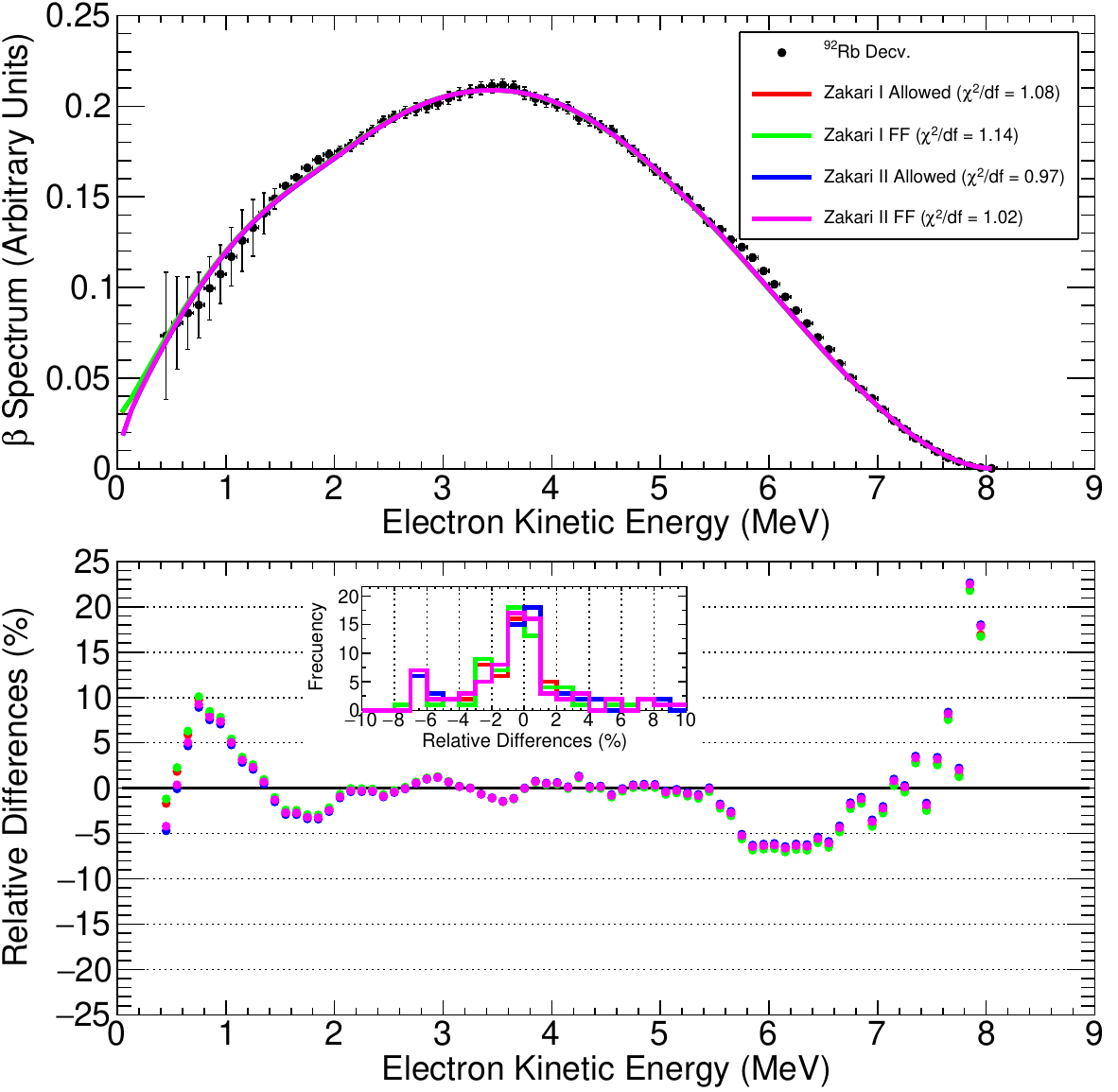}}}
    \caption{Comparison of the deduced beta spectrum of the $^{92}$Rb $\rightarrow$ $^{92}$Sr beta decay ($^{92}$Rb Decv.) with Zakari-Issoufou \textit{et al.} TAGS beta decay feedings \cite{ZAKARI}. The allowed shape corrections from Hayen \textit{et al.} \cite{HAYEN_Allowed} and Huber \cite{HUBER} without and with the the ground state to ground state first-forbidden shape correction factor from Hayen \textit{et al.} \cite{HAYEN_1F} (Zakari I Allowed, Zakari I FF, Zakari II Allowed and Zakari II FF, respectively) are included.}
    \label{fig_Decv92RbEndMatter}
  \end{figure}

  Fig. \ref{fig_Decv92RbEndMatter} shows the comparison of the deconvoluted spectrum of $^{92}$Rb with predictions based on the beta feedings from Zakari-Issoufou et al. \cite{ZAKARI}. The predictions include the allowed corrections from Hayen et al. \cite{HAYEN_Allowed} and Huber \cite{HUBER}, both without and with the first-forbidden shape correction factor for the ground state to ground state transition (0$^-$ → 0$^+$) proposed by Hayen et al. \cite{HAYEN_1F}. This comparison highlights the relatively small impact of the corresponding first-forbidden correction term on the spectral shape.

\end{document}